\documentclass[onecolumn,footinbib,floatfix,aps,prx,superscriptaddress,showpacs,showkeys, notitlepage]{revtex4-1}
\usepackage[usenames,dvipsnames]{xcolor} 
\usepackage[normalem]{ulem}

\usepackage{amsmath,amssymb}
\usepackage{amsfonts}
\usepackage{bm}
\usepackage{bbm}
\usepackage{color}
\usepackage{latexsym}
\usepackage{epstopdf}
\usepackage[english]{babel}
\usepackage{times}
\usepackage{stmaryrd}
\usepackage{psfrag,graphicx}
\usepackage{subfigure}
\usepackage{appendix}
\usepackage{enumitem}
\usepackage{bbold}
\usepackage{placeins}

\newcommand{\be}{\begin{equation}}
\newcommand{\ee}{\end{equation}}
\newcommand{\ben}{\begin{eqnarray}}
\newcommand{\een}{\end{eqnarray}}
\def\tr{{\rm{Tr}}}

\definecolor{mygrey}{gray}{0.35}
\definecolor{myblue}{rgb}{0.2,0.2,0.8}
\definecolor{myzard}{cmyk}{0,0,0.05,0}
\definecolor{mywhite}{rgb}{1,1,1}
\definecolor{myred}{rgb}{0.9,0.1,0.}
\usepackage[colorlinks=true,citecolor=myblue,linkcolor=myred,urlcolor=myblue]{hyperref}

\newcommand{\e}{\operatorname{e}}
\renewcommand{\i}{\mathrm{i}}

\begin{document}

\title{Reaching for the quantum limits in the simultaneous estimation of phase and phase diffusion}

\author{Magdalena Szczykulska}
\thanks{Corresponding author: magdalena.szczykulska@physics.ox.ac.uk}
\affiliation{Clarendon Laboratory, Department of Physics, University of Oxford, OX1 3PU Oxford, United Kingdom}

\author{Tillmann Baumgratz}
\affiliation{Department of Physics, University of Warwick, CV4 7AL Coventry, United Kingdom}
\affiliation{Clarendon Laboratory, Department of Physics, University of Oxford, OX1 3PU Oxford, United Kingdom}

\author{Animesh Datta}
\affiliation{Department of Physics, University of Warwick, CV4 7AL Coventry, United Kingdom}

\begin{abstract}
\vspace{2mm}
\line(1,0){390}
\section*{abstract}

Phase diffusion invariably accompanies all phase estimation strategies -- quantum or classical. A precise estimation of the former can often provide valuable understanding of the physics of the phase generating phenomena itself. We theoretically examine the performance of fixed-particle number probe states in the simultaneous estimation of phase and collective phase diffusion.
We derive analytical quantum limits associated with the simultaneous local estimation of phase and phase diffusion within the quantum Cram{\'e}r-Rao bound framework in the regimes of large and small phase diffusive noise. The former is for a general fixed-particle number state and the latter for Holland Burnett states, for which we show quantum-enhanced estimation of phase as well as phase diffusion. We next investigate the simultaneous attainability of these quantum limits using projective measurements acting on a single copy of the state in terms of a trade-off relation. In particular, we are interested how this trade-off varies as a function of the dimension of the state. We derive an analytical bound for this trade-off in the large phase diffusion regime for a particular form of the measurement, and show that the maximum of $2,$ set by the quantum Cram{\'e}r-Rao bound, is attainable. Further, we show numerical evidence that as diffusion approaches zero, the optimal trade-off relation approaches $1$ for Holland-Burnett states. These numerical results are valid in the small particle number regime and suggest that the trade-off for estimating one parameter with quantum-limited precision leads to a complete lack of precision for the other parameter as the diffusion strength approaches zero. Finally, we provide numerical results showing behaviour of the trade-off for a general value of phase diffusion when using Holland-Burnett probe states.
\vspace{2mm}
\line(1,0){400}
\vspace{1mm}
\end{abstract}


\maketitle

\section{Introduction} 

The central role of unitary evolution in quantum mechanics makes phase estimation a salient problem in precision metrology including the sensing of magnetic and electric fields, changes in refractive index, and measurements of time and displacements. This includes some of the most challenging tasks in physics such as gravitational wave detection~\cite{CavesRMP1980, Caves1981, Blair2016} as well as highly desirable applications such as biological tracking and imaging~\cite{Taylor2014}. Therefore, the understanding of the fundamental limits in the precision of phase estimation set by quantum mechanics is a worthwhile endeavour. Single phase estimation is well understood~\cite{Giovannetti2004, Giovannetti2011,Demkowicz-Dobrzanski2014} and if appropriate quantum probe states are chosen then the associated quantum limits provide a better scaling of the precision in the number of particles or energy of the probe. Additionally, these enhanced scalings can always be attained by a suitable measurement on the evolved probe state.


However, in real scenarios, phase estimation schemes are invariably affected by noise that results in the loss of quantum enhancement. This reverts quantum-enhanced scalings back down to that given by the standard quantum limit~\cite{EscherNatPhys2011, Demkowicz-DobrzanskiNatComm2012, Demkowicz-Dobrzanski2014}. Typical noise channels include particle loss and phase diffusion. Substantial effort has been invested in studying the effect of noise on phase estimation, both theoretically and experimentally in the case of loss~\cite{Demkowicz-Dobrzanski2009, Dorner2009, Kacprowicz2009, Datta2011} as well as phase diffusion~\cite{Brivio2010, Genoni2011, Genoni2012,Degen2016}. The quantum limit (although in general not tight) on the phase estimation precision in the presence of diffusion has also been obtained using a variational approach~\cite{Escher2012}.

Phase diffusion is a dissipationless noise channel that describes phase drifts within the measurement time due to interaction with the environment. Instead of being a nuisance, phase diffusion can, in fact, provide useful information about the physical system that the quantum probe interacts with. This has been labelled as decoherence microscopy~\cite{Cole2009} and has found numerous applications~\cite{Schirhagl2014} including thermometry~\cite{Stace2010,Kucsko2013, Toyli2013, Xie2016}, optomechanical temperature measurements within and beyond the linearised regime~\cite{Aspelmeyer2014,Brawley2016}, and fundamental physics~\cite{Cronin2009}. Phase fluctuations between modes of an optical interferometer are inexorable, and can be caused by mechanical strains and thermal fluctuations inside an optical fiber~\cite{Musha1982, Du2013}. Phase diffusion also arises in matter-wave optics, particularly in confined geometries where its precise estimation can provide information on atom-atom interaction strength~\cite{Berrada2013}. The precision of cold-atom accelerometers is limited by phase diffusion resulting from rotational fluctuations and gravitational forces, whose precise estimation might help alleviate their effect~\cite{Fang2012}, or be used in precise measurement of physical constants~\cite{Rosi2014}. At a more fundamental level, phase diffusion is a parameter of interest in gravitational decoherence models~\cite{Pikovski2015, Oniga2016,Genoni2016} and dark matter detection~\cite{Riedel2013}. Moreover, in circumstances where these parameters vary in time, joint estimation scheme can provide more satisfactory results. 

Numerous scenarios thus exist where a precise knowledge of phase diffusion in addition to the phase parameter can furnish a better understanding of the system under study. Schemes to estimate phase and phase diffusion simultaneously have been studied recently. Theoretical studies on the joint quantum-limited estimation of phase and phase diffusion using a quantum state comprised of a large number of qubits~\cite{Knysh2013} have been undertaken, as well as experimental studies on single qubit systems~\cite{Vidrighin2014,Altorio2015}. While our work deals with collective dephasing, models considering independent dephasing have also been studied in~\cite{Ragy2016}.

The central challenge of joint quantum-limited estimation schemes is the possible non-commutativity of measurements in quantum mechanics. As a result, the quantum limits may not always be attainable, even in principle. The joint estimation scenarios thus inevitably lead to trade-off relations in the attainable precision for the different parameters. Such trade-off relations tell us how much information can be gained about one parameter in the expense of the knowledge about the other parameters~\cite{Szczykulska2016}. This is unlike the case for single parameter estimation, where the quantum-enhanced limit can always be attained~\cite{Helstrom1976, Braunstein1994}. Gill and Massar derived a bound on the trade-off relation in the joint estimation of all the parameters characterising a quantum state of which one possesses many copies~\cite{Gill2000}. Their bound for mixed states is derived for measurements acting on a single copy of the state at a time. Recently, a symmetric measurement saturating the Gill-Massar bound was found~\cite{Li2016}. This measurement is local and equally extracts optimal information about all the parameters, but applies to pure states only. It is important to stress that the Gill-Massar bounding strategy is only tight when the number of parameters is the same as the dimension of the Hilbert space in which the quantum state resides. Since we are interested in the joint estimation of phase and phase diffusion -- only two parameters -- using a multi-dimensional quantum probe state, the Gill-Massar strategy is excessively lax for our purposes. We thus develop new methods to seek tighter trade-off relations.

Our main goal is to understand how this trade-off relation depends on the dimension of the state. Holland-Burnett (HB)~\cite{Holland1993} states form an apt example of FPN states for this purpose. Our strategy for obtaining the trade-off relations for the simultaneous quantum-limited estimation of phase and phase diffusion applies to fixed-particle number (FPN) states. It relies on a Taylor expansion of the full density matrix in the regimes of large and small diffusion. This leads to a tridiagonal density matrix and a low-rank density matrix in the large and small diffusion regimes respectively which enables their analytic diagonalisation. This allows us to overcome the fundamental challenge for calculating the quantum Fisher information (QFI) for mixed quantum states analytically -- their explicit diagonalisation. It is also important to stress that our method differs from strategies undertaken in other works, for instance~\cite{Jarzyna2015} where the diffusion parameter is a constant independent of the dimension of the state. In this work, the validities of the approximations put some dependence between phase diffusion and the total number of particles. Our main results are as follows:
\begin{enumerate}
\item\label{1.} In the large phase diffusion regime, for any FPN state, we provide an analytically closed-form expression for the QFI matrix. See Sec.~(\ref{sec:4.1}).

\item\label{2.} In the large phase diffusion regime, for any FPN state, we provide an analytically closed-form expression for the trade-off quantity 
\be
\tr[\textbf{F}\textbf{H}^{-1}]=\frac{F_{11}}{H_{11}}+\frac{F_{22}}{H_{22}},
\ee
where $\textbf{H}$ and $\textbf{F}$ are the quantum and classical Fisher information matrices respectively, the latter for a particular choice of separable positive operator-valued measurements (POVMs). The maximum value of this quantity is 2, when both parameters are estimated with quantum-limited precision. We show that, in the large diffusion regime, it is possible to attain the optimal trade-off relation of $\tr[\textbf{F}\textbf{H}^{-1}] =2$ for certain FPN states. It therefore follows that the globally optimal trade-off relation with respect to all diffusion values also approaches $2$ for such states. HB states are an example of such states in the large particle number regime. See Sec.~(\ref{sec:4.2}).

\item\label{3.} In the small phase diffusion regime, for any HB states, we provide an analytically closed-form expression for the QFI matrix. Most importantly, these states allow a quantum-enhanced estimation of phase as well as phase diffusion. See Sec.~(\ref{sec:5.1}).

\item\label{4.} In the regime of phase diffusion approaching zero, for HB states of small particle numbers, we present numerical evidence that the maximum value of $\tr[\textbf{F}\textbf{H}^{-1}]$ approaches 1. This implies that, in the simultaneous estimation of phase and diffusion in this regime, when we estimate one parameter at the quantum-limit we gain no precision about the other parameter. See Sec.~(\ref{sec:5.2}).
\end{enumerate} 

In Sec.~(\ref{sec:2}), we give the necessary background for the Cram\'er-Rao inequality and define the setting of our estimation scheme. Sec.~(\ref{sec:3}) provides the structure of the QFI matrix for our problem and the associated bound in the joint estimation of phase and phase diffusion. Importantly, it shows that the QFI matrix and the optimal bound are phase independent, and that the QFI matrix is diagonal. Sec.~(\ref{sec:6}) outlines the numerical results based on a simulated annealing algorithm~\cite{Bohachevsky1986} for the joint information bound valid in all phase diffusion regimes in the case of HB states (small particle numbers). This bound is peaked at an intermediate value of diffusion which, as shown in Sec.~(\ref{sec:4.2}), will approach the upper limit of $2$ set by the quantum Cram{\'e}r-Rao bound for a large number of particles. Finally, Sec.~(\ref{sec:7}) concludes our work.

\section{\label{sec:2} Background}

Throughout the paper, bold quantities will refer to vectors of numbers or more generally to matrices, whereas the circumflex diacritic will denote operators. 

\subsection{\label{sec:2.1} Formalism}

A general estimation process consists of three stages: probe state preparation, interaction with the system containing the parameters of interest and probe readout. After the probe readout stage, the gathered statistics can be analysed by invoking an estimator function to yield an estimate of the parameters. In particular, we consider unbiased estimators for local estimation schemes. It is desirable that an estimator $\tilde\lambda_{i}$ of the $i^{\text{th}}$ parameter $\lambda_{i}$ should have low variance. In the multi-parameter case, variance of a single estimator is replaced by the covariance matrix $\textbf{Cov}$ of the estimators with elements $\operatorname{Cov}_{ij} = \langle\tilde{\lambda}_{i}\tilde{\lambda}_{j}\rangle-\langle\tilde{\lambda}_{i}\rangle\langle\tilde{\lambda}_{j}\rangle$. For $M$ repeated experiments, $\textbf{Cov}$ is lower bounded by
\begin{equation}
M\textbf{Cov}\geq{(\textbf{F})^{-1}}\geq{(\textbf{H})^{-1}},
\label{eqn:CramerRaoBound}
\end{equation}
where $\textbf{F}$ and $\textbf{H}$ are classical and quantum Fisher information matrices, and the first and second inequalities give the classical Cram{\'e}r-Rao bound (CRB) and the quantum Cram{\'e}r-Rao bound (QCRB) respectively ~\cite{Cover2006}. While the multi-parameter CRB is guaranteed to be saturated by the maximum likelihood estimator, the multi-parameter QCRB is not necessarily so due to the possible non-commutativity of optimal measurements~\cite{Szczykulska2016}. Matrices $\textbf{F}$ and $\textbf{H}$ have the elements
\begin{equation}
{F}_{ij}=\sum_{y}\frac{\partial_{\lambda_i}\!\left({p}_{y}\right)\partial_{\lambda_j}\!\left({p}_{y}\right)}{p_y}
\label{eqn:ClassicalFisherInformation}
\end{equation}
and
\begin{equation}
{H}_{ij} = {\operatorname{Re}}[{\operatorname{Tr}}[\hat\varrho_{\bm{\lambda}}\hat{L}_{i}\hat{L}_{j}]]
\label{eqn:QFI}
\end{equation}
respectively, where $\hat\varrho_{\bm{\lambda}}$ is the evolved probe state which depends on the vector of parameters $\bm\lambda$ and ${p}_{y}=\operatorname{Tr}[\hat\Pi_{y}\hat\varrho_{\bm{\lambda}}]$ is the probability associated with the $y^{\text{th}}$ measurement outcome given by the POVM element $\hat\Pi_{y}$. $\operatorname{Re}$ is the real part and $\hat{L}_{i}$ denotes the symmetric logarithmic derivative (SLD) corresponding to parameter $\lambda_{i}$ obeying 
\begin{equation}
2\partial_{\lambda_i}\hat\varrho_{\bm{\lambda}}=\hat{L}_{i}\hat\varrho_{\bm{\lambda}}+\hat\varrho_{\bm{\lambda}}\hat{L}_{i}.
\label{eqn:SLD}
\end{equation}
The elements of $\textbf{H}$ can be written in terms of the eigenbasis of the density matrix $\hat\varrho_{\bm{\lambda}}=\sum_{k}{E_{k}|e_{k}\rangle\langle{e}_{k}|}$ with eigenvalues $E_{k}$ and eigenvectors $|e_k\rangle$, leading to 
\begin{equation}
{H}_{ij}={\operatorname{Re}}\left[\sum_{n}\frac{\partial_{\lambda_{i}}\!(E_n)\partial_{\lambda_{j}}\!(E_n)}{E_n}+4\sum_{n,m}E_m\right.
\left.\frac{(E_n-E_m)^2}{(E_n+E_m)^2}\langle{e}_n|\partial_{\lambda_i}{e_m}\rangle\langle\partial_{\lambda_j}e_{m}|e_n\rangle\right].
\label{eqn:QFIEigenbasis}
\end{equation}
Therefore, calculating QFI requires diagonalising the density matrix which can be a challenging task for high rank matrices. See Appendix~(\ref{app:QFIInTermsOfEigenbasis}) for a derivation of Eqn.~\eqref{eqn:QFIEigenbasis}.

\subsection{\label{sec:2.2} Probe system} 

\begin{figure}[t!]
\begin{center}
\includegraphics[trim=2cm 17cm 2cm 2cm,width=0.9\columnwidth]{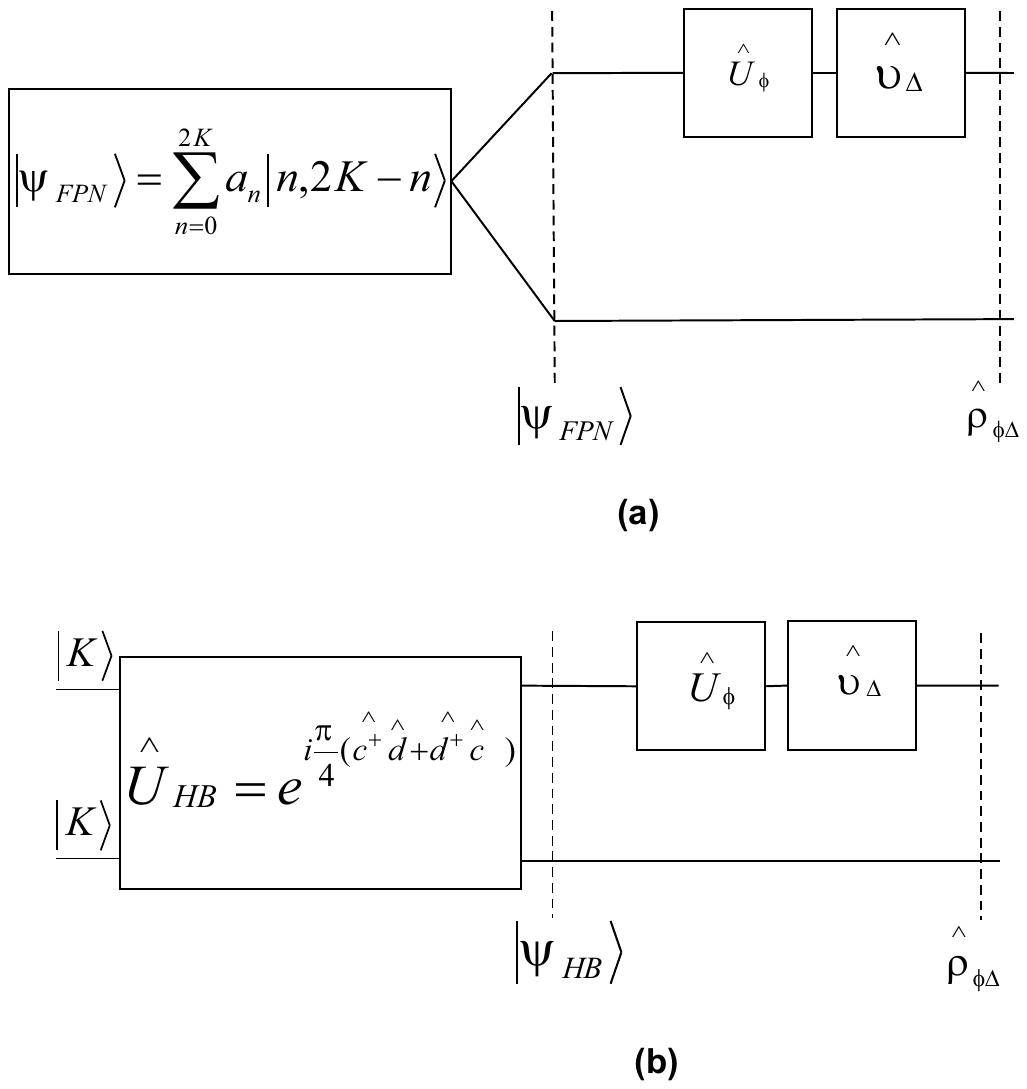}
\end{center}
\caption{Schematic diagram of the investigated system $\textbf{(a)}$ with a general FPN state -- $\hat\varrho_{\phi\Delta}$ has the form of  Eqn.~\eqref{eqn:MixedFixedPhotonNumberState} and $\textbf{(b)}$ with the special case of HB states -- $\hat\varrho_{\phi\Delta}$ has the form of  Eqn.~\eqref{eqn:MixedHBstate2} or  Eqn.~\eqref{eqn:MixedHBstate}. HB states are generated by acting with the unitary $\hat{U}_{HB}$ on two modes (associated with annihilation operators $\hat{c}$ and $\hat{d}$) containing equal number of particles $K$. In photonic systems, this corresponds to interfering two equal Fock states on a 50/50 beam splitter. It is also worth noting that ${\hat{U}}_{\phi}$ and $\hat\nu_{\Delta}$ commute and therefore the order in which phase shift and diffusion channel are applied is irrelevant.}
\label{fig:FPNdiagram}
\end{figure}

In this work, we consider joint estimation of two parameters, phase $\phi$ and collective phase diffusion $\Delta$. The investigated system is shown in Fig.~(\ref{fig:FPNdiagram}). In our studies, the probe system is a  pure FPN state with a total number of $2K$ particles. The mathematical representations of such a state and the special case of HB states are
\begin{equation}
|{\psi_{FPN}}\rangle = \sum_{n=0}^{2K}{a}_{n}|{n},{2K}-{n}\rangle
\label{eqn:FixedPhotonNumberState}
\end{equation}
and
\begin{equation}
|{\psi_{HB}}\rangle = \sum_{n=0}^{2K}{b}_{\frac{n}{2}}\delta_{n,2q}|{n},{2K}-{n}\rangle
\label{eqn:PureHBstate}
\end{equation}
respectively, where $q\in\mathbb{Z}\{0:K\}$. The normalised complex amplitudes, $a_n=|a_n|\e^{\i\theta_n}$, satisfy $\sum_{n=0}^{2K}{|{a}_{n}|}^2=1$ and $b_n=\sqrt{(2n)!(2K-2n)!}/\left(2^{K}n!(K-n)!\right)$ is a special case of $a_n$ which takes non-zero values only for even terms.

The propagator of $\phi$ is a unitary transformation of the form ${\hat{U}}_{\phi}=\e^{\i\phi\hat{n}}$, where $\hat{n}$ is the number operator. The collective phase diffusion channel, $\hat\nu_{\Delta}$, is a random phase shift distributed according to a normal distribution of width $\Delta$ which acting on the probe state $\hat\varrho_{\text{in}}$ gives
\begin{equation}
\hat\varrho_{\Delta}=\hat\nu_{\Delta}(\hat\varrho_{\text{in}})=\frac{1}{\sqrt{2\pi}\Delta}\int_{-\infty}^{+\infty}\operatorname{d}\epsilon\e^{-\frac{\epsilon^2}{2\Delta^2}}\hat{U}_{\epsilon}\hat\varrho_{\text{in}}\hat{U}_{\epsilon}^{\dagger}.
\end{equation}
Its effect is to exponentially erase the off-diagonal elements of the density matrix which also means reducing the information content about the parameters from the probe state.   

The $n^{\text{th}}$ component of the input FPN state acquires the phase term of $\e^{\i{n}\phi}$ and after passing through the phase diffusion channel, the overall state becomes mixed. The resulting density matrix is of the form,
\begin{equation}
\label{eqn:MixedFixedPhotonNumberState}
\hat{\varrho} = \hat{\varrho}_{FPN} = \sum_{n,n'=0}^{2 K}{{a}_{n}{a}_{n'}^*\e^{\i\phi(n-n'){-}\frac{\Delta^{2}}{2}(n-n')^2}}
|{n},{2}{K}-{n}\rangle\langle{n'},{2}{K}-{n'}|.
\end{equation}
The particular instance of HB states can be written as
\begin{equation}
\label{eqn:MixedHBstate2}
\hat{\varrho}_{HB} = \sum_{n,n'=0}^{2K}{\delta_{n,2q}\delta_{n',2q'}{b}_{\frac{n}{2}}{b}_{\frac{n'}{2}}\e^{\i\phi(n-n'){-}{\frac{\Delta^{2}}{2}}(n-n')^2}}
|{n},{2}{K}-{n}\rangle\langle{n'},{2}{K}-{n'}|
\end{equation}
or equivalently
\begin{equation}
\label{eqn:MixedHBstate}
\hat{\varrho}_{HB} = \sum_{n,n'=0}^{ K}{{b}_{n}{b}_{n'}\e^{2\i\phi(n-n'){-}2{\Delta^{2}}(n-n')^2}}
|{2n},{2}{K}-{2n}\rangle\langle{2n'},{2}{K}-{2n'}|,
\end{equation}
where $\{q, q'\}\in{\mathbb{Z}}\{0:K\}$.

\section{\label{sec:3} Attaining the quantum limit}

When solving the SLD equation, Eqn.~\eqref{eqn:SLD}, for a general evolved FPN state, $\hat{\varrho}$, it is useful to work in the basis that contains all the phase information, so that the resulting density matrix contains only real elements. This representation is given in Appendix~(\ref{app:FPNstateNewBasis}). In this basis representation, $|\Gamma_{n,K,\phi,\theta}\rangle \equiv\e^{\i\left(\phi{n}+\theta_{n}\right)}|n, 2K-n\rangle$ where $\theta_n$ is the phase of each component of the input FPN state, the derivatives of $\hat\varrho$ with respect to $\phi$ and $\Delta$ are imaginary and real respectively. The SLD equation is a Lyapunov equation and since $\hat{\varrho}$ is a positive matrix, it has a unique continuous solution\cite{Paris2009} 
\begin{equation}
\hat{L}_{i}=2\int_{0}^{\infty}dt\e^{-\hat\varrho{t}}\partial_{\lambda_i}\hat\varrho\e^{-\hat\varrho{t}},
\label{eqn:LyapunovSolution}
\end{equation}
where $\lambda_i$ labels the different parameters and in our case, $i = 1$ and $i = 2$ correspond to $\phi$ and $\Delta$ respectively. From this form of the solution, we can conclude that $\hat{L}_1$ and $\hat{L}_2$ are imaginary and real respectively in the basis of $|\Gamma_{n,K,\phi,\theta}\rangle$. With this and using Hermiticity of SLDs, $\hat{L}_1$ and $\hat{L}_2$ can be expressed as,
\begin{equation}
\hat{L}_1=\i\sum_{n,n'=0}^{2K}{f_{n,n',\Delta}}|\Gamma_{n,K,\phi,\theta}\rangle\langle{\Gamma_{n',K,\phi,\theta}}|
\label{eqn:SLDPhi}
\end{equation}
and
\begin{equation}
\hat{L}_2=\sum_{n,n'=0}^{2K}{g_{n,n',\Delta}}|\Gamma_{n,K,\phi,\theta}\rangle\langle{\Gamma_{n',K,\phi,\theta}}|,
\label{eqn:SLDDelta}
\end{equation}
where $\{f_{n,n',\Delta}$, $g_{n,n',\Delta}\}$ $\in\Re$, and $f_{n,n',\Delta}=-f_{n',n,\Delta}$ and $g_{n,n',\Delta}=g_{n',n,\Delta}$. Since all the phases can be absorbed in the basis and according to Eqn.~\eqref{eqn:QFI}, the QFI elements are basis independent, the whole QFI matrix is $\phi$ independent. Additionally, the off-diagonal elements of the QFI matrix are necessarily zero since trace of the product of two real matrices and one imaginary is imaginary as a result of which the real part is zero. Therefore, the QFI matrix for phase and phase diffusion for any input pure FPN state is of the form
\begin{equation}                                                                                                                                                                                                                                             
\textbf{H}= \begin{pmatrix}
H_{11} & 0\\
0 & H_{22}
\label{eqn:QFImatrix}
\end{pmatrix},
\end{equation}
where subscripts $1$ and $2$ refer to $\phi$ and $\Delta$, as noted earlier. $H_{11}$ and $H_{22}$ correspond to the maximum information content in the individual estimation of $\phi$ and $\Delta$ respectively. Obtaining a general closed-form of the quantum limits for this system is a difficult task because the evolved density matrix has the rank of $(2K+1)$ for FPN and $(K+1)$ for the special case of HB states, and the diagonalisation of such matrices is a challenging problem for arbitrary $K$. To obtain a better understanding, we derive analytical expressions in the regimes of large and small phase diffusion. For intermediate values of phase diffusion, we employ a numerical approach to solve for the SLDs and calculate the QFI (see Appendix~(\ref{app:QFInumerics})).

The limits on the covariance matrix set by the QFI are not necessarily attainable in multi-parameter estimation scenarios due to the possible non-commutativity of optimal measurements. A sufficient condition for saturating the QCRB is $[\hat{L}_{1},\hat{L}_{2}]=0$ which does not hold for a general FPN state or the special case of HB states. This implies that the optimal POVMs in the individual parameter estimation constructed from the eigenvectors of the SLDs do not commute in general and will not saturate the two-parameter QCRB. A weaker condition, which is sufficient and necessary, for saturating the multi-parameter QCRB is $\operatorname{Tr}\left[\hat\varrho\left[\hat{L}_{1},\hat{L}_{2}\right]\right]=0$. However, we numerically found that this quantity is non-zero for some fixed particle number pure states. We found numerically that it is zero for HB states for particle numbers up to 158, but we lack a general analytical proof. Additionally, to fullfil the multi-parameter saturability condition given by the expectation value of the commutator of the SLDs, it is necessary to consider multiple copies of the state and collective measurements. Although such strategy can bring further enhancements in the attainable joint precision~\cite{Vidrighin2014, Ragy2016}, it is harder to treat theoretically and implement experimentally~\cite{Roccia2017}. Therefore, as a first step, we restrict  ourselves to single copies of the state only. Additionally, we choose the class of projective measurements as they optimise the FI of the associated probability distribution~\cite{Gill2000,Dempster1977}.


The figure of merit that we consider in this work is the trace of the covariance matrix which for our two parameter case, $\phi$ and $\Delta$, is minimised when the off-diagonal elements of the FI matrix are zero. We can therefore compute how close the minimum variances of the estimators for phase and phase diffusion parameters (${\text{Var}_{\text{min}}(\phi)}$ and ${\text{Var}_{\text{min}}(\Delta)}$ respectively) associated with a given measurement compare with the QCRB by considering the ratios of the corresponding diagonal elements of the FI and QFI matrices. This is given by
 \be
 \operatorname{Tr}[\textbf{F}\textbf{H}^{-1}]=\frac{F_{11}}{H_{11}}+\frac{F_{22}}{H_{22}}
=\frac{H_{11}^{-1}}{M\text{Var}_{\text{min}}(\phi)}+\frac{H_{22}^{-1}}{M\text{Var}_{\text{min}}(\Delta)}
 \label{eqn:Trade-offRelationGeneral}
 \ee
which is the bound studied in this work. The higher the bound, the higher precision we can attain about the two parameters simultaneously. Its upper limit is set by the QCRB which in the case of our two-parameter estimation problem gives $\operatorname{Tr}[\textbf{F}\textbf{H}^{-1}]\leq{2}$. This quantity has been considered in previous works such as~\cite{Gill2000, Vidrighin2014,Li2016}.

The optimal bound $\operatorname{Tr\left[\textbf{F}\textbf{H}^{-1}\right]}_{\text{max}}$ with respect to all projective measurements, similarly to the QFI, is $\phi$ insensitive. The reason is the following: the phase dependence of the bound arises because the input state $\hat\varrho_{in}$ acquires phase in the unitary transformation $\hat{U}_{\phi}=\e^{\i\phi\hat{n}}$. If some optimal value of phase which maximises the bound in Eqn.~\eqref{eqn:Trade-offRelationGeneral} is changed by an amount $\zeta$ due to a unitary transformation $\hat{U}_{\zeta}=\e^{\i\zeta\hat{n}}$ we can always reverse the action of this unitary by applying $\hat{U}_{\zeta}^{-1}$ in our optimal measurement. As a result, the optimal bounds depend only on the particle number and the phase diffusion parameter.

Though highly desirable, the maximisation of $\operatorname{Tr}[\textbf{F}\textbf{H}^{-1}]$ over all projective measurements is in general exigent. This is partially due to the lack of knowledge of the closed form of the QFI for a general phase diffusion parameter and particle number, and partially due to the need to optimise over a complex space containing $2d-2$ parameters, where $d$ is the dimension of the Hilbert space of the input probe state, factor of two accounts for the fact that we have real and imaginary parameters and subtraction of two accounts for normalisation and global phase. We therefore, explore this aspect analytically in the regime of large (for a general FPN state) and small (for the special case of HB states) $\Delta$, and provide numerical results for the general $\Delta$ regime (for HB states). 

\section{\label{sec:4}Large diffusion regime}

The off-diagonal elements of the FPN density matrix in Eqn.~\eqref{eqn:MixedFixedPhotonNumberState} decay exponentially in $\Delta$. If $\Delta$ is sufficiently large, we can neglect all the off-diagonals except for the one ($k^{\text{th}}$ off-diagonal) that is closest to the main diagonal and contributes non-zero elements. This approximation corresponds to Taylor expansion of $\hat\varrho$ to the order of $2k^2$ in $x=\e^{-{\Delta^{2}}{/}{4}}$ about $x$ = 0 which results in the approximate, tridiagonal density matrix,
\begin{equation}
\label{eqn:TridiagFPNState}
\hat{\varrho}_{\text{L}\Delta} = \sum_{n,n'=0}^{2K}{{a}_{n}{a}_{n'}^*\e^{\i\phi(n-n')}x^{2(n-n')^2}}
(\delta_{n,n'}+\delta_{n,n'\pm k})
|{n},{2}{K}-{n}\rangle\langle{n'},{2}{K}-{n'}|,
\end{equation} 
where subscript $\text{L}\Delta$ identifies the large phase diffusion approximation. The case of $k = 1$ corresponds to an FPN state, where the first off-diagonal contributes non-zero elements and this is the last off-diagonal of the density matrix that we keep. The instance of $k = 2$ corresponds to a state, where the first off-diagonal has all zero elements and the second off-diagonal contributes non-zero elements, and therefore this is the last one that we keep. HB states are a good example of the $k = 2$ case, where in fact only even numbered off-diagonals contribute non-zero elements. Also, in the case of HB states, $a_n$ = $b_{\frac{n}{2}}\delta_{n,2q}$ as defined in Eqn.~\eqref{eqn:PureHBstate}. The approximation in Eqn.~\eqref{eqn:TridiagFPNState} is valid when 
\begin{equation}
\label{eqn:LargeDeltaValidityFPN}
\Delta^{(FPN)}\geq\sqrt{\frac{2}{\left(k+1\right)^2}\ln{\left(\frac{2K}{f}\right)}}
\end{equation}
and 
\begin{equation}
\label{eqn:LargeDeltaValidityHB}
\Delta^{(HB)}\geq\sqrt{\frac{1}{8}\ln{\left(\frac{K}{f}\right)}}
\end{equation}

\begin{figure}[!t]
\begin{center}
\includegraphics[trim=1.5cm 7.0cm 1.5cm 6.5cm,width=0.5\columnwidth]{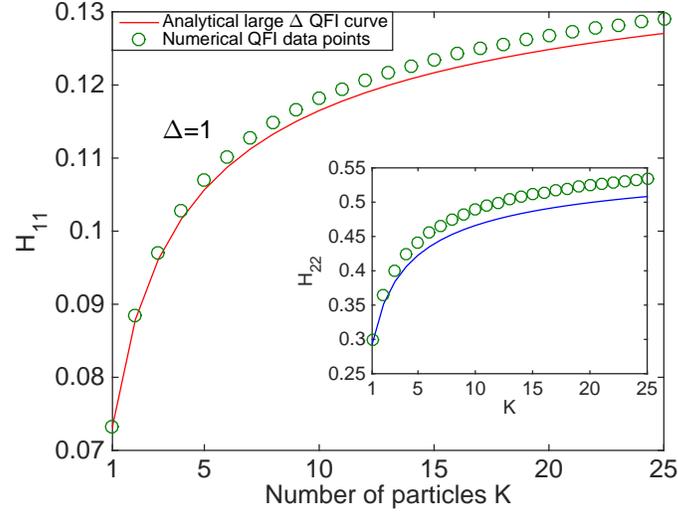}
\end{center}
\caption{Diagonal QFI matrix elements as a function of $K$ for HB states in the large phase diffusion approximation. $K$ is the particle number in each of the two input modes of the unitary generating the HB state (see Fig.~(\ref{fig:FPNdiagram})). Both $H_{11}$ and $H_{22}$ are increasing functions of $K$ as expected from Eqn.~\eqref{eqn:H11fpn} and Eqn.~\eqref{eqn:H22fpn}. The accuracy of the analytical expressions gets worse as $K$ increases which is expected from the validity constraint in Eqn.~\eqref{eqn:LargeDeltaValidityHB}. In general, the approximation for $H_{22}$ is less accurate than for $H_{11}$ due to the factor of $\Delta^2$ appearing in Eqn.~\eqref{eqn:H22fpn} as explained in the main text.}
\label{fig:QFIvsK_LargeDelta}
\end{figure}
\begin{figure}[!ht]
\begin{center}
\includegraphics[trim=1.5cm 7cm 1.5cm 6.5cm,width=0.5\columnwidth]{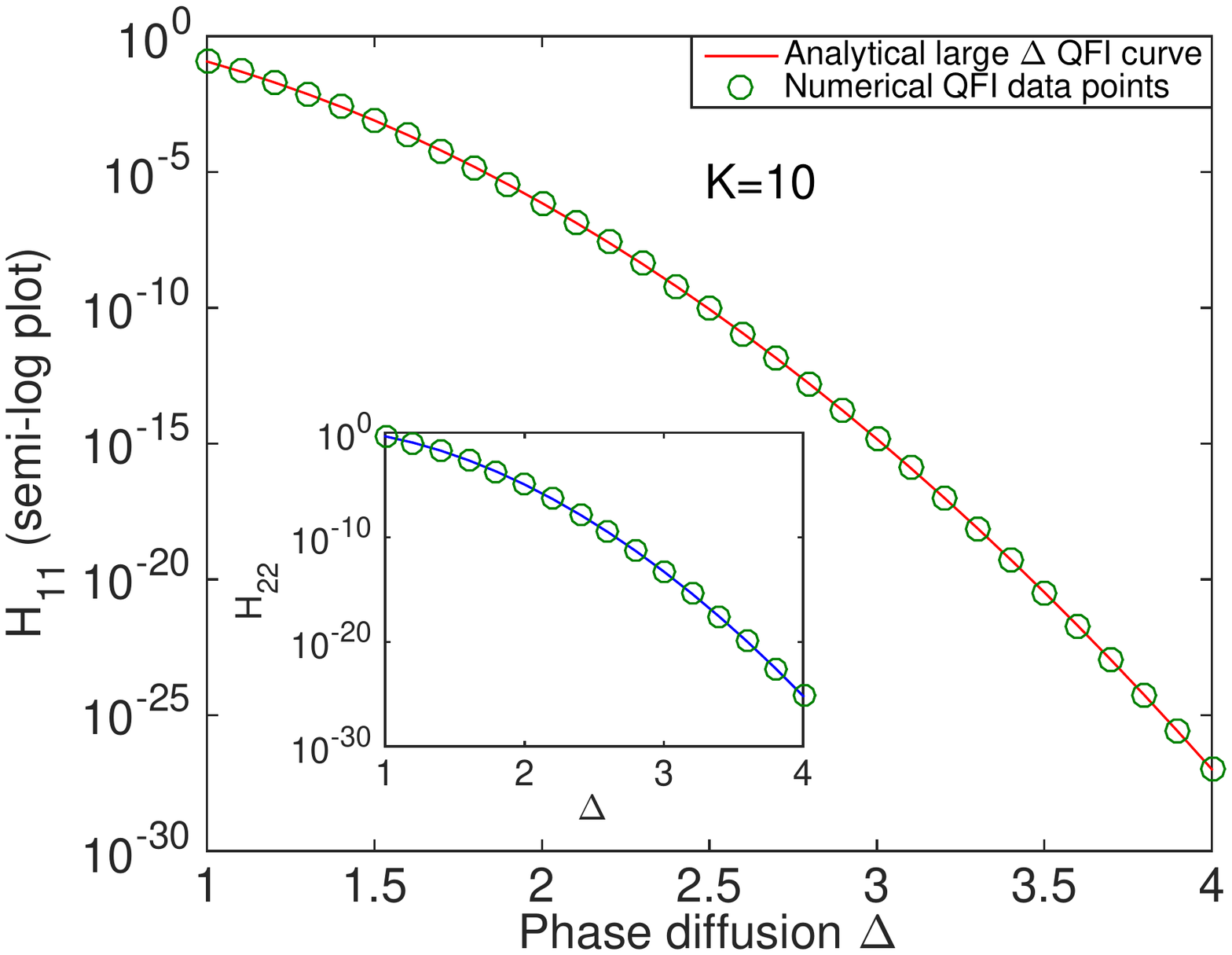}
\end{center}
\caption{Semi-log plot of the diagonal QFI matrix elements as a function of $\Delta$ for HB states in the large phase diffusion approximation. $H_{11}$ and $H_{22}$ show an exponential decrease with respect to $\Delta$ as expected due to the exponential erasing of the off-diagonal elements of the density matrix. We present the y-axis on the logarithmic scale to show that the QFI elements go to zero at large $\Delta$. A complementary, linear plot of the QFI as a function of $\Delta$ is shown in Fig.~(\ref{fig:QFI_LargeDelta2}) of Appendix~(\ref{app:LargeDeltaPlot}) which magnifies the calculation error at smaller phase diffusion values. The analytical approximations show a good agreement with the numerical data. At $K=25$ and $\Delta=1$, the relative error of the analytical values with respect to the numerical values of $H_{11}$ and $H_{22}$ are $1.6\%$ and $4.9\%$ respectively. This behaviour improves at larger $\Delta$ values and at $K=25$ and $\Delta=1.2$, the relative errors reduce to $0.3\%$ for $H_{11}$ and $0.9\%$ for $H_{22}$.}
\label{fig:QFIvsDelta_LargeDelta}
\end{figure}

for a general FPN and the special case of HB states respectively, where $f$ is the allowed relative error on the density matrix (see Appendix~(\ref{app:LargeDeltaValidity}) for details). This shows the $\sqrt{\ln{\left(K\right)}}$ dependence on the threshold value of $\Delta$ for the desired relative error.

\subsection{\label{sec:4.1}Quantum limit}

In order to get an intuition for the form of the SLDs in the large $\Delta$ approximation, the SLD equation described in Eqn.~\eqref{eqn:SLD} can be solved for small values of $K$. This gives us an ansatz for the SLDs for $\phi$ and $\Delta$, $\hat{L}_{1,\text{L}\Delta}$ and $\hat{L}_{2,\text{L}\Delta}$ respectively, shown in Eqn.~\eqref{eqn:TridiagFPNsld1} and Eqn.~\eqref{eqn:TridiagFPNsld2} of Appendix~(\ref{app:SLDsLargeDelta}). The SLDs are accurate to the order of $x^{2k^2}$ which can be proven by substituting them into the left and right hand sides of the SLD equation to verify that the equality holds within the approximation. This is shown in Appendix~(\ref{app:SLDsLargeDelta}). It is worth noting that the presence of the factor of $\Delta$ in $\hat{L}_{2,\text{L}\Delta}$ decreases its order of accuracy from $x^{2k^2}$ to $x^{2k^2-{4}{\Delta^{-2}}\ln{\left(\Delta\right)}}$.

The resulting QFI elements, $H_{11,\text{L}\Delta}$ and $H_{22,\text{L}\Delta}$, are given by
\begin{equation}
\label{eqn:H11fpn}
{H}_{11,\text{L}\Delta}=4k^2{x^{4k^2}}\sum_{n=k}^{2K}\frac{|{a}_{n}|^2|{a}_{n-k}|^2}{|{a}_{n}|^2+|{a}_{n-k}|^2},
\end{equation}
\begin{equation}
\label{eqn:H22fpn}
{H}_{22,\text{L}\Delta}=4k^4\Delta^2{x}^{4k^2}\sum_{n=k}^{2K}\frac{|{a}_{n}|^2|{a}_{n-k}|^2}{|{a}_{n}|^2+|{a}_{n-k}|^2},
\end{equation}
where the sums above are upper-bounded by $1/2$ (see Appendix~(\ref{app:QFIsLargeDelta})). For HB states, this can be attained in the regime of large $K$. This upper bound implies the loss of quantum enhancement in the large phase diffusion regime as at some point increasing the particle number in the input probe state will have no influence on the precision of our estimation scheme. The exponential decrease of $H_{11,\text{L}\Delta}$ and $H_{22,\text{L}\Delta}$ with respect to $\Delta$ is to be expected since the effect of phase diffusion is to exponentially erase the off-diagonal elements of the density matrix. Again, the accuracy of $H_{22,\text{L}\Delta}$ is decreased from $x^{4k^2}$ to $x^{4k^2-{4}{\Delta^{-2}}\ln{\left(\Delta^2\right)}}$ due to the presence of the ${\Delta}^2$ factor. The calculation of the $H_{11,\text{L}\Delta}$ term is shown in Appendix~(\ref{app:QFIsLargeDelta}). 

The analytical quantum limits given in Eqn.~\eqref{eqn:H11fpn} and Eqn.~\eqref{eqn:H22fpn} are compared with the values obtained using a numerical routine developed in MATLAB (see Appendix~(\ref{app:QFInumerics})) for HB states. The resulting curves are shown in Fig.~(\ref{fig:QFIvsK_LargeDelta}) and Fig.~(\ref{fig:QFIvsDelta_LargeDelta}). Further, we provide a numerical phase diffusion validity plot in Fig.~(\ref{fig:QFI_LargeDeltaValidity}) of Appendix~(\ref{app:LargeDeltaValidity}), also for HB states. The plot shows the threshold values of $\Delta$ for which the approximation is valid as a function of particle number ($K$) when the maximum allowed relative error ($\delta{H_{ii}}/{H_{ii}}$) is $0.05$.

\subsection{\label{sec:4.2}Trade-off relation} 

\begin{figure}[t!]
	\begin{center}
		\includegraphics[trim=1.5cm 7.0cm 1.5cm 6.5cm,width=0.5\columnwidth]{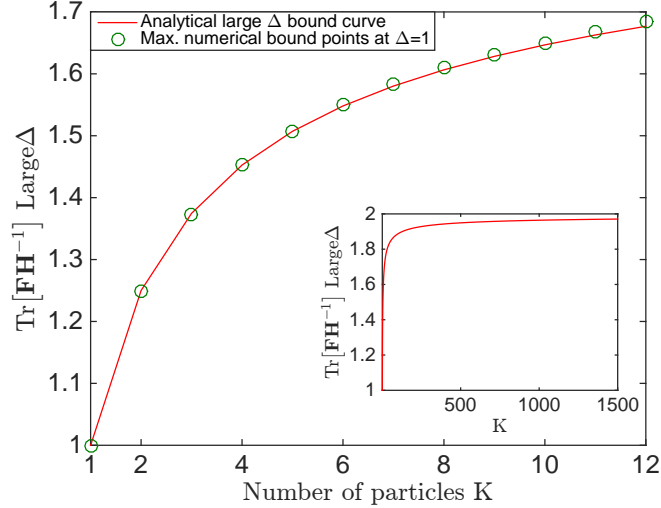}
	\end{center}
	\caption{Joint information bound in estimating $\phi$ and $\Delta$ simultaneously as a function of $K$ for HB states in the large phase diffusion regime. The main plot compares the analytical bound in Eqn.~\eqref{eqn:LargeDeltaTradeoff} with the optimal bound at $\Delta=1$ found using the simulated annealing algorithm. The two methods show a good agreement which suggests that the POVM defined in Eqn.~\eqref{eqn:LargeDeltaPOVM1} and Eqn.~\eqref{eqn:LargeDeltaPOVM2} is indeed optimal in the special case of HB states for the investigated small $K$ regime. The inlay plot displays the analytical bound for a higher range of $K$ values in the case of HB states. It shows that the bound approaches the upper limit of $2$ set by the QCRB for large $K$. At $K=1500$, the analytical bound is $1.97$.}
	\label{fig:TradeOffvsK_LargeDelta}
\end{figure}

To find the maximal joint information bound for our problem, we are required to maximise $\operatorname{Tr}\left[\textbf{F}\textbf{H}^{-1}\right]$ with respect to all projective measurements $|v_y\rangle=\sum_{n=0}^{2K}r_{y,n}\e^{\i{X}_{y,n}}|n, 2K-n\rangle$ which form a POVM set $\{|v_y\rangle\langle{v_y}|\}$ for the considered Hilbert space. We perform such an optimisation in Appendix~(\ref{app:LargeDeltaTradeoff}) in the large $\Delta$ regime. For any FPN state with coefficients $a_{n}=|a_{n}|\e^{\i\theta_{n}}$, the phase $X_{y,n}$ of the optimal projectors must satisfy
\begin{equation}
\label{eqn:LargeDeltaPOVM2}
X_{y,n}=X_{y,n-k-1}-2X_{y,n-k}-\theta_{n-k-1}+2\theta_{n-k}-\theta_{n}.
\end{equation}
Additionally, we assume that the magnitudes $r_{y,n}$ of the optimal projectors are equally weighted which gives
\begin{equation}
\label{eqn:LargeDeltaPOVM1}
r_{y,n}=\frac{1}{\sqrt{\frac{2K}{k}+1}}.
\end{equation}
This last choice in Eqn.~\eqref{eqn:LargeDeltaPOVM1} is motivated by the results from the simulated annealing algorithm which searches through the space of projective measurements to maximise our bound for HB states. The numerics is performed in the small particle number regime and shows that the bound becomes $\Delta$ independent  for large values of $\Delta$ (see Fig.~(\ref{fig:BoundMax_vs_DeltaGeneral}) of Sec.~(\ref{sec:6})). The associated optimal projectors have the structure as given by Eqn.~\eqref{eqn:LargeDeltaPOVM2} and Eqn.~\eqref{eqn:LargeDeltaPOVM1}.

The resulting bound, $\operatorname{Tr}[\textbf{F}\textbf{H}^{-1}]$, in the simultaneous estimation of $\phi$ and $\Delta$ in the large phase diffusion regime is
\begin{equation}
\label{eqn:LargeDeltaTradeoff}
\operatorname{Tr}[\textbf{F}\textbf{H}^{-1}]=\frac{(\sum_{n=k}^{2 K}|{a}_{n}||{a}_{n-k}|)^2}{\sum_{n=k}^{2 K}\frac{|{a}_{n}|^2|{a}_{n-k}|^2}{|{a}_{n}|^2+|{a}_{n-k}|^2}}.
\end{equation}
See Appendix~(\ref{app:LargeDeltaTradeoff}) for derivation. This bound reaches the QCRB for two-parameter estimation, i.e. $\operatorname{Tr}\left[\textbf{F}\textbf{H}^{-1}\right]=2$, when $|a_{n}|=|a_{n-k}|$ for $n\in\mathbb{Z}\{k:2K\}$. This shows that the globally optimal bound with respect to all diffusion values will also approach the QCRB for such states. For HB states, the condition $|a_{n}|=|a_{n-k}|$ coincides with the large particle number ($K$) regime. Plot of the bound in Eqn.~\eqref{eqn:LargeDeltaTradeoff} as a function of $K$ in the case of HB states is shown in Fig.~(\ref{fig:TradeOffvsK_LargeDelta}). The inlay plot shows the asymptotic attainability of the QCRB at large $K$, whereas the main plot shows the comparison with the numerical optimization routine based on the simulated annealing algorithm. At large values of $\Delta$, the analytical bound in Eqn.~\eqref{eqn:LargeDeltaTradeoff}, obtained using the POVM defined in Eqn.~\eqref{eqn:LargeDeltaPOVM2} and Eqn.~\eqref{eqn:LargeDeltaPOVM1}, provides a good agreement with the numerical data in the case of HB states. However, an analytical proof of the optimality of this POVM for a general FPN state in the large phase diffusion regime remains an open question.

\section{\label{sec:5} Small diffusion regime}

Our analysis for the small phase diffusion regime is restricted to the special case of HB states. In this regime, we perform a Taylor expansion of the HB density matrix in Eqn.~\eqref{eqn:MixedHBstate} to the second order in $\Delta$ for $H_{11}$ and to the fourth order in $\Delta$ for $H_{22}$ near $\Delta=0$. The reason for the fourth order Taylor expansion in $\Delta$ in the case of $H_{22}$ is to ensure that the corresponding SLD equation, $2\partial_{\Delta}\hat\varrho=\hat{L}_{2}\hat\varrho+\hat\varrho\hat{L}_{2}$, is accurate to at least second order in $\Delta$ as we lose one order in $\Delta$ on the left hand side due to the differentiation with respect to $\Delta$.

The Taylor-expanded forms of the evolved HB density matrix to the second and fourth order in $\Delta$ are respectively
\begin{equation}
\label{eqn:MixedHBStateTaylor2}
\hat{\varrho}_{1,\text{S}\Delta}\! =\!\sum_{n,n'=0}^{K}{{b}_{n}{b}_{n'}\e^{{2}\i\phi(n-n')}}(1-2\Delta^2(n-n')^2)
|{2}{n},{2}{K}-{2}{n}\rangle\langle{2}{n'},{2}{K}-{2}{n'}|
\end{equation}

and
\begin{equation}
\label{eqn:MixedHBStateTaylor4}
\hat{\varrho}_{2,\text{S}\Delta}\!=\!\sum_{n,n'=0}^{K}{{b}_{n}{b}_{n'}\e^{{2}\i\phi(n-n')}}
(1-2\Delta^2(n-n')^2+2\Delta^4(n-n')^4)
|{2}{n},{2}{K}-{2}{n}\rangle\langle{2}{n'},{2}{K}-{2}{n'}|,
\end{equation}
where subscript $\text{S}\Delta$ identifies the small phase diffusion approximation. The validity of the approximation is evaluated by comparing the last kept non-zero term with the first neglected non-zero term. When expressing $\hat\varrho$ as $\hat{\varrho}_{1,\text{S}\Delta}$ this corresponds to $\Delta^{2}\ll{1}{/}{\left(n-n'\right)^2}$. This can be further tightened by taking $\left(n-n'\right)=K$ which gives 
\begin{equation}
\label{eqn:SmallDeltaValidity1}
\Delta^{2}\ll\frac{1}{K^2}.
\end{equation}
The equivalent validity constraint can be derived for $\hat{\varrho}_{2,\text{S}\Delta}$ which has the same form as Eqn.~\eqref{eqn:SmallDeltaValidity1} but with an extra factor of $3/2$ in front.


Expanding Eqn.~\eqref{eqn:MixedHBStateTaylor2} and Eqn.~\eqref{eqn:MixedHBStateTaylor4}, we get a set ($\mathcal{W}_{1}$) of 3 linearly independent vectors describing $\hat\varrho_{1,\text{S}\Delta}$ and a set ($\mathcal{W}_{2}$) of 5 linearly independent vectors describing $\hat\varrho_{2,\text{S}\Delta}$. The elements of sets $\mathcal{W}_{1}$ and $\mathcal{W}_{2}$ are defined as follow,
\begin{equation}
\label{eqn:W1}
\mathcal{{W}}_{1}=\{|w_{1}\rangle\ , |w_{2}\rangle , |w_{3}\rangle\}
\end{equation}
and
\begin{equation}
\label{eqn:W2}
\mathcal{W}_{2}=\{|w_{1}\rangle\ , |w_{2}\rangle , |w_{3}\rangle , |w_{4}\rangle\ , |w_{5}\rangle\}
\end{equation}
where $|w_k\rangle=\sum_{n=0}^{K}n^{k-1}|\varphi_n\rangle$ with $|\varphi_n\rangle={b}_{n}\e^{2\i\phi{n}}|2n,2K-2n\rangle$. See Appendix~(\ref{app:LinearIndependenceSmallDelta}) for the proof of linear independence of the elements of sets $\mathcal{W}_1$ and $\mathcal{W}_2$. Since $\hat\varrho_{1,\text{S}\Delta}$ and $\hat\varrho_{2,\text{S}\Delta}$ are spanned by 3 and 5 linearly independent vectors respectively, they can be expressed as $3\times 3$ and $5\times 5$ dimensional matrices if an appropriate orthonormal basis set is chosen. This is crucial since it overcomes the problem of diagonalising a ($K$$+$$1$) dimensional matrix. The orthonormal sets can be found by orthonormalising the vectors $\{|w_{1}\rangle , |w_{2}\rangle ,..., |w_{5}\rangle\}$ using the Gram-Schmidt procedure and they are denoted by ${\mathcal{V}}_{1}=\{|v_{1}\rangle, |v_{2}\rangle, |v_{3}\rangle\}$ and ${\mathcal{V}}_{2}=\{|v_{1}\rangle, |v_{2}\rangle, |v_{3}\rangle, |v_{4}\rangle, |v_{5}\rangle\}$ for $\hat\varrho_{1,\text{S}\Delta}$ and $\hat\varrho_{2,\text{S}\Delta}$ respectively. The Gram-Schmidt procedure and its results are shown in Appendix~(\ref{app:GramSchmidt}). Given $\mathcal{V}_1$ and $\mathcal{V}_2$, we can now perform the basis change of $\hat\varrho_{1,\text{S}\Delta}$ and $\hat\varrho_{2,\text{S}\Delta}$ by applying
\begin{equation}
\label{eqn:BasisChange}
\hat\varrho'=\sum_{i,j=0}^{N}\operatorname{Tr}\left[|{v}_{j}\rangle\langle{v}_{i}|\hat\varrho\right]|{v}_{i}\rangle\langle{v}_{j}|,
\end{equation}
where $N$ is the number of basis vectors forming a complete set. The resulting $3\times 3$ and $5\times 5$ density matrices, written in the basis $\mathcal{V}_1$ and $\mathcal{V}_2$, are
\begin{equation}\hat{\varrho}'_{1,\text{S}\Delta}=
\begin{pmatrix}
\label{eqn:SmallDelta3x3}  
{b}_{11} & {0} & {b}_{13} \\
{0} & {b}_{22} & {0} \\
{b}_{13} & {0} & {0} \\
\end{pmatrix}
\end{equation}
and
\begin{equation}\hat{\varrho}'_{2,\text{S}\Delta}=
\begin{pmatrix}  
\label{eqn:SmallDelta5x5}
{c}_{11} & {0} & {c}_{13} & {0} & {c}_{15} \\
{0} & {c}_{22} & {0} & {c}_{24} & {0} \\
{c}_{13} & {0} & {c}_{33} & {0} & {0} \\
{0} & {c}_{24} & {0} & {0} & {0}\\
{c}_{15} & {0} & {0} & {0} & {0}\\
\end{pmatrix},
\end{equation}
where $b_{ij}\in\Re$ and $c_{ij}\in\Re$ and these matrix elements are functions of $K$ and $\Delta$ only. The exact forms of the entries in terms of $K$ and $\Delta$ are given in Appendix~(\ref{app:SmallDeltaReducedMatricesAndDiagonalisation}). It is worth noting that the reduced matrix elements have no $\phi$ dependence, whereas the orthonormal basis vectors defining these matrices have no $\Delta$ dependence. 

\begin{figure}[!t]
	\begin{center}
		\includegraphics[trim=1.5cm 7cm 1.5cm 6.5cm,width=0.5\columnwidth]{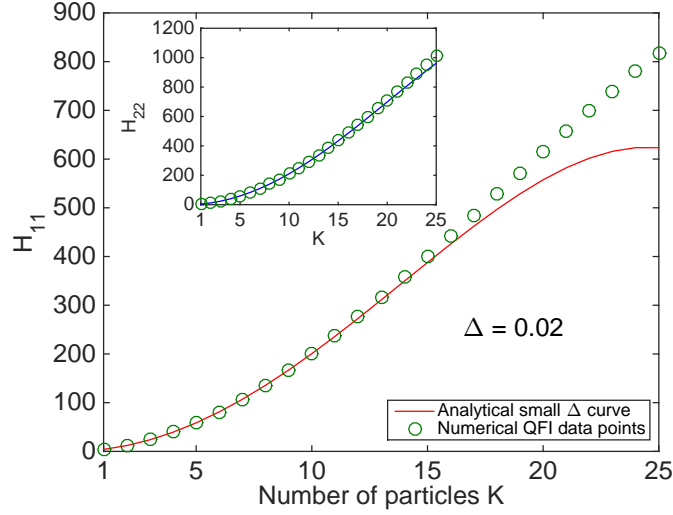}
	\end{center}
	\caption{Diagonal QFI matrix elements as a function of $K$ for HB states in the small phase diffusion approximation. As expected, $H_{11}$ and $H_{22}$ are increasing functions of $K$ showing the approximate Heisenberg scaling of $K^2$. The approximation gets worse at larger $K$ values for a given $\Delta$ as expected from the validity constraints in Eqn.~\eqref{eqn:SmallDeltaValidity1}. In general, $H_{22}$ is more accurate than $H_{11}$ because in this case the density matrix of the probe state is Taylor expanded to the fourth order in $\Delta$ rather than second order in $\Delta$.}
	\label{fig:QFIvsK_SmallDelta}
\end{figure}

\begin{figure}[!ht]
	\begin{center}
		\includegraphics[trim=1.5cm 7cm 1.5cm 6.5cm,width=0.5\columnwidth]{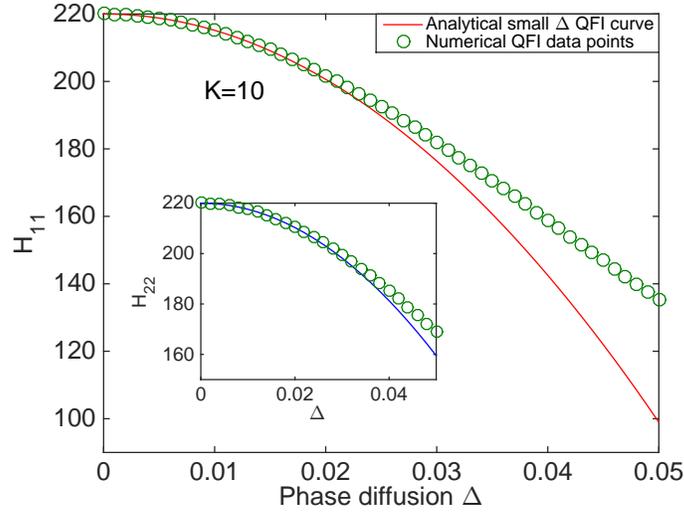}
	\end{center}
	\caption{Diagonal QFI matrix elements as a function of $\Delta$ for HB states in the small phase diffusion approximation. Both $H_{11}$ and $H_{22}$ are decreasing functions of $\Delta$ in this regime. For a given value of $K$ the approximation gets worse at larger $\Delta$ which agrees with the validity constraints in Eqn.~\eqref{eqn:SmallDeltaValidity1}. At $K=20$ and $\Delta=0.02$, the relative error of the analytical values with respect to the numerical values of $H_{11}$ and $H_{22}$ are $9\%$ and $2\%$ respectively. The accuracy improves at smaller $\Delta$ values and at $K=20$ and $\Delta=0.018$, the relative errors reduce to $6\%$ for $H_{11}$ and $1\%$ for $H_{22}$.}
	\label{fig:QFIvsDelta_SmallDelta}
\end{figure}

\begin{figure}[!t]
	\begin{center}
		\includegraphics[trim=1.5cm 7cm 1.5cm 6.5cm,width=0.5\columnwidth]{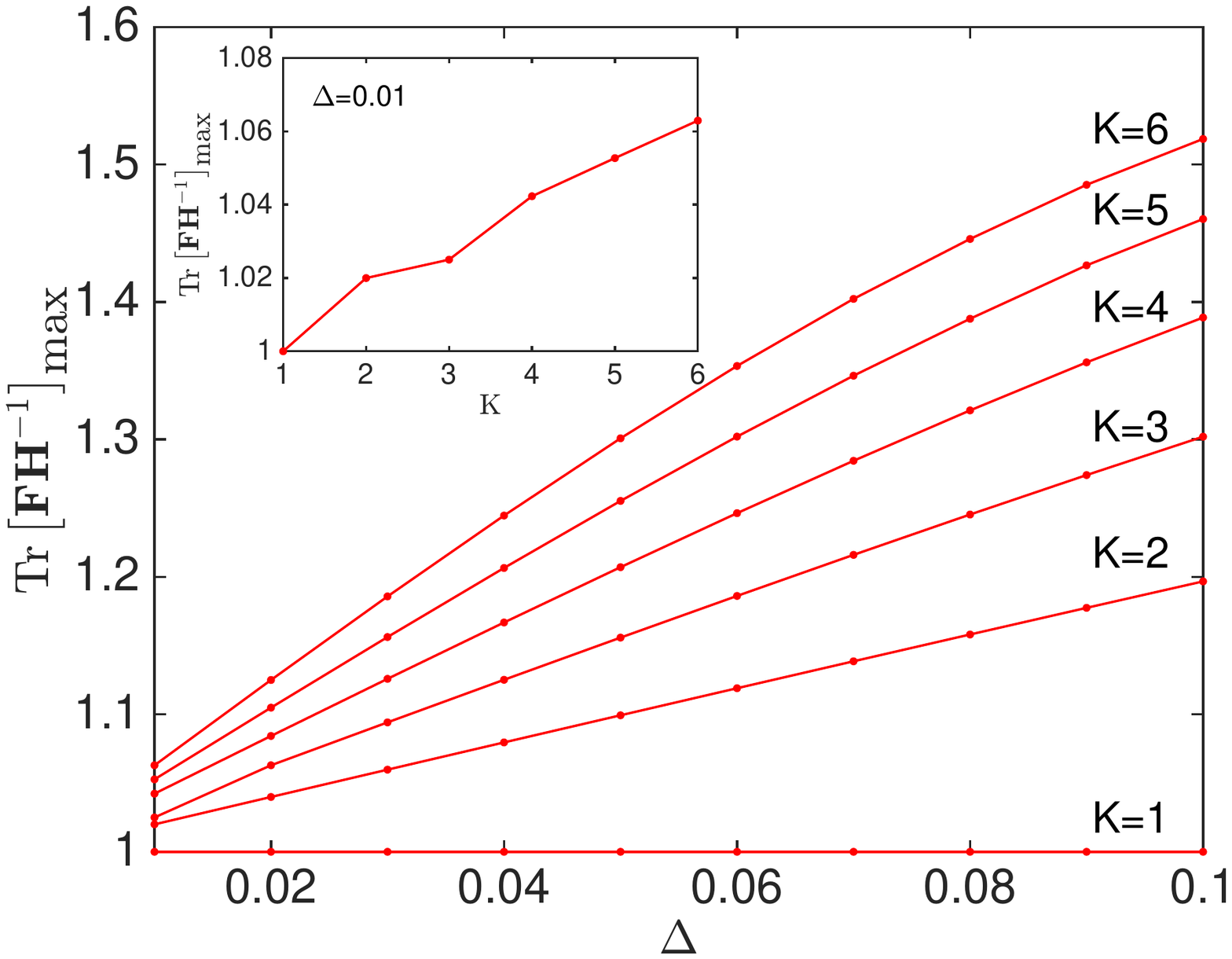}
	\end{center}
	\caption{Bound on the joint estimation of phase and diffusion for HB states, numerically optimised with respect to orthonormal proejctive measurements. $K$ labels different particle numbers. The main plot shows that the bound $\operatorname{Tr}\left[\textbf{F}\textbf{H}^{-1}\right]$ as $\Delta\rightarrow{0}$ for the considered $K$ values. The inlay plot shows the optimal bound as a function of $K$ for the smallest sampled $\Delta$ of $0.01$.}
	\label{fig:BoundMaxSmallDelta}
\end{figure}

\subsection{\label{sec:5.1}Quantum limit}

Diagonalisation of matrices in Eqn.~\eqref{eqn:SmallDelta3x3} and Eqn.~\eqref{eqn:SmallDelta5x5} is greatly simplified due to their sparse structure. In particular, $\hat{\varrho}'_{1,\text{S}\Delta}$ can be expressed as a direct sum of $2\times 2$ and $1\times 1$ matrices, whereas $\hat{\varrho}'_{2,\text{S}\Delta}$ as a direct sum of $3\times 3$ and $2\times 2$ matrices. The diagonalisation of these matrices reveals that one of the eigenvalues of $\hat{\varrho}'_{1,\text{S}\Delta}$ and two eigenvalues of $\hat{\varrho}'_{2,\text{S}\Delta}$ are negative up to the order of $\Delta^{4}$ and $\Delta^{6}$ respectively. Therefore, they are zero within our small phase diffusion approximation and they do not contribute towards reconstructing our Taylor expanded density matrices. Consequently, $\hat{\varrho}'_{1,\text{S}\Delta}$ is a rank 2 matrix and $\hat{\varrho}'_{2,\text{S}\Delta}$ rank 3 matrix within the small phase diffusion approximation. See Appendix~(\ref{app:SmallDeltaReducedMatricesAndDiagonalisation}) for details of the diagonalisation procedure and the results. The alternative form of the QFI expression in terms of the eigenbasis of the density matrix, given in Eqn.~\eqref{eqn:QFIEigenbasis}, can be used to find $H_{11}$ and $H_{22}$ within the small diffusion approximation. The resulting diagonal QFI matrix elements (when keeping second order in $\Delta$ terms only) corresponding to the HB probe state are
\begin{equation}
{H}_{11,\text{S}{\Delta}}=2K(K+1)-(2K(K+1))^2\Delta^2,
\label{eqn:H11SmallDelta}
\end{equation}
and
\begin{equation}
{H}_{22,\text{S}{\Delta}}=2K(K+1)-\frac{1}{2}(2K(K+1))^2\Delta^2.
\label{eqn:H22SmallDelta}
\end{equation}
The first of these expressions shows that the presence of phase diffusion diminishes the precision attainable in the estimation of phase. This has been anticipated before~\cite{Demkowicz-Dobrzanski2014,Brivio2010, Genoni2011, Genoni2012, Vidrighin2014}, but we now provide an exact expression for HB states. The second expression shows that it is possible to achieve a quantum-enhanced estimation of the phase diffusion parameter. This is because the QFI for $\Delta$ scales quadratically with $K,$ while any classical strategy with $K$ particles will only lead to a QFI scaling at most linearly in $K.$ 
Figures~(\ref{fig:QFIvsK_SmallDelta}) and~(\ref{fig:QFIvsDelta_SmallDelta}) show the comparison of the obtained analytical expressions in Eqn.~\eqref{eqn:H11SmallDelta} and Eqn.~\eqref{eqn:H22SmallDelta} with the numerical data points. Additionally, numerical phase diffusion validity plot for this approximation is shown in Fig.~(\ref{fig:QFI_SmallDeltaValidity}) of Appendix~(\ref{app:SmallDeltaReducedMatricesAndDiagonalisation}).

\begin{figure}[!t]
	\begin{center}
		\includegraphics[trim=1.5cm 7cm 1.5cm 6.5cm,width=0.5\columnwidth]{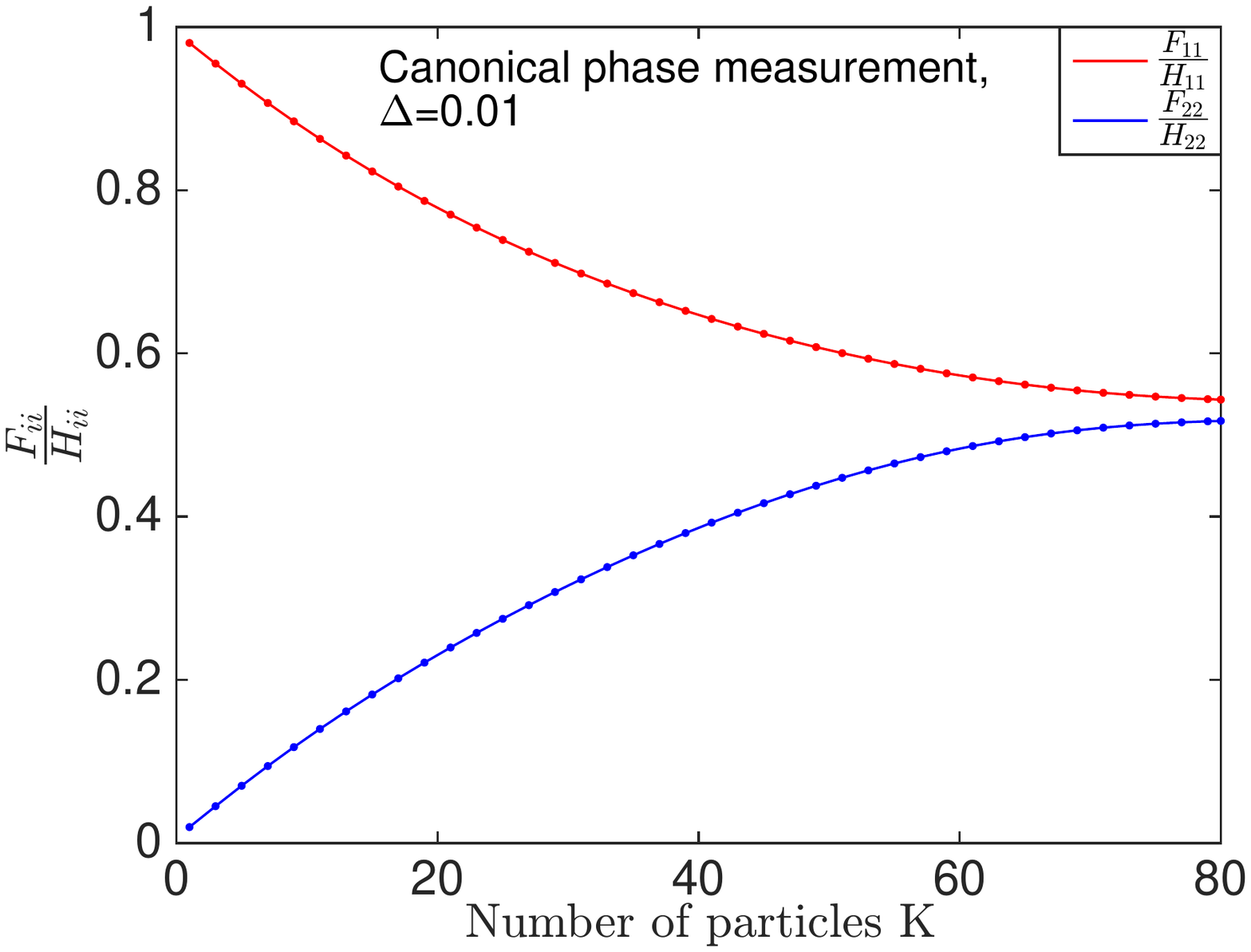}
	\end{center}
	\caption{Numerical results with the canonical phase measurement POVM at $\Delta=0.01$. In the high particle number regime (considered within the simulation), the POVM gives significant information about both phase and diffusion with the total trade-off $\operatorname{Tr}\left[\textbf{F}\textbf{H}^{-1}\right]$ of approximately 1.06.}
	\label{fig:CanonicalPhaseMeasurementSmallDelta}
\end{figure}

\subsection{\label{sec:5.2}Trade-off relation}

The trade-off relation in the limit of \hspace{0.5mm} $\Delta\rightarrow{0}$ \hspace{0.5mm} for an arbitrary POVM set \hspace{0.5mm} $\{\pi_y\}$ \hspace{0.5mm} composed of projectors \hspace{2.5mm} $|v_y\rangle=\sum_{n=0}^{K}r_{y,n}\e^{\i{X_{y,n}}}|2n,2K-2n\rangle$ is
\begin{widetext}
\begin{equation}
\begin{split}
&\operatorname{Tr}\left[\textbf{F}\textbf{H}^{-1}\right]=\\
&=\frac{2}{K(K+1)}\sum_{y=0}^{K}\frac{\left(\sum_{n,n'=0}^{K}R_{y,n,n'}\sin{\left(\Theta_{y,n,n'}\right)}\left(n-n'\right)\right)^2+4\Delta^2\left(\sum_{n,n'=0}^{K}R_{y,n,n'}\cos{\left(\Theta_{y,n,n'}\right)}\left(n-n'\right)^2\right)^2}{\sum_{n,n'=0}^{K}R_{y,n,n'}\cos{\left(\Theta_{y,n,n'}\right)}},\\
\label{eqn:SmallDeltaTradeOffMainText}
\end{split}
\end{equation}
\end{widetext}
where $\Theta_{y,n,n'}=\left(X_{y,n'}-X_{y,n}+2\phi{(n-n')}\right)$ and $R_{y,n,n'}=r_{y,n}r_{y,n'}b_{n}b_{n'}$. The term independent of $\Delta$ corresponds to $F_{11}/H_{11}$, whereas the remaining term containing the factor of $\Delta^2$ corresponds to $F_{22}/H_{22}$. Details of the derivation are given in Appendix~(\ref{app:SmallDeltaTradeoff}). Equation~\eqref{eqn:SmallDeltaTradeOffMainText} shows that in the regime of $\Delta\rightarrow{0}$, to attain high precision about $\Delta$, we require some $\Delta$ dependence in the POVM to reduce the effect of the $\Delta^2$ factor appearing in the $F_{22}$ term. Without knowing this further structure of the optimal POVM, standard maximisation seems to be a challenging task.

To gain some intuition of what the trade-off relation might be as $\Delta\rightarrow{0}$, we present numerical results based on a simulated annealing algorithm performed in the small particle number regime. The results suggest that the bound in the joint estimation of phase and phase diffusion using HB states approaches $1$ as $\Delta$ approaches $0$. Figure~(\ref{fig:BoundMaxSmallDelta}) shows this tendency with the smallest sampled $\Delta$ of $0.01$.

Although the diagonal elements of the QFI matrix ($H_{11}$ and $H_{22}$) reach maximum at $\Delta=0$, the above mentioned numerical results (performed for small particle numbers) suggest that the opposite is the case for the joint information bound.

As an example, we consider the canonical phase measurement POVM ~\cite{Wiseman1998,Schultz2013}, corresponding to a projection onto the phase state $|\psi\rangle=\sum_{n=0}^{\infty}\e^{\i{n}{y}}|n\rangle$,  which for our two mode HB probe state has the form,
\begin{equation}
|v_{y}\rangle=\frac{1}{\sqrt{2\pi}}\sum_{n=0}^{K}\e^{\i{n}{y}}|n,2K-n\rangle.
\end{equation}
This POVM was suggested in Ref.~\cite{Knysh2013} as an optimal measurement for estimating both phase and diffusion in the large particle number regime using the cosine state. 

The canonical phase measurement has no $\Delta$ dependence and therefore for our scheme, according to Eqn.~\eqref{eqn:SmallDeltaTradeOffMainText}, the FI of the diffusion parameter gets vanishingly small as $\Delta\rightarrow{0}$. However, it is still possible to extract some information about $\Delta$ using this POVM if $\Delta$ is not too small. We numerically simulate the performance of this measurement in our estimation scheme for $\Delta=0.01$ and particle number ($K$) regime between $1$ and $80$. The results presented in Fig.~(\ref{fig:CanonicalPhaseMeasurementSmallDelta}) show that in the high particle number regime (considered within the simulation), we can extract significant information about both $\phi$ and $\Delta$. 

\section{\label{sec:6}Trade-off for intermediate diffusion}

\begin{figure}[!t]
\begin{center}
\includegraphics[trim=1.5cm 7cm 1.5cm 6.5cm,width=0.5\columnwidth]{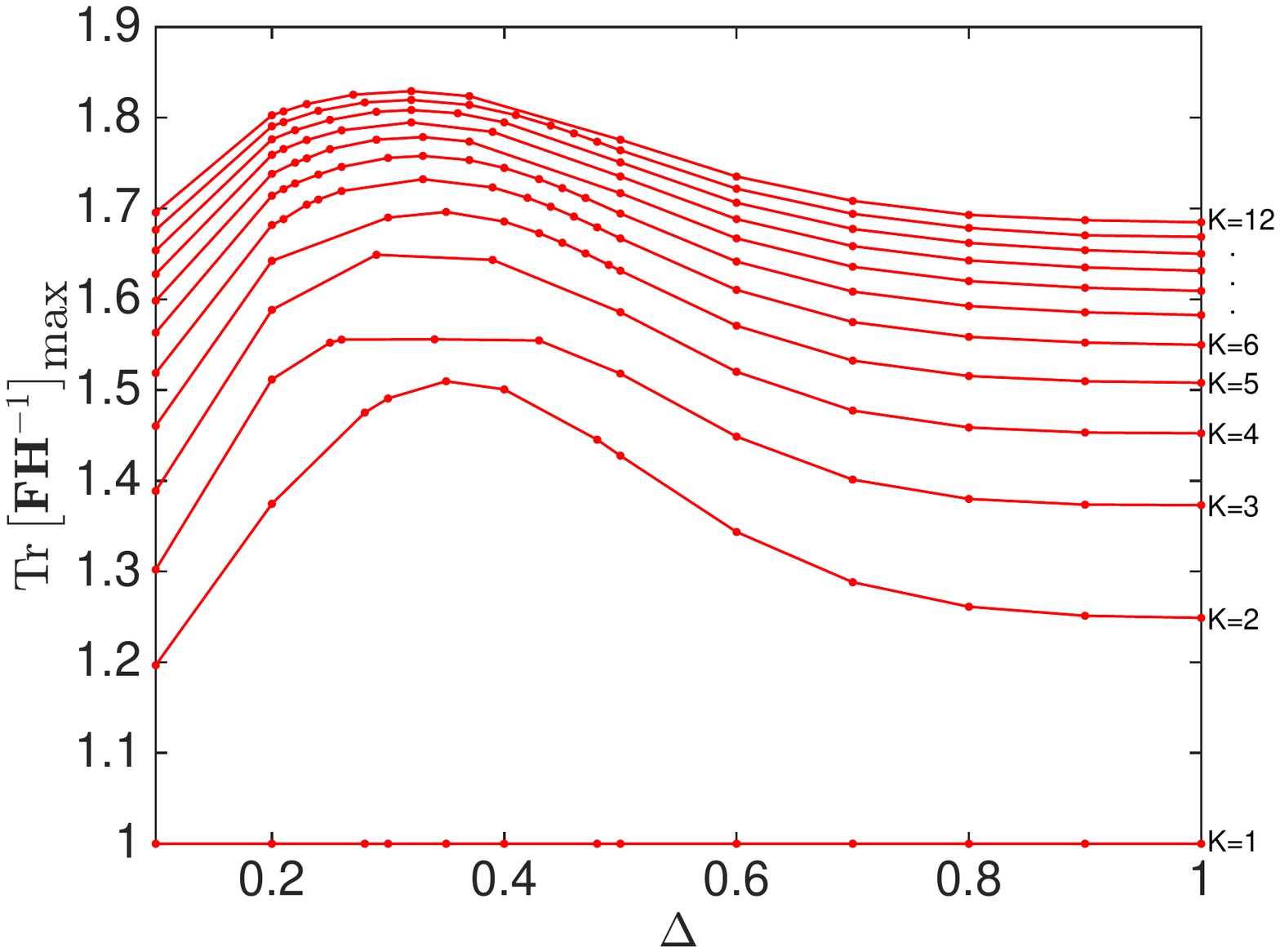}
\end{center}
\caption{Numerical optimal join information bound for HB probe states for different $\Delta$ and $K$ values obtained using the simulated annealing search algorithm. The bound peaks at some intermediate diffusion value (between $0.3$ and $0.4$) for the considered particle numbers.}
\label{fig:BoundMax_vs_DeltaGeneral}
\end{figure}

\begin{figure}[!ht]
\begin{center}
\includegraphics[trim=1.5cm 7cm 1.5cm 6.5cm,width=0.5\columnwidth]{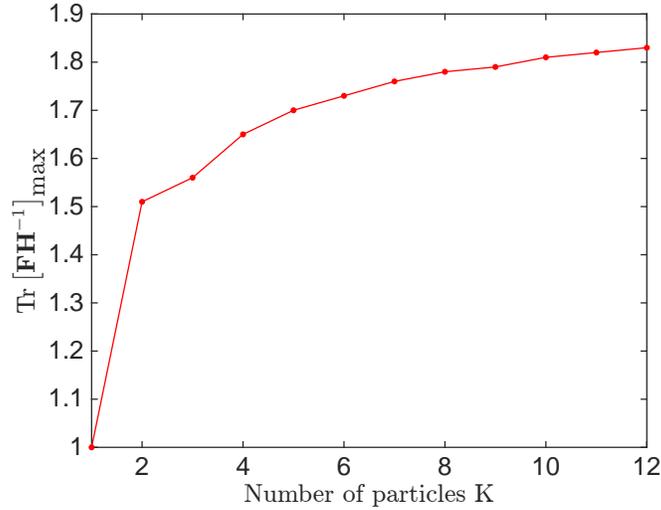}
\end{center}
\caption{Globally maximal joint information bound with respect to all $\Delta$ values plotted as a function of $K$ for HB states.}
\label{fig:BoundMax_vs_KGeneral}
\end{figure}

The analytical understanding of the trade-off (valid for arbitrary $\Delta$ values) in the joint estimation of phase and diffusion for a general FPN state and in the special case of HB states remains an open question. However, the large diffusion analysis shows that for FPN states with equally weighted magnitudes the joint information bound approaches $2$ which is the upper bound set by the QCRB. For HB states, this corresponds to the large particle number regime. Since this is true in the large diffusion regime and $\operatorname{Tr}\left[\textbf{F}\textbf{H}^{-1}\right]\leq{2}$ then it has to also apply to the globally maximal bound across all diffusion values for such states.

We used a simulated annealing algorithm to perform a numerical search on the space of projective measurements to maximise the joint information bound at a given $\Delta$ and $K$ value in the case of HB states. The results are shown in Fig.~(\ref{fig:BoundMax_vs_DeltaGeneral}) and Fig.~(\ref{fig:BoundMax_vs_KGeneral}). The bound peaks at intermediate values of $\Delta$ (between $0.3$ and $0.4$) for the considered $K$ values. As $K$ increases, the numerics suggests that this maximal bound also increases but at a decreasing rate for higher $K$ values. As explained at the beginning of this section, we expect this maximal bound to reach the upper limit of $2$ set by the QCRB.

\section{\label{sec:7}Conclusion}

In any physical system, noise processes are always present and to be able to include them in the estimation schemes is an important step towards real-life, robust metrology. In fact, in many circumstances, noise itself may be the parameter that we require to gain the knowledge of. This brings us to the situation, where the simultaneous estimation of unitary and non-unitary parameters is of interest. The presence of non-unitary parameters (noise) typically makes the analysis of the system much harder since one has to deal with mixed states as opposed to pure states.

In this work, we studied the problem of quantum-limited joint estimation of phase and collective dephasing in the framework of QCRB using FPN states. The QFI for our system is a diagonal matrix whose elements are derived in the regimes of large (for a general FPN state) and small (for the special case of HB states) diffusion by performing Taylor expansion of the evolved density matrix in the appropriate limits. The results are shown in Eqn.~\eqref{eqn:H11fpn}, Eqn.~\eqref{eqn:H22fpn}, Eqn.~\eqref{eqn:H11SmallDelta} and Eqn.~\eqref{eqn:H22SmallDelta}. 

We investigated how closely the precision promised by the QFI can be attained when one tries to estimate phase and diffusion simultaneously by performing projective measurements acting on a single copy of the state. In particular, we were interested how this attainability is affected by the dimension of the state which in the instance of HB states is equal to $(K+1)$. Although the QFI is attainable in the individual estimation of parameters~\cite{Szczykulska2016}, it is not necessarily so in the joint estimation schemes due to the possible non-compatibility of optimal measurements. This gives rise to bounds quantifying how much information can be gained about one parameter in the expense of the knowledge about the other parameters. Such bounds are of fundamental importance in quantum mechanics and therefore worth pursuing. They were studied in previous works for different schemes~\cite{Vidrighin2014,Gill2000,Li2016} and defined by $\operatorname{Tr}\left[\textbf{F}\textbf{H}^{-1}\right]$. In general, the QFI as well as the optimal bound are phase independent.

We analytically derived such joint information bound for phase and collective diffusion in the large diffusion regime (for a general FPN state) by assuming the optimal POVM to be composed of projectors with components of equally weighted magnitudes. The choice of this POVM was motivated by the results from the simulated annealing algorithm which searched through the space of projective measurements to maximise the bound for the special case of HB states. The POVM and the associated bound are shown in Eqn.~\eqref{eqn:LargeDeltaPOVM2},  Eqn.~\eqref{eqn:LargeDeltaPOVM1} and  Eqn.~\eqref{eqn:LargeDeltaTradeoff}.

The analytical bound in the small phase diffusion regime is more difficult to derive due to the more complex structure of the optimal POVM. We expect this POVM to contain some diffusion dependence in order to obtain significant FI for the vanishing diffusion parameter as $\Delta\rightarrow{0}$ (see Sec.~(\ref{sec:5.2})). We therefore used our numerical results for HB states (performed in the small particle number regime) and found that, in this regime, the bound $\operatorname{Tr}\left[\textbf{F}\textbf{H}^{-1}\right]$ gets close to $1$ as $\Delta\rightarrow{0}$. We also found that if diffusion is not too small then it is possible to obtain significant FI for both $\phi$ and $\Delta$ with a POVM which is $\Delta$ independent. This is the case for the canonical phase measurement where our numerical results, performed at $\Delta=0.01$ and for particle number ($K$) between $1$ and $80$, show exactly this for higher $K$ values.

Our main conclusions are:
\begin{enumerate}
\item\label{1.} The QFI matrix in the small phase diffusion regime shows that HB probe states can achieve nearly Heisenberg like scaling not only for phase, but also for diffusion in the instances where the second order in diffusion terms are very small. This is promising for areas where estimation of small diffusive noise parameters is of interest such as thermometry and optomechanics.

\item\label{2.} In principle, quantum-limited simultaneous estimation of phase and phase diffusion is possible in the large phase diffusion regime for FPN states whose components are of equal magnitude. However, the quantum limits themselves decrease exponentially with diffusion and therefore, this is of practical importance in the instances where we are limited to such states. Nevertheless, this result advances the analytical understanding of this simultaneous estimation scheme. Additionally, it shows that the globally maximal bound across all diffusion values, which may or may not coincide with the large phase diffusion regime, will also saturate the QCRB for such states.  

\item\label{3.} As diffusion approaches zero, our numerical simulations for HB states suggest that one cannot achieve the quantum-limited precision for both parameters and the associated trade-off relation, $\operatorname{Tr}\left[\textbf{F}\textbf{H}^{-1}\right]$, approaches one. Therefore, the individual estimation of parameters is always favourable in such instances. However, the simultaneous estimation is sometimes unavoidable, for instance in imaging, and it is important to have the knowledge of the associated limitations. This numerical evidence still awaits an analytical understanding.

\end{enumerate} 

While this work provides several results on the simultaneous estimation of phase and phase diffusion, and the relevant framework for further studies of quantum-limited estimation schemes in the presence of noise, we close with some immediate open questions:
\begin{enumerate}
\item\label{1.} The QFI matrix for HB states and general FPN states for arbitrary diffusion strength. 
\item\label{2.} Analytical joint estimation bound in the small, and ideally, in arbitrary diffusion regimes.
\item\label{3.} Structure of the optimal POVM in the small, and ideally, in the general diffusion case.
\item Proof of optimality of the POVM in Eqn.~\eqref{eqn:LargeDeltaPOVM1} in the large diffusion regime.
\item\label{4.} Translation of the optimal POVMs into feasible experiments enabling optimal joint estimation of phase and diffusion. 
\item Finally, the problem of estimating phase diffusion in the presence of loss and phase~\cite{Crowley2014} is identifiable with the estimation of $T_2$ and $T_1$ times in spin systems~\cite{Degen2016}, while the quantum-limited estimation of phase diffusion simultaneously with multiple phases~\cite{Ballester2004, Humphreys2013, Baumgratz2016, Gagatsos2016, Szczykulska2016} also remains unaddressed.
\end{enumerate}

\section*{Acknowledgements}

MS thanks Mihai Vidrighin, Christos N. Gagatsos, Dominic Branford and Helen Chrzanowski for useful discussions. MS also thanks Ian A. Walmsley for discussions at the early stage of the project. MS is supported by the DSTL (DSTLX-1000091903). This work was also supported by the EPSRC (EP/K04057X/2) and the UK National Quantum Technologies Programme (EP/M01326X/1, EP/M013243/1).

\FloatBarrier
\appendix


\section{\label{app:QFIInTermsOfEigenbasis}QFI matrix elements in terms of the eigenbasis of the density matrix of the input probe state}
The following is a proof for the alternative formula of the QFI matrix elements ($H_{ij}$), given in Eqn.~\eqref{eqn:QFIEigenbasis}, in terms of the eigenbasis of the density matrix of the input probe state. The SLD equation given by Eqn.~\eqref{eqn:SLD} is the so called Lyapunov matrix equation which has a solution
\begin{equation}
\hat{L}_{i}=2\int_{0}^{\infty}dt\e^{-\hat\varrho_{\bm{\lambda}}t}\partial_{\lambda_i}\hat\varrho_{\bm{\lambda}}\e^{-\hat\varrho_{\bm{\lambda}}t}.
\end{equation}
Writing $\hat\varrho_{\bm{\lambda}}$ in its eigenbasis, $\hat\varrho_{\bm\lambda}=\sum_{k}E_{k}|e_{k}\rangle\langle{e}_{k}|$, and applying the definition of an exponential function, $\e^{-\hat\varrho_{\bm{\lambda}}t}=\sum_{k=0}^{\infty}\frac{(-\hat\varrho_{\bm{\lambda}}{t})^k}{k!}$ we get
\begin{equation}
\hat{L}_{i}=2\sum_{n,m}\frac{\langle{e}_{m}|\partial_{\lambda_{i}}\hat\varrho_{\bm{\lambda}}|{e}_{n}\rangle}{E_n+E_m}|e_m\rangle\langle{e}_{n}|.
\end{equation}
$\hat{L}_{i}\hat{L}_{j}$ can therefore be written as
\begin{equation}
\hat{L}_{i}\hat{L}_{j}=4\sum_{n,m,n'}\frac{\langle{e}_{m}|\partial_{\lambda_{i}}\hat\varrho_{\bm{\lambda}}|{e}_{n}\rangle}{E_n+E_m}
\frac{\langle{e}_{n}|\partial_{\lambda_{j}}\hat\varrho_{\bm{\lambda}}|{e}_{n'}\rangle}{E_{n'}+E_n}|{e}_{m}\rangle\langle{e}_{n'}|.
\end{equation}
Writing $\hat\varrho_{\bm\lambda}$ in its eigenbasis form, the following expression for $\hat\varrho_{\bm\lambda}{L}_{i}{L}_{j}$ can be obtained,
\begin{equation}
\hat\varrho_{\bm\lambda}\hat{L}_{i}\hat{L}_{j}=4\sum_{n,m,n'}{E}_{m}\frac{\langle{e}_{m}|\partial_{\lambda_{i}}\hat\varrho_{\bm{\lambda}}|{e}_{n}\rangle}{E_n+E_m}
\frac{\langle{e}_{n}|\partial_{\lambda_{j}}\hat\varrho_{\bm{\lambda}}|{e}_{n'}\rangle}{E_{n'}+E_n}|{e}_{m}\rangle\langle{e}_{n'}|.
\end{equation}
Taking the trace of the above expression and using the eigenbasis form for $\partial_{\lambda_{i}}\hat\varrho_{\bm{\lambda}}$, i.e. $\partial_{\lambda_{i}}\hat\varrho_{\bm{\lambda}}=\sum_{k}\partial_{\lambda_{i}}E_{k}|e_{k}\rangle\langle{e}_{k}|+E_{k}|\partial_{\lambda_{i}}{e}_{k}\rangle\langle{e}_{k}|+E_{k}|e_{k}\rangle\langle\partial_{\lambda_{i}}{e}_{k}|$, we get
\begin{equation}
\begin{split}
\operatorname{Tr}[\hat\varrho_{\bm\lambda}\hat{L}_{i}\hat{L}_{j}]=
4\sum_{n,m}\frac{E_m}{(E_n+E_m)^2}
(\delta_{n,m}\partial_{\lambda_{i}}E_{n}+(E_n-E_m)\langle\partial_{\lambda_{i}}e_{n}|e_{m}\rangle)
(\delta_{n,m}\partial_{\lambda_{j}}E_{m}+(E_m-E_n)\langle\partial_{\lambda_{j}}e_{m}|e_{n}\rangle).
\end{split}
\end{equation}
Multiplying the brackets, using $\partial_{\lambda_{i}}\langle{e}_{n}|e_m\rangle\equiv\langle\partial_{\lambda_{i}}{e}_{n}|e_m\rangle+\langle{e}_{n}|\partial_{\lambda_{i}}{e}_{m}\rangle=0$ and taking the real part of the resulting expression, the required formula for $H_{ij}$ is
\begin{equation}
{H}_{ij}={\operatorname{Re}}\left[\sum_{n}\frac{\partial_{\lambda_{i}}(E_n)\partial_{\lambda_{j}}(E_n)}{E_n}
+4\sum_{n,m}E_m\frac{(E_n-E_m)^2}{(E_n+E_m)^2}
\langle{e}_n|\partial_{\lambda_i}{e_m}\rangle\langle\partial_{\lambda_j}e_{m}|e_n\rangle\right].
\end{equation}


\section{\label{app:FPNstateNewBasis} FPN state in the phase dependent basis}
Complex amplitudes, $a_n$, of $\hat{\varrho}$ can be written in the exponential form as $a_n$ = $|a_{n}|\e^{\i\theta_{n}}$, where $\theta_{n}\in\{0,2\pi\}$. With this, $\hat{\varrho}$ can be expressed as,
\begin{equation}
\begin{split}
\hat{\varrho} = \sum_{n,n'=0}^{2 K}{|{a}_{n}||{a}_{n'}|\e^{\i\left[\phi(n-n')+\theta_n-\theta_{n'}\right]}}
\e^{{-}\frac{\Delta^{2}}{2}(n-n')^2}
|{n},{2}{K}-{n}\rangle\langle{n'},{2}{K}-{n'}|.
\label{eqn:MixedFPN2} 
\end{split}
\end{equation}
All the phases can be absorbed into the basis of $\hat{\varrho}$ and the same is true for its derivatives. Therefore, operating in this new basis, $|\Gamma_{n,K,\phi,\theta}\rangle = e^{\i(\phi{n}+\theta_n)}|n, 2K-n\rangle$, all the matrix elements of $\hat{\varrho}$ and $\partial_{\Delta}\hat\varrho$ are positive, and all the elements of $\partial_{\phi}\hat\varrho$ are imaginary. The corresponding matrix representations are,
\begin{equation}
\hat{\varrho} = \sum_{n,n'=0}^{2 K}{|{a}_{n}||{a}_{n'}|\e^{{-}\frac{\Delta^{2}}{2}(n-n')^2}}|\Gamma_{n,K,\phi,\theta}\rangle\langle{\Gamma}_{n',K,\phi,\theta}|,
\label{eqn:MixedFPN3}
\end{equation}
\begin{equation}
\begin{split}
\partial_{\phi}\hat{\varrho}  = \i\sum_{n,n'=0}^{2 K}{{(n-n')}|{a}_{n}||{a}_{n'}|\e^{{-}\frac{\Delta^{2}}{2}(n-n')^2}}
|\Gamma_{n,K,\phi,\theta}\rangle\langle{\Gamma}_{n',K,\phi,\theta}|,
\label{eqn:MixedFPNdphi} 
\end{split}
\end{equation}
\begin{equation}
\begin{split}
\partial_{\Delta}\hat{\varrho}  = -\Delta\sum_{n,n'=0}^{2 K}{{(n-n')^2}|{a}_{n}||{a}_{n'}|\e^{{-}\frac{\Delta^{2}}{2}(n-n')^2}}
|\Gamma_{n,K,\phi,\theta}\rangle\langle{\Gamma}_{n',K,\phi,\theta}|,.
\label{eqn:MixedFPNddelta} 
\end{split}
\end{equation}


\section{\label{app:QFInumerics}Quantum Fisher information numerical routine}
The numerical routine for calculating QFI relies on vectorising the SLD equation, given in Eqn.~\eqref{eqn:SLD}, so that it becomes
\begin{equation}
2|\partial_{\lambda_i}\hat\varrho\rangle=[\hat\varrho\otimes\hat{I}+\hat{I}\otimes\hat\varrho^{\operatorname{T}}]|L_{i}\rangle,
\label{eqn:SLDnum1}
\end{equation}
where $\hat{I}$ is the identity matrix. $\hat\varrho$ can be written in its eigenbasis representation as $\hat\varrho=VDV^{\dagger}$, where $V$ is the matrix of eigenvectors of $\hat\varrho$ and $D$ is the corresponding diagonal matrix of eigenvalues. Similarly, $\hat\varrho^{\operatorname{T}}$ can be expressed as $V^{\star}{D}V^{\operatorname{T}}$. Therefore, the right hand side of Eqn.~\eqref{eqn:SLDnum1} becomes
\begin{equation}
\begin{split}
&[\hat\varrho\otimes\hat{I}+\hat{I}\otimes\hat\varrho^{T}]|L_{i}\rangle=\\
&=[(V\otimes{V}^{\star})(D\otimes\hat{I})(V^{\dagger}\otimes{V}^{T})\\
&\hspace{5 mm}+(V\otimes{V}^{\star})(\hat{I}\otimes{D})(V^{\dagger}\otimes{V}^{T})]|L_{i}\rangle\\
&=(V\otimes{V}^{\star})(D\otimes\hat{I}+\hat{I}\otimes{D})(V\otimes{V}^{\star})^{\dagger}|L_{i}\rangle.
\label{eqn:SLDnum2}
\end{split}
\end{equation}
$|L_{i}\rangle$ can therefore be found by taking the inverse of $D\otimes\hat{I}+\hat{I}\otimes{D}$ so that it can be written as
\begin{equation}
|L_{i}\rangle=2(V\otimes{V}^{\star})(D\otimes\hat{I}+\hat{I}\otimes{D})^{-1}(V\otimes{V}^{\star})^{\dagger})|\partial_{\lambda_i}\hat\varrho\rangle.
\label{eqn:SLDnum}
\end{equation}
The numerical QFIs can then be found by putting $|L_{i}\rangle$ back into its matrix form and substituting it into Eqn.~\eqref{eqn:QFI}. 


\section{\label{app:LargeDeltaValidity}Validity of the large $\Delta$ approximation}
We evaluate the error associated with the large phase diffusion approximation by considering norm one of the matrix composed of the neglected elements of the original density matrix $\hat{\varrho}$ (Eqn.~\eqref{eqn:MixedFixedPhotonNumberState}). Therefore, the error $\epsilon$ can be expressed as
\begin{equation}
\begin{split}
\epsilon=\sum_{n,n'=0}^{2K}\left|\hat{\varrho}_{n,n'} - \hat{\varrho}_{L\Delta {n,n'}}\right|
=\sum_{\substack{n,n'=0,\\n\neq{n'},\\n\neq{n'\pm{k}}}}^{2K}|a_n||a_{n'}|{x}^{2(n-n')^2},
\end{split}
\end{equation}
where we used the fact that the magnitude of the complex exponential is $1$. We can overestimate the error by setting the power of $x$ to the lowest possible number, $(n-n')_{min}=k+1$, which corresponds to the first off-diagonal that is negligible within the approximation. For the special case of HB states, $k=2$, but also $(n-n')_{min}=k+2$ since only even numbered off-diagonals contribute non-zero elements to the density matrix. With this, the errors for a general FPN state and for the special case of HB states (where $a_n = b_{\frac{n}{2}}\delta_{n,2q}$ as defined in Eqn.~\eqref{eqn:PureHBstate}) are respectively
\begin{equation}
\label{eqn:ErrorFPN1}
\epsilon^{\left(FPN\right)}\leq{x}^{2\left(k+1\right)^2}\left(-1+\sum_{\substack{n,n'=0,\\n\neq{n'\pm{k}}}}^{2K}\left|a_n\right|\left|a_{n'}\right|\right)
\end{equation}
and
\begin{equation}
\begin{split}
\label{eqn:ErrorHB1}
\epsilon^{\left(HB\right)}\leq{x}^{2\left(k+2\right)^2}\left(-1+\sum_{\substack{n,n'=0,\\n\neq{n'\pm{k}}}}^{2K}\left|a_n\right|\left|a_{n'}\right|\right)
=x^{32}\left(-1+\sum_{\substack{n,n'=0,\\n\neq{n'\pm{1}}}}^{K}b_{n}{b}_{n'}\right).
\end{split}
\end{equation}
The summation term in Eqn.~\eqref{eqn:ErrorFPN1} and Eqn.~\eqref{eqn:ErrorHB1} consists of two sums, one positive $\sum_{n,n'=0}^{2K}\left|a_n\right|\left|a_{n'}\right|$ and one negative $2\sum_{\substack{n=k}}^{2K}\left|a_n\right|\left|a_{n-k}\right|$. To further overestimate the error, we can set the negative part of the summation to zero. Additionally, to get a closed form of the error in terms of $K$, we can overestimate the error even further by setting $\left|a_n\right|=\left|a_{n'}\right|$. This is due to the arithmetic-harmonic mean inequality for positive numbers, $\left({\left|a_n\right|+\left|a_{n'}\right|}\right){/}{2}\ge{2\left|a_n\right|\left|a_{n'}\right|}{/}\left({\left|a_n\right|+\left|a_{n'}\right|}\right)$, which can be re-written as $\left|a_n\right|\left|a_{n'}\right|\leq\left({\left|a_n\right|^2+\left|a_{n'}\right|^2}\right){/}{2}$ with equality when $\left|a_n\right|=\left|a_{n'}\right|$. The amount by which the errors are overestimated gets worse when $k>{1}$ since some of the coefficients $a_n$ are zero. We can account for this in the case of HB states by reducing the Hilbert space from $2K$ to $K$ as shown in Eqn.~\eqref{eqn:ErrorHB1}. We therefore get the following errors
\begin{equation}
\begin{split}
\label{eqn:ErrorFPN2}
\epsilon^{\left(FPN\right)}\leq{x}^{2\left(k+1\right)^2}\left(-1+\sum_{\substack{n,n'=0}}^{2K}\left|a_n\right|^2\right)
=2K{x}^{2\left(k+1\right)^2}
\end{split}
\end{equation}
and
\begin{equation}
\label{eqn:ErrorHB2}
\epsilon^{(HB)}\leq{K}{x}^{32}.
\end{equation}
We evaluate the validity of the approximation by defining the error $\epsilon$ (equations ~\eqref{eqn:ErrorFPN1}, ~\eqref{eqn:ErrorHB1}, ~\eqref{eqn:ErrorFPN2} and ~\eqref{eqn:ErrorHB2}) as some fraction $f$ of the sum of the elements of the original density matrix $\hat{\varrho}$. Therefore,
\begin{equation}
\epsilon={f}\sum_{n,n'=0}^{2K}\left|\varrho_{n,n'}\right|=\sum_{n,n'=0}^{2K}|a_n||a_{n'}|{x}^{2(n-n')^2}.
\label{eqn:ErrorDef}
\end{equation}
We can tighten our error bound by using the lowest possible value of $\sum_{n,n'=0}^{2K}\left|\varrho_{n,n'}\right|$ which is $1$. This effectively demands our error to be smaller than that given in  Eqn.~\eqref{eqn:ErrorDef} and re-defines it as,
\begin{equation}
\epsilon={f}.
\end{equation}
This translates to the following regimes of $\Delta$ within which our approximation is valid,
\begin{equation}
\begin{split}
\label{eqn:LargeDeltaValidityFPN1}
\Delta^{(FPN)}\geq\sqrt{\frac{2}{\left(k+1\right)^2}\ln\left(\frac{2K}{f}\right)}
\end{split}
\end{equation}
or in terms of the coefficients of the state $a_n$
\begin{equation}
\begin{split}
\label{eqn:LargeDeltaValidityFPN2}
\Delta^{(FPN)}\geq\sqrt{\frac{2}{\left(k+1\right)^2}}
\sqrt{\ln\left[\frac{1}{f}\left(-1+\sum_{\substack{n,n'=0,\\n\neq{n'\pm{k}}}}^{2K}\left|a_n\right|\left|a_{n'}\right|\right)\right]}
\end{split}
\end{equation}
and
\begin{equation}
\begin{split}
\label{eqn:LargeDeltaValidityHB1}
\Delta^{(HB)}\geq\sqrt{\frac{1}{8}\ln\left(\frac{K}{f}\right)}
\end{split}
\end{equation}
or in terms of the coefficients of the HB state $b_n$
\begin{equation}
\begin{split}
\label{eqn:LargeDeltaValidityHB2}
\Delta^{(HB)}\geq\sqrt{\frac{1}{8}\ln\left[\frac{1}{f}\left(-1+\sum_{\substack{n,n'=0,\\n\neq{n'\pm{k}}}}^{K}b_n{b}_{n'}\right)\right]}.
\end{split}
\end{equation}
Eqn.~\eqref{eqn:LargeDeltaValidityFPN1} and Eqn.~\eqref{eqn:LargeDeltaValidityHB1} show the $\sqrt{\ln{(K)}}$ dependence on the threshold value of $\Delta$ for the considered relative error $f$.

\begin{figure}[t!]
\begin{center}
\includegraphics[trim=1.5cm 7cm 1.5cm 6.5cm,width=0.5\columnwidth]{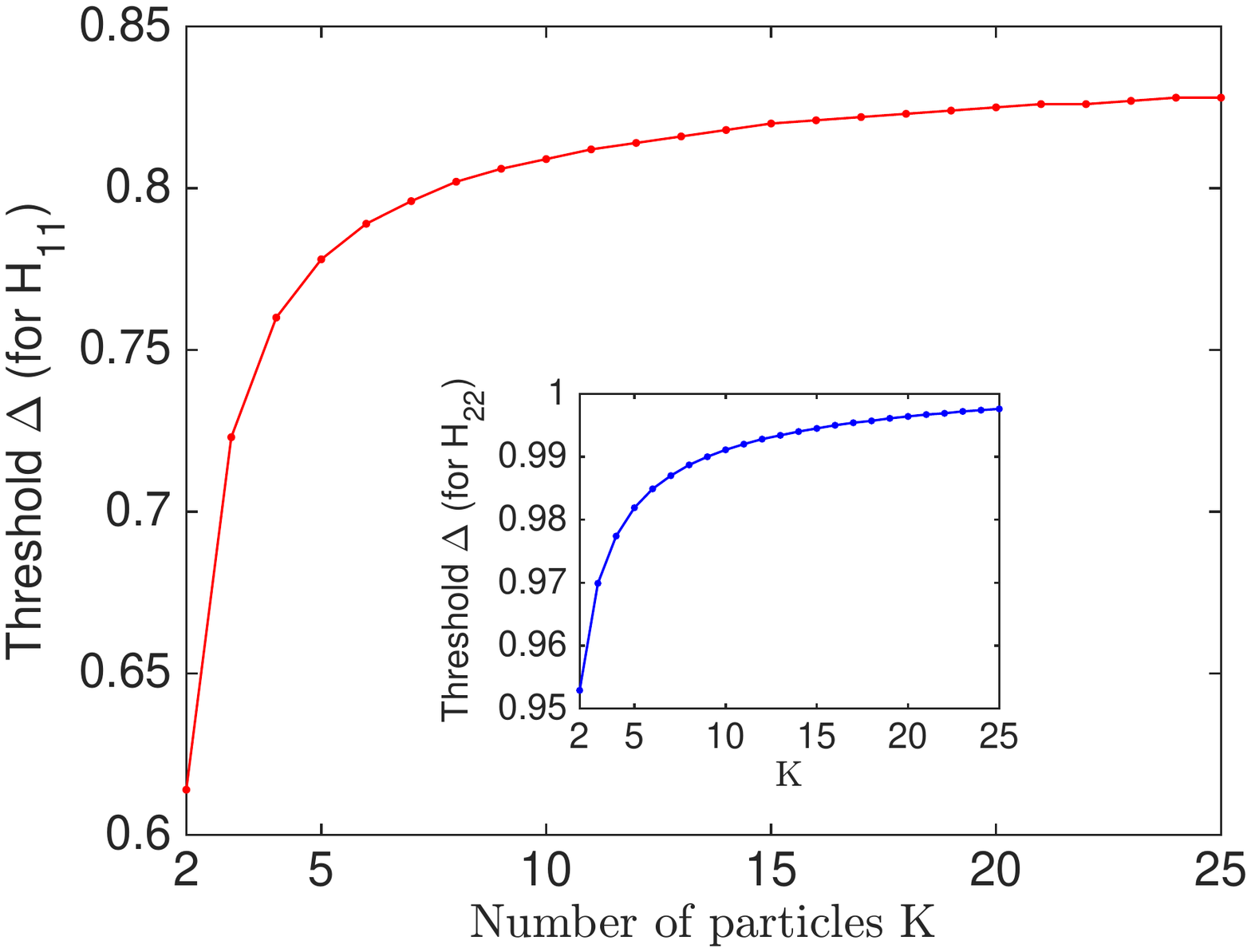}
\end{center}
\caption{Phase diffusion numerical validity plot within the large phase diffusion approximation for HB states. The figure shows the threshold values of $\Delta$ as a function of particle number ($K$) when the highest allowed relative error on the QFI is $0.05$. $K=1$ case is not shown as it already gives a tridiagonal density matrix and therefore an exact solution valid for all $\Delta$.}
\label{fig:QFI_LargeDeltaValidity}
\end{figure}

The exact numerical validity plot for the actual QFI is shown in Fig.~(\ref{fig:QFI_LargeDeltaValidity}) in the case of HB states. It shows the minimum values of $\Delta$ as a function of particle number ($K$) when the highest allowed relative error ($\delta{H_{ii}}/H_{ii}$) is $0.05$.

\section{\label{app:SLDsLargeDelta}SLD matrices in the large diffusion regime}
In the large $\Delta$ approximation, the SLDs for a general FPN state are given by the following expressions,
\begin{equation}
\begin{split}
\label{eqn:TridiagFPNsld1}
\hat{L}_{1,\text{L}\Delta}=2\i{x}^{2k^2}\sum_{n,n'=0}^{2K}(n-n')\frac{{a}_{n}{a}_{n'}^*}{|{a}_{n}|^{2}+|{a}_{n'}|^{2}}
\e^{\i\phi(n-n')}\delta_{n,n'\pm k}
|{n},{2}{K}-{n}\rangle\langle{n'},{2}{K}-{n'}|
\end{split}
\end{equation}
and
\begin{equation}
\begin{split}
\label{eqn:TridiagFPNsld2}
\hat{L}_{2,\text{L}\Delta}=
-2k^2\Delta{x}^{2k^2}\sum_{n,n'=0}^{2K}\frac{{a}_{n}{a}_{n'}^*}{|{a}_{n}|^{2}+|{a}_{n'}|^{2}}
\e^{\i\phi(n-n')}\delta_{n,n'\pm k}
|{n},{2}{K}-{n}\rangle\langle{n'},{2}{K}-{n'}|,
\end{split}
\end{equation}
where subscripts 1 and 2 refer to $\phi$ and $\Delta$ respectively. The above SLDs can be substituted into the SLD equation, Eqn.~\eqref{eqn:SLD}, to check that these are the correct solutions within the approximation. Firstly, we perform such verification for $\hat{L}_{1,\text{L}\Delta}$. The left hand side of the SLD equation is,
\begin{equation}
\begin{split}
\label{eqn:TridiagSLD1eqnLHS}
2\partial_{\phi}{\hat\varrho}_{\text{L}\Delta}=2\i\sum_{n,n'=0}^{2K}&{a}_{n}{a}_{n'}^*(n-n')\e^{\i\phi(n-n')}
{x}^{2({n}-{n'})^2}(\delta_{n,n'}+\delta_{n,n'\pm{k}})
|{n},{2}{K}-{n}\rangle\langle{n'},{2}{K}-{n'}|\\
&\hspace{-17.5 mm}=2\i{x}^{2k^2}\sum_{n,n'=0}^{2K}{a}_{n}{a}_{n'}^*(n-n')
\e^{\i\phi(n-n')}\delta_{n,n'\pm{k}}
|{n},{2}{K}-{n}\rangle\langle{n'},{2}{K}-{n'}|,
\end{split}
\end{equation}
where $\delta_{n,n'}$ does not contribute any non-zero terms due to the factor of $(n-n')$ appearing in the summation. To calculate the right hand side of the SLD equation, we require expressions for $\hat\varrho_{\text{L}\Delta}\hat{L}_{{1,\text{L}\Delta}}$ and $\hat{L}_{1,\text{L}\Delta}\hat\varrho_{\text{L}\Delta}$,
\begin{equation}
\begin{split}
\label{eqn:TridiagSLD1eqnRHS1}
&\hat\varrho_{\text{L}\Delta}\hat{L}_{1,\text{L}\Delta}=
2\i{x}^{2k^2}\sum_{n,n',m'=0}^{2K}(n-n')\frac{a_{n}|a_{n'}|^2{a}_{m'}^*}{|a_{n}|^2+|a_{n'}|^2}\e^{i\phi(n-m')}{x}^{2(n'-m')^2}\\
&\hspace{41 mm}\times(\delta_{n',m'}+\delta_{n',m'\pm k})
\delta_{n,n'\pm k}|{n},{2}{K}-{n}\rangle\langle{m'},{2}{K}-{m'}|\\
&\hspace{14.5 mm}=2\i{x}^{2k^2}\sum_{n,n'=0}^{2K}(n-n')
\frac{a_{n}{a}_{n'}^*|{a_{n'}}|^2}{|a_{n}|^2+|a_{n'}|^2}\e^{\i\phi(n-n')}\delta_{n,n'\pm k}
|{n},{2}{K}-{n}\rangle\langle{n'},{2}{K}-{n'}|
\end{split}
\end{equation}
and
\begin{equation}
\begin{split}
\label{eqn:TridiagSLD1eqnRHS2}
&\hat{L}_{1,\text{L}\Delta}\hat\varrho_{\text{L}\Delta}=(\hat\varrho_{\text{L}\Delta}\hat{L}_{1,\text{L}\Delta})^{\dagger}=
2\i{x}^{2k^2}\sum_{n,n'=0}^{2K}(n-n')
\frac{a_{n}{a}_{n'}^*|{a_{n}}|^2}{|a_{n}|^2+|a_{n'}|^2}\e^{\i\phi(n-n')}\delta_{n,n'\pm k}
|{n},{2}{K}-{n}\rangle\langle{n'},{2}{K}-{n'}|.
\end{split}
\end{equation}
In obtaining Eqn.~\eqref{eqn:TridiagSLD1eqnRHS1} and Eqn.~\eqref{eqn:TridiagSLD1eqnRHS2}, we keep terms of order up to $x^{2k^2}$ only. The right hand side of the SLD equation is a sum of Eqn.~\eqref{eqn:TridiagSLD1eqnRHS1} and Eqn.~\eqref{eqn:TridiagSLD1eqnRHS2} and hence
\begin{equation}
\begin{split}
\label{eqn:TridiagSLD1eqnRHS}
\hat\varrho_{\text{L}\Delta}\hat{L}_{1,\text{L}\Delta}+\hat{L}_{1,\text{L}\Delta}\hat\varrho_{\text{L}\Delta}&=
2\i{x}^{2k^2}\sum_{n,n'=0}^{2K}({n}-{n'})\frac{{a}_{n}{a}_{n'}^*(|{a}_{n}|^2+|{a}_{n'}|^2)}{|{a}_{n}|^2+|{a}_{n'}|^2}
\e^{\i\phi(n-n')}\delta_{n,n'\pm k}
|{n},{2}{K}-{n}\rangle\langle{n'},{2}{K}-{n'}|\\
&=2\i{x}^{2k^2}\sum_{n,n'=0}^{2K}({n}-{n'}){a}_{n}{a}_{n'}^*
\e^{\i\phi(n-n')}\delta_{n,n'\pm k}
|{n},{2}{K}-{n}\rangle\langle{n'},{2}{K}-{n'}|\\
\end{split}
\end{equation}
which is equal to Eqn.~\eqref{eqn:TridiagSLD1eqnLHS} i.e. $2\partial_{\phi}{\hat\varrho}_{\text{L}\Delta}$. Therefore, Eqn.~\eqref{eqn:TridiagFPNsld1} gives the correct expression for $\hat{L}_{1,\text{L}\Delta}$ up to the order of $x^{2k^2}=\e^{-{2k^{2}\Delta^{2}}{/}{4}}$. A similar proof can be constructed for $\hat{L}_{2,\text{L}\Delta}$, however, here the order of accuracy is attenuated from $x^{2k^2}$ to $x^{2k^2-{4}{\Delta^{-2}}\ln{\left(\Delta\right)}}$ due to the presence of the factor of $\Delta$.


\section{\label{app:QFIsLargeDelta}QFI elements in the large diffusion regime}
The following is the calculation of $H_{11,\text{L}\Delta}$, QFI corresponding to the individual estimation of phase, in the large diffusion regime for a general FPN state. Firstly, we calculate $\hat{L}_{1,\text{L}\Delta}^2$,
\begin{equation}
\begin{split}
\label{eqn:L1L1LargeDelta}
&(\hat{L}_{1,\text{L}\Delta})^2 =
-4{x}^{4k^2}\sum_{n,n',m'=0}^{2K}(n-n')(n'-m')
\frac{{a}_{n}{a}_{m'}^*|{a}_{n'}|^2}{(|{a}_{n}|^2+|{a}_{n'}|^2)(|{a}_{n'}|^2+|{a}_{m'}|^2)}
\e^{\i\phi(n-m')}\delta_{n,n'\pm {k}}\delta_{n',m'\pm {k}}\\
&\hspace{40 mm}\times{|}{n},{2}{K}-{n}\rangle\langle{m'},{2}{K}-{m'}|.
\end{split}
\end{equation}
With this, 
\begin{equation}
\begin{split}
\label{eqn:RhoL1L1LargeDelta}
\hat\varrho_{\text{L}\Delta}(\hat{L}_{1,\text{L}\Delta})^2 &=
-4{x}^{4k^2}\sum_{l,n,n',m'=0}^{2K}(n-n')(n'-m')
\frac{{a}_{l}|{a}_{n}|^2|{a}_{n'}|^2{a}_{m'}^*}{(|{a}_{n}|^2+|{a}_{n'}|^2)(|{a}_{n'}|^2+|{a}_{m'}|^2)}
\e^{\i\phi(l-m')}{x}^{2(l-n)^2}\\
&\hspace{30 mm}\times\delta_{n,n'\pm {k}}\delta_{n',m'\pm {k}}(\delta_{l,n}+\delta_{l,n\pm k})
|{l},{2}{K}-{l}\rangle\langle{m'},{2}{K}-{m'}|\\
&=-4{x}^{4k^2}\sum_{n,n',m'=0}^{2K}(n-n')(n'-m')
\frac{{a}_{n}|{a}_{n}|^2|{a}_{n'}|^2{a}_{m'}^*}{(|{a}_{n}|^2+|{a}_{n'}|^2)(|{a}_{n'}|^2+|{a}_{m'}|^2)}\e^{\i\phi(n-m')}\\
&\hspace{29 mm}\times\delta_{n,n'\pm {k}}\delta_{n',m'\pm {k}}
{|}{n},{2}{K}-{n}\rangle\langle{m'},{2}{K}-{m'}|,
\end{split}
\end{equation}
where in Eqn.~\eqref{eqn:RhoL1L1LargeDelta}, terms of order higher than $x^{4k^2}$ are neglected as explained in the main text. The total neglected term is given by $\delta{q}_{\phi}$,
\begin{equation}
\begin{split}
\label{eqn:QFIneglectedTerms}
&\delta{q}_{\phi}=
-4x^{4k^2}\sum_{l,n,n',m=0}^{2K}(n-n')(n'-m)
\frac{a_{l}|a_{n}|^2|a_{n'}|^2a_{m}^*}{(|a_n|^2+|a_{n'}|^2)(|a_{n'}|^2+|a_{m}|^2)}
\e^{\i\phi(l-m)}x^{2(l-n)^2}\\
&\hspace{34 mm}\times\delta_{n,n'\pm{k}}\delta_{n',m\pm{k}}\delta_{l,n\pm{k}}
|l,2K-l\rangle\langle{m,2K-m}|.
\end{split}
\end{equation}
Taking the trace of the neglected term in Eqn.~\eqref{eqn:QFIneglectedTerms} gives zero and therefore neglecting terms of order higher than $x^{4k^2}$ contributes no error to the actual QFI. Taking the trace of Eqn.~\eqref{eqn:RhoL1L1LargeDelta} gives,
\begin{equation}
\begin{split}
\label{eqn:TrRhoL1L1LargeDelta}
\operatorname{Tr}[\hat\varrho_{\text{L}\Delta}(\hat{L}_{1,\text{L}\Delta})^2 ]&=
4{x}^{4k^2}\sum_{n,n'=0}^{2K}(n-n')^2\frac{|{a}_{n}|^4|{a}_{n'}|^2}{(|{a}_{n}|^2+|{a}_{n'}|^2)^2}\delta_{n,n'\pm k}\\
&=4k^2{x}^{4k^2}\left(\sum_{n=k}^{2K}\frac{|{a}_{n}|^4|{a}_{n-k}|^2}{(|{a}_{n}|^2+|{a}_{n-k}|^2)^2}\right.
\left.+\sum_{n'=k}^{2K}\frac{|{a}_{n'-k}|^4|{a}_{n'}|^2}{(|{a}_{n'-k}|^2+|{a}_{n'}|^2)^2}\right)\\
&=4k^2{x}^{4k^2}\sum_{n=k}^{2K}\frac{|{a}_{n}|^2|{a}_{n-k}|^2}{|{a}_{n}|^2+|{a}_{n-k}|^2}.
\end{split}
\end{equation}
As all the terms in the above expression are real then this is also equal to $H_{11,\text{L}\Delta}$. The same procedure can be performed to calculate $H_{22,\text{L}\Delta}$, QFI corresponding to the individual estimation of $\Delta$. In this case, the neglected term also contributes zero to the actual QFI.

We can upper-bound the summation term, $A$ = $\sum_{n=k}^{2K}{|{a}_{n}|^2|{a}_{n-k}|^2}{/}\left({|{a}_{n}|^2+|{a}_{n-k}|^2}\right)$, appearing in Eqn.~\eqref{eqn:TrRhoL1L1LargeDelta} by noting that the terms inside the sum have the form of the harmonic mean and therefore,
\begin{equation}
A\leq\frac{1}{2}\sum_{n=k}^{2K}\frac{|{a}_{n}|^2+|{a}_{n-k}|^2}{2}
\end{equation}
with the equality when $|a_n|^2$ = $|a_{n-k}|^2$ for $n\in\mathbb{Z}\{k:2K\}$. Further, we can expand the sum on the right hand side and re-group its elements to give
\begin{equation}
A\leq\frac{1}{4}(|a_0|^2+...+|a_{2K}|^2+|a_k|^2+...+|a_{2K-k}|^2).
\end{equation}
Using $\sum_{n=0}^{2K}{|a_n|^2}=1$ we get
\begin{equation}
A\leq\frac{1}{2}-\frac{1}{4}\sum_{n=0}^{k-1}(|a_n|^2+|a_{2K-n}|^2)
\end{equation}
which gives the upper bound of ${1}{/}{2}$ when $\sum_{n=0}^{k-1}(|a_n|^2+|a_{2K-n}|^2){/}{4}=0$.

\begin{figure}[!t]
\begin{center}
\includegraphics[trim=2cm 0.6cm 2.4cm 0.005cm,width=0.33\columnwidth]{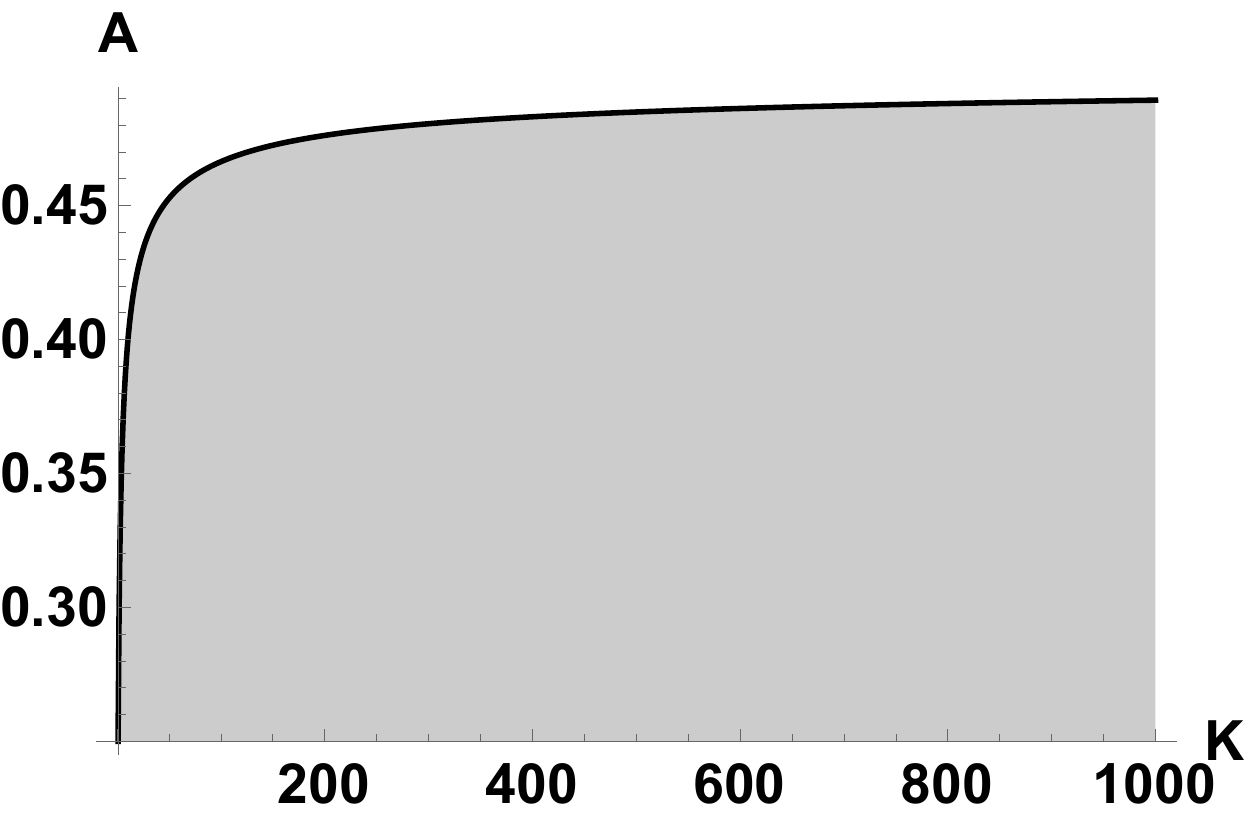}
\end{center}
\caption{The summation term ($A$) in Eqn.~\eqref{eqn:TrRhoL1L1LargeDelta} gets close to ${1}/{2}$ at large $K$. For $K$ = $1000$, $A$ = $0.49$.}
\label{fig:HBcoeff}
\end{figure}

For HB states, the upper bound of ${1}{/}{2}$ is reached when $\left(|b_0|^2+|b_{K}|^2\right){/}4=0$. This condition and hence the upper bound is asymptotically reached when $K$ gets large as shown in Fig.~(\ref{fig:HBcoeff}).  

\section{\label{app:LargeDeltaPlot}Complementary QFI plot to Fig.~(\ref{fig:QFIvsDelta_LargeDelta})}
\begin{figure}[!t]
\begin{center}
\includegraphics[trim=1.5cm 7cm 1.5cm 6.5cm,width=0.5\columnwidth]{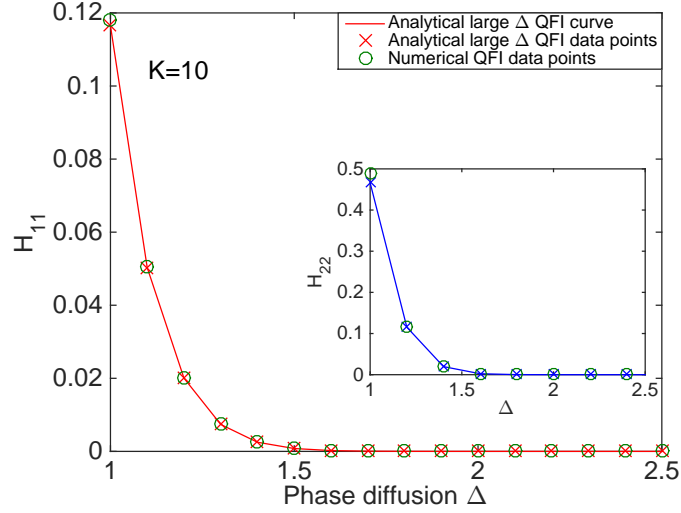}
\end{center}
\caption{Complementary figure to Fig.~(\ref{fig:QFIvsDelta_LargeDelta}) showing a linear plot of the QFI as a function of $\Delta$ for large phase diffusion values. Additionally to the curve, we plotted the sampled $\Delta$ points for the analytical QFI expression (crosses) to magnify the calculation error for smaller $\Delta$ values.}
\label{fig:QFI_LargeDelta2}
\end{figure}

Fig.~(\ref{fig:QFI_LargeDelta2}) shows a linear plot of the QFI as a function of $\Delta$ for large phase diffusion values.

\section{\label{app:LargeDeltaTradeoff}Trade-off in the large diffusion regime}
We consider orthonormal projective measurments acting on a single copy of the evolved FPN state, given in Eqn.~\eqref{eqn:MixedFixedPhotonNumberState}. Such a measurement can be realized by a POVM set $\{|v_{y}\rangle\langle{v_y}|\}$, where
\begin{equation}
\label{eqn:Projector}
|v_y\rangle=\sum_{n=0}^{2K}r_{y,n}\e^{\i{X}_{y,n}}|n,2K-n\rangle
\end{equation}
with $r_{y,n}\in\Re\{0:1\}$ and $X_{y,n}\in\Re\{0:2\pi\}$. The completeness relation, $\sum_{y=0}^{2K}|v_y\rangle\langle{v_y}|=\mathbb{\hat{1}}$, and the normalisation give the following constraints respectively,
\begin{equation}
\label{eqn:CompletenessRelation}
\sum_{y=0}^{2K}r_{y,n}r_{y,n'}\e^{\i{(X_{y,n}-X_{y,n'})}}=\delta_{n,n'}
\end{equation} 
and 
\begin{equation}
\label{eqn:Normalisation}
\sum_{n=0}^{2K}r_{y,n}^2=1.
\end{equation}
The probability associated with the $y^{\textit{th}}$ outcome and its corresponding derivatives with respect to $\phi$ and $\Delta$ are given by,
\vskip-0.1in
\begin{align}
\begin{split}
\label{eqn:Probability}
p_y&=\langle{v_y}|\hat{\varrho}_{L\Delta}|v_y\rangle
=\sum_{n=0}^{2K}r_{y,n}^{2}|a_n|^2+
2x^{2k^2}\sum_{n=k}^{2K}r_{y,n}r_{y,n-k}|a_n||a_{n-k}|\cos(\Theta_{y,n,k})
\end{split}
\end{align}
\begin{align}
\begin{split}
\label{eqn:DerivativeProbPhi}
\partial_{\phi}{p_y}=&-2kx^{2k^2}\sum_{n=k}^{2K}r_{y,n}r_{y,n-k}|a_n||a_{n-k}|
\sin{\left(\Theta_{y,n,k}\right)}
\end{split}
\end{align}
and
\begin{align}
\begin{split}
\label{eqn:DerivativeProbDelta}
\partial_{\Delta}{p_y}=&-2k^2\Delta{x}^{2k^2}\sum_{n=k}^{2K}r_{y,n}r_{y,n-k}|a_n||a_{n-k}|
\cos{\left(\Theta_{y,n,k}\right)},
\end{split}
\end{align}
where $\Theta_{y,n,k}=\left(X_{y,n-k}-X_{y,n}+\theta_{n}-\theta_{n-k}+k\phi\right)$.
Using Eqn.~\eqref{eqn:ClassicalFisherInformation}, the resulting elements of the Fisher information matrix, $F_{11}$ and $F_{22}$, corresponding to the individual estimation of $\phi$ and $\Delta$ are given by, 
\begin{widetext}
\begin{eqnarray}
\begin{split}
\label{eqn:LargeDeltaF11}
F_{11}&=4k^2{x}^{4k^2}\sum_{y=0}^{2K}\left(\frac{\left(\sum_{n=k}^{2K}r_{y,n}r_{y,n-k}|a_n||a_{n-k}|\sin{\left(\Theta_{y,n,k}\right)}\right)^2}{\sum_{n=0}^{2K}\left({r_{y,n}^2}|a_n|^2\right)+2x^{2k^2}\sum_{n=k}^{2K}\left(r_{y,n}r_{y,n-k}|a_n||a_{n-k}|\cos{\left(\Theta_{y,n,k}\right)}\right)}\right)\\
&=4k^2{x}^{4k^2}\sum_{y=0}^{2K}\left(\frac{\left(\sum_{n=k}^{2K}r_{y,n}r_{y,n-k}|a_n||a_{n-k}|\sin{\left(\Theta_{y,n,k}\right)}\right)^2}{\sum_{n=0}^{2K}{r_{y,n}^2}|a_n|^2}\right)
\end{split}
\end{eqnarray}
and
\begin{eqnarray}
\begin{split}
\label{eqn:LargeDeltaF22}
F_{22}&=4k^4\Delta^2{x}^{4k^2}\sum_{y=0}^{2K}\left(\frac{\left(\sum_{n=k}^{2K}r_{y,n}r_{y,n-k}|a_n||a_{n-k}|\cos{\left(\Theta_{y,n,k}\right)}\right)^2}{\sum_{n=0}^{2K}\left({r_{y,n}^2}|a_n|^2\right)+2x^{2k^2}\sum_{n=k}^{2K}\left(r_{y,n}r_{y,n-k}|a_n||a_{n-k}|\cos{\left(\Theta_{y,n,k}\right)}\right)}\right)\\
&=4k^4\Delta^2{x}^{4k^2}\sum_{y=0}^{2K}\left(\frac{\left(\sum_{n=k}^{2K}r_{y,n}r_{y,n-k}|a_n||a_{n-k}|\cos{\left(\Theta_{y,n,k}\right)}\right)^2}{\sum_{n=0}^{2K}{r_{y,n}^2}|a_n|^2}\right),
\end{split}
\end{eqnarray}
where we neglect orders higher than $x^{4k^2}$ to obtain the final expressions. The trade-off, $\operatorname{Tr}\left[\textbf{F}\textbf{H}^{-1}\right]={F_{11}}{/}{H_{11}}+{F_{22}}{/}{H_{22}}$, is calculated using quantum limits found in Eqn.~\eqref{eqn:H11fpn} and Eqn.~\eqref{eqn:H22fpn}. It is
\begin{eqnarray}
\begin{split}
\label{eqn:LargeDeltaTradeoff2}
\operatorname{Tr}[\textbf{F}\textbf{H}^{-1}]=\frac{1}{s}\sum_{y=0}^{2K}(\frac{1}{{\sum_{n=0}^{2K}{r_{y,n}^2}|a_n|^2}}\sum_{n,n'=k}^{2K}&r_{y,n}r_{y,n-k}r_{y,n'}r_{y,n'-k}|a_n||a_{n-k}||a_{n'}||a_{n'-k}|\\
&\times\left(\sin{\left(\Theta_{y,n,k}\right)}\sin{\left(\Theta_{y,n',k}\right)}+\cos{\left(\Theta_{y,n,k}\right)}\cos{\left(\Theta_{y,n',k}\right)}\right))\\
&\hspace{-47mm}=\frac{1}{s}\sum_{y=0}^{2K}\left(\frac{\sum_{n,n'=k}^{2K}r_{y,n}r_{y,n-k}r_{y,n'}r_{y,n'-k}|a_n||a_{n-k}||a_{n'}||a_{n'-k}|\cos{\left(\Theta_{y,n,k}-\Theta_{y,n',k}\right)}}{{\sum_{n=0}^{2K}{r_{y,n}^2}|a_n|^2}}\right),\\
\end{split}
\end{eqnarray}
\end{widetext}
where $s=\sum_{n=k}^{2K}{|a_n|^2|a_{n-k}|^2}{/}\left({|a_n|^2+|a_{n-k}|^2}\right)$ and to get the final expression, we used the identity $\cos{\left(a-b\right)}=\cos{\left(a\right)}\cos{\left(b\right)}+\sin{\left(a\right)}\sin{\left(b\right)}$.
Since $s$, $r_{y,n}$ and $|a_n|$ in Eqn.~\eqref{eqn:LargeDeltaTradeoff2} are positive then the trade-off is maximised when $\cos{\left(\Theta_{y,n,k}-\Theta_{y,n',k}\right)}=1$. Further, assuming projectors whose coefficients have equally weighted magnitudes i.e. $r_{y,n}={1}{/}{\sqrt{\frac{2K}{k}+1}}$, we get the trade-off as given by Eqn. ~\eqref{eqn:LargeDeltaTradeoff}.


\section{\label{app:LinearIndependenceSmallDelta}Linear independence of the elements of sets $\mathcal{W}_1$ and $\mathcal{W}_2$}

Sets $\mathcal{W}_{1}$ and $\mathcal{W}_{2}$ are described in Eqn.~\eqref{eqn:W1} and Eqn.~\eqref{eqn:W2}. Their elements consist of vectors $|w_{k}\rangle=\sum_{n=0}^{K}n^{k-1}|\varphi_{n}\rangle$ for $k=\{1,...,N\}$, where $N$ (number of vector elements) is 3 for $\mathcal{W}_1$ and 5 for $\mathcal{W}_2$. We are interested in a situation, where $(K+1)\geq{N}$ since the size of the density matrix in the number basis is $(K+1)\times{(K+1)}$. A set is linearly dependent if one of the vectors in the set can be expressed as a linear combination of the other vectors. If none of the vectors is linearly dependent then the set is said to be linearly independent. Mathematically, the linear independence can be stated as
\begin{equation}
\begin{split}
&\sum_{k=1}^{N}c_{k-1}\sum_{n=0}^{K}n^{k-1}|\varphi_{n}\rangle=0\\
&\operatorname{iff}{c}_{k-1}=0\operatorname{for}{k}=\{1,...,N\}.
\label{eqn:LinearDependence1}
\end{split}
\end{equation}
The above condition can be re-written as 
\begin{equation}
\begin{split}
&\sum_{n=0}^{K}\left(\sum_{k=1}^{N}{c}_{k-1}n^{k-1}\right)|\varphi_{n}\rangle=0\\
&\operatorname{iff}{c}_{k-1}=0\operatorname{for}{k}=\{1,...,N\}
\label{eqn:LinearDependence2}
\end{split}
\end{equation}
and since vectors $\varphi_{n}=b_{n}\e^{2\i\phi{n}}|2n,2K-2n\rangle$, $n=\{0,...,K\}$ are orthogonal then the condition for linear independence can be further simplified to
\begin{equation}
\begin{split}
&\sum_{k=1}^{N}{c}_{k-1}n^{k-1}={c}_{0}+{c}_{1}{n}+{c}_{2}{n}^{2}+...+{c}_{N}{n}^{N}=0\\
&\operatorname{for}{n}=\{0,...,K\}\\
&\operatorname{iff}{c}_{k-1}=0\operatorname{for}{k}=\{1,...,N\}.
\label{eqn:LinearDependence3}
\end{split}
\end{equation}
We, therefore, obtain $(K+1)$ distinct, simultaneous equations, corresponding to cases $n=\{0,...,K\}$, with $N$ unknowns. However, the first equation ($n=0$ case) imposes condition $c_0=0$. Therefore, after solving for the first $N$ equations we will obtain two equations relating two unknowns, but there will be a contradiction because the two unknowns (in each of the two equations) will be related by a different constant factor because the coefficients in front of $c_{k-1}$ take different values for each equation. Therefore, if $(K+1)\geq{N}$ then $\sum_{k=1}^{N}{c}_{k-1}n^{k-1}$ can only be satisfied if ${c}_{k-1}=0\operatorname{for}{k}=\{1,...,N\}$ as required by the linear independence condition.  

To see this, we can consider an example of $\mathcal{W}_1$ set containing $N=3$ vectors and $K=2$. We get the following 3 simultaneous equations,

\begin{equation}
\begin{split}
&{c}_{0}=0\\
&{c}_{0}+{c}_{1}+{c}_{2}=0\\
&{c}_{0}+2{c}_{1}+4{c}_{2}=0.
\label{eqn:SimEq}
\end{split}
\end{equation}
It is clear that if $c_0=0$, as required by the first equation in ~\eqref{eqn:SimEq}, then the remaining two equations, will be in contradiction.


\section{\label{app:GramSchmidt}Gram Schmidt procedure and the orthonormal sets $\mathcal{V}_{1}$ and $\mathcal{V}_{2}$}

The first vector $|w_1\rangle$ of sets $\mathcal{W}_1$ and $\mathcal{W}_2$, given in equations~\eqref{eqn:W1} and~\eqref{eqn:W2}, is already normalised and therefore, it can be taken as the first element $|v_1\rangle$ of the orthonormal sets $\mathcal{V}_1$ and $\mathcal{V}_2$. The remaining vectors are found using 
\begin{equation}
\label{eqn:GramSchmidt}
|v_{k+1}\rangle=\frac{|w_{k+1}\rangle - \sum_{i=1}^{K}\langle{v}_{i}|{w}_{k+1}\rangle|v_{i}\rangle}{\|{|}w_{k+1}\rangle-\sum_{i=1}^{K}\langle{v}_{i}|{w}_{k+1}\rangle{|}v_{i}\rangle\|}.
\end{equation}
The Gram Schmidt procedure produces the following orthonormal basis set,
\begin{equation}
\begin{split}
\label{eqn:OrthonormalSet}
&|{v}_{1}\rangle=\sum_{n=0}^{K}|\varphi_{n}\rangle,\\
&|{v}_{2}\rangle=\frac{2\sqrt{2}\sum_{n=0}^{K}(n-\frac{K}{2})|\varphi_{n}\rangle}{\sqrt{\prod_{n=0}^{1}(n+K)}},\\
&|{v}_{3}\rangle=\frac{8\sqrt{2}\sum_{n=0}^{K}((n-\frac{K}{2})^2-\frac{K}{8}(K+1))|\varphi_{n}\rangle}{\sqrt{\prod_{n=0}^{3}(n+K-1)}},\\
&|{v}_{4}\rangle=\frac{-\sqrt{2}}{\sqrt{\prod_{n=0}^{5}(n+K-2)}}\sum_{n=0}^{K}(K-2n)
(2+(-3+K)K+16n(n-K))|\varphi_n\rangle,\\
&|{v}_{5}\rangle=\frac{\sqrt{2}}{\sqrt{\prod_{n=0}^{7}(n+K-3)}}\sum_{n=0}^{K}(\prod_{n'=0}^{3}(K-n')
-32K(2+(K-1)K)n\\
&\hspace{47 mm}+32(2+K(5K-1))n^{2}
-256Kn^3+128n^4)|\varphi_{n}\rangle.
\end{split}
\end{equation}


\section{\label{app:SmallDeltaReducedMatricesAndDiagonalisation}Small diffusion regime: reduced HB density matrix entries and diagonalisation procedure}

In the small $\Delta$ approximation, the HB density matrix can be Taylor expanded to the second and fourth order in $\Delta$ about $\Delta=0$, as shown in Eqn.~\eqref{eqn:MixedHBStateTaylor2} and Eqn.~\eqref{eqn:MixedHBStateTaylor4}. The entries of $\hat{\varrho}'_{1,\text{S}\Delta}$ (see Eqn.~\eqref{eqn:SmallDelta3x3}) in terms of $K$ and $\Delta$ are
\begin{equation}
\begin{split}
&b_{11}=1-\frac{\Delta^2}{2}{K}{(K+1)},\\
&b_{13}=-\frac{\Delta^2}{{4\sqrt{2}}}\sqrt{\prod_{n=0}^{3}(n+K-1)},\\
&b_{22}=\frac{\Delta^2}{2}{K}{(K+1)},
\end{split}
\end{equation}
and the entries of $\hat{\varrho}^{\prime}_{2,\text{S}\Delta}$ (see Eqn.~\eqref{eqn:SmallDelta5x5}) are
\begin{equation}
\begin{split}
&c_{11}=1+\frac{\Delta^2}{32}{K}{(K+1)}
(-16+\Delta^2(-2+9K(K+1))),\\
&c_{13}=\frac{1}{8\sqrt{2}}\sqrt{\prod_{n=0}^{3}(n+K-1)}
(-2\Delta^2+\Delta^4(-1+2K(K+1))),\\
&c_{15}=\frac{\Delta^4}{64\sqrt{2}}\sqrt{\prod_{n=0}^{7}(n+K-3)},\\
&c_{22}=-\frac{\Delta^2}{8}K(K+1)
(-4+\Delta^2(-2+3K(K+1))),\\
&c_{24}=-\frac{\Delta^4}{16}\sqrt{K(K+1)\prod_{n=0}^{5}(n+K-2)},\\
&c_{33}=\frac{3\Delta^4}{32}\prod_{n=0}^{3}(n+K-1).
\end{split}
\end{equation}

The eigenvalue equation corresponding to $\hat{\varrho}'_{1,\text{S}\Delta}$ can be written as

\begin{equation}
\begin{pmatrix} b_{11} & 0 & b_{13} \\ 0 & b_{22} & 0 \\ b_{13} & 0 & 0\end{pmatrix}\left(
\begin{array}{c} x_{1} \\ x_{2} \\ x_{3} \end{array} \right)={E}^{(1)}\left(
\begin{array}{c} x_1 \\ x_2 \\ x_3\end{array}\right),
\label{eqn:EigenvalueEqn1}
\end{equation}
where $E^{(1)}$ and $|e^{(1)}\rangle=x_1|v_1\rangle+x_2|v_2\rangle+x_3|v_3\rangle$ are the eignenvalues and eigenvectors (the basis vectors are given in Eqn.~\eqref{eqn:OrthonormalSet}) of $\hat{\varrho}'_{1,\text{S}\Delta}$. The matrix equation gives 3 simultaneous equations of the form
\begin{equation}
\begin{split}
&b_{11}x_{1}+b_{13}x_{3}=E^{(1)}x_{1}\\
&b_{22}x_{2}=E^{(1)}x_{2}\\
&b_{13}x_1=E^{(1)}x_{3}.
\label{eqn:SimEqnRho1}
\end{split}
\end{equation}
The second equation in~\eqref{eqn:SimEqnRho1} gives the first eigenvalue and eigenvector of 
\begin{equation}
E_1^{(1)}=b_{22}
\end{equation}
and
\begin{equation}
|e_1^{(1)}\rangle=|v_2\rangle.
\end{equation}
The remaining two equations form a $2\times 2$ matrix with eigenvalue equation
\begin{equation}
\begin{pmatrix}b_{11} & b_{13} \\ b_{13} & 0\end{pmatrix}\left(
\begin{array}{c} x_{1} \\ x_{3} \end{array} \right)={E}^{(1)}\left(
\begin{array}{c} x_1 \\ x_3\end{array}\right)
\label{2x2EigenvalueEqn1}
\end{equation}
and characteristic polynomial of $(E^{(1)})^2-b_{11}E^{(1)}-b_{13}^2$. The resulting eigenvalues and the corresponding eigenvectors are 
\begin{equation}
\begin{split}
&E_2^{(1)}=\frac{b_{11}-\sqrt{b_{11}^{2}+4b_{13}^2}}{2}\\
&E_3^{(1)}=\frac{b_{11}+\sqrt{b_{11}^{2}+4b_{13}^2}}{2}\\
\end{split}
\end{equation}
and
\begin{equation}
\begin{split}
&|e_2^{(1)}\rangle=\frac{E_2^{(1)}}{b_{13}}|v_1\rangle+|v_3\rangle\\
&|e_3^{(1)}\rangle=\frac{E_3^{(1)}}{b_{13}}|v_1\rangle+|v_3\rangle.\\
\end{split}
\end{equation}
Eigenvectors $|e_{2}^{(1)}\rangle$ and $|e_{3}^{(1)}\rangle$ need to be normalised by dividing them by the magnitude of the vectors. 
$E_1^{(1)}$ and  $|e_1^{(1)}\rangle$ contain terms of order up to $\Delta^2$, however, $E_2^{(1)}$ and  $|e_2^{(1)}\rangle$, and $E_3^{(1)}$ and  $|e_3^{(1)}\rangle$ contain also terms of higher order. Additionally, eigenvalue $E_2^{(1)}$ appears to be negative, however, further Taylor expanding this eigenvalue reveals that it is in fact zero within the approximation. Taylor expansion of the eigenvalues and eigenvectors to the second order in $\Delta$ about $\Delta=0$ results in the following eigenvalues and normalised eigenvectors,
\begin{equation}
\begin{split}
&E_{1}^{\left(1\right)}=b_{22}\\
&E_{2}^{\left(1\right)}=0\\
&E_{3}^{\left(1\right)}=b_{11}\\
\label{eqn:SmallDeltaEigenvalues1}
\end{split}
\end{equation}
and
\begin{equation}
\begin{split}
&|e_1^{(1)}\rangle=|v_2\rangle\\
&|e_2^{(1)}\rangle=-b_{13}|v_1\rangle+|v_3\rangle\\
&|e_3^{(1)}\rangle=-|v_1\rangle{-}b_{13}|v_3\rangle.\\
\label{eqn:SmallDeltaEigenvectors1}
\end{split}
\end{equation}
These eigenvectors and eigenvalues reconstruct the HB density matrix accurate to the second order in $\Delta$. Therefore, Taylor expanding the evolved HB density matrix to the second order in $\Delta$ about $\Delta=0$ in fact produces a rank 2 matrix within the approximation which can be used when calculating $H_{11}$. Note that neglecting the higher order terms fits within the original constraint of $\Delta^{2}\ll{1}{/}{K^2}$.

The same procedure is applied to $\hat{\varrho}'_{2,\text{S}\Delta}$ which can  be described by the following 5 simultaneous equations,
\begin{equation}
\begin{split}
&c_{11}y_{1}+c_{13}y_{3}+c_{15}y_{5}=E^{(2)}y_{1}\\
&c_{22}y_{2}+c_{24}y_{4}=E^{(2)}y_{2}\\
&c_{13}y_{1}+c_{33}y_{3}=E^{(2)}y_{3}\\
&c_{24}y_{2}=E^{(2)}y_{4}\\
&c_{15}y_{1}=E^{(2)}y_{5},
\label{eqn:SimEqnRho2}
\end{split}
\end{equation}
where $E^{(2)}$ and $|e^{(2)}\rangle=y_1|v_1\rangle+y_2|v_2\rangle+y_3|v_3\rangle+y_4|v_4\rangle+y_5|v_5\rangle)$ are the eignenvalues and eigenvectors of $\hat{\varrho}'_{2,\text{S}\Delta}$. Since $y_{2}$ and $y_{4}$ occur only in the second and fourth equation of ~\eqref{eqn:SimEqnRho2} and $y_{1}$, $y_{3}$ and $y_{5}$ in the first, third and fifth equation of ~\eqref{eqn:SimEqnRho2} then $\hat{\varrho'}_{2,\text{S}\Delta}$ can be written as a direct sum of $2\times 2$ and $3\times 3$ matrices with the following eigenvalue equations,
\begin{equation}
\begin{pmatrix}c_{22} & c_{24} \\ c_{24} & 0\end{pmatrix}\left(
\begin{array}{c} y_{2} \\ y_{4} \end{array} \right)={E}^{(2)}\left(
\begin{array}{c} y_2 \\ y_4\end{array}\right)
\label{eqn:2x2EigenvalueEqn2}
\end{equation}
and
\begin{equation}
\begin{pmatrix}c_{11} & c_{13} & c_{15} \\ c_{13} & c_{33} & 0 \\ c_{15} & 0 & 0\end{pmatrix}\left(
\begin{array}{c} y_{1} \\ y_{3} \\ y_{5} \end{array} \right)={E}^{(2)}\left(
\begin{array}{c} y_1 \\ y_3 \\ y_5\end{array}\right).
\label{eqn:3x3EigenvalueEqn2}
\end{equation}
The $2\times 2$ matrix gives eigenvalues
\begin{equation}
\begin{split}
&E_1^{(2)}=\frac{c_{22}-\sqrt{c_{22}^2+4c_{24}^2}}{2}\\
&E_2^{(2)}=\frac{c_{22}+\sqrt{c_{22}^2+4c_{24}^2}}{2}
\label{eqn:2x2EigenvalueSolution}
\end{split}
\end{equation}
with the corresponding eigenvectors
\begin{equation}
\begin{split}
&|e_1^{(2)}\rangle=\frac{E_1^{(2)}}{c_{24}}|v_2\rangle+|v_4\rangle,\\
&|e_2^{(2)}\rangle=\frac{E_2^{(2)}}{c_{24}}|v_2\rangle+|v_4\rangle,
\label{eqn:2x2EigenvectorSolution}
\end{split}
\end{equation}
where the eigenvectors need to be normalised by dividing them by the vector magnitude.

The $3\times 3$ matrix has a characteristic polynomial $\alpha$ of the form of a cubic equation
\begin{equation}
\alpha=k(E^{(2)})^3+l(E^{(2)})^2+mE^{(2)}+n,
\label{eqn:CharacteristicPolynomial3x3}
\end{equation}
where $k=-1$, $l=(c_{33}+c_{11})$, $m=(-c_{11}c_{33}+c_{15}^2+c_{13}^2)$ and $n=-c_{15}^2c_{33}$. The resulting eigenvalue solutions and the corresponding eigenvectors are
\begin{equation}
\begin{split}
&E_i^{(2)}=-\frac{1}{3k}\left(l+u_{i}C+\frac{\zeta_{0}}{u_{i}C}\right)\\
\label{eqn:3x3EigenvalueSolution}
\end{split}
\end{equation}
and
\begin{equation}
\begin{split}
|e_i^{(2)}\rangle=&\frac{E_i^{(2)}}{c_{15}}|v_1\rangle
+\frac{1}{c_{13}}\left(\frac{(E_i^{(2)})^2}{c_{15}}-c_{15}-\frac{c_{11}E_i^{(2)}}{c_{15}}\right)|v_3\rangle
+|v_5\rangle,
\label{eqn:3x3EigenvectorSolution}
\end{split}
\end{equation}
where $i=\{3,4,5\}$, $u_1=1$, $u_2=\left({-1+\i\sqrt{3}}\right){/}{2}$, $u_3=\left({-1-\i\sqrt{3}}\right){/}{2}$, $C=\left(\left({\zeta_{1}+\sqrt{-27k^2\zeta}}\right){/}{2}\right)^{{1}{/}{3}}$, $\zeta_0=(l^2-3km)$, $\zeta_1=(2l^3-9klm+27k^2n)$ and $\zeta=(18klmn-4l^3n+l^2m^2-4km^3-27k^2n^2)$. This time, eigenvalues $E_1^{(2)}$ and $E_4^{(2)}$ appear to be negative, however, Taylor expanding the eigenvalues to the fourth order in $\Delta$ about $\Delta=0$, we get
\begin{equation}
\begin{split}
&E_1^{(2)}=0\\
&E_2^{(2)}=c_{22}\\
&E_3^{(2)}=c_{11}+\frac{K(1+K)(-2+K(1+K))\Delta^4}{32}\\
&E_4^{(2)}=0\\
&E_5^{(2)}=\frac{2}{3}c_{33}.\\
\label{eqn:4thOrderSmallDeltaEigenvalues}
\end{split}
\end{equation}
The corresponding normalised eigenvectors expanded to the fourth order in $\Delta$ are not shown since they are complicated functions of $K$ and $\Delta$. The zero eigenvalues and the corresponding eigenvectors do not contribute when reconstructing the density matrix accurate to the fourth order in $\Delta$. The remaining non-zero eigenvalues ($E_2^{(2)}$, $E_3^{(2)}$ and $E_5^{(2)}$) and the corresponding eigenvectors reconstruct the density matrix accurate to the fourth order in $\Delta$ when using their full forms given in Eqn.~\eqref{eqn:2x2EigenvalueSolution}, Eqn.~\eqref{eqn:2x2EigenvectorSolution}, Eqn.~\eqref{eqn:3x3EigenvalueSolution} and Eqn.~\eqref{eqn:3x3EigenvectorSolution}. The reason to use the full forms is that the eigenvectors (corresponding to $E_2^{(2)}$, $E_3^{(2)}$ and $E_5^{(2)}$) contain $\Delta^{-4}$ terms and therefore higher order terms in the expansion are required to ensure that we do not lose the fourth order terms in $\Delta$ due to cancellations. 

These eigenvalues and eigenvectors can then be used to calculate the diagonal elements of the QFI matrix by using Eqn.~\eqref{eqn:QFIEigenbasis}. These elements are given in Eqn.~\eqref{eqn:H11SmallDelta} and Eqn.~\eqref{eqn:H22SmallDelta} of the main text.

\begin{figure}[t!]
\begin{center}
\includegraphics[trim=1.5cm 7cm 1.5cm 6.5cm,width=0.5\columnwidth]{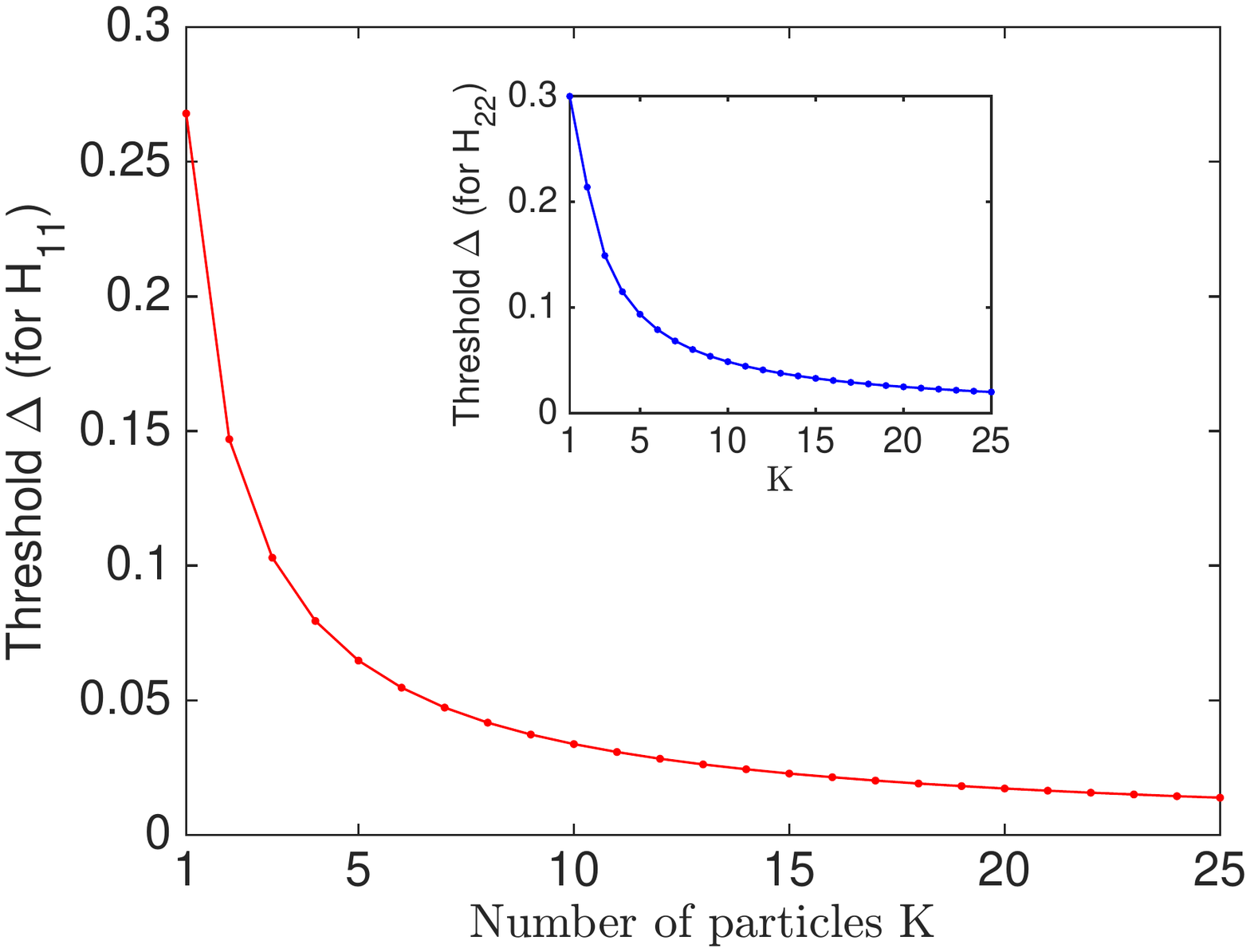}
\end{center}
\caption{Phase diffusion validity plot for the small phase diffusion approximation for HB states. The figure shows the threshold values of $\Delta$ as a function of particle number ($K$) when the highest allowed relative error on the QFI ($\delta{H}_{ii}/H_{ii}$) is $0.05$.}
\label{fig:QFI_SmallDeltaValidity}
\end{figure}

Fig.~(\ref{fig:QFI_SmallDeltaValidity}) represents the phase diffusion validity plot for the small $\Delta$ approximation. It shows the maximum allowed phase diffusion values as a function of $K$ when the highest allowed relative error on the QFI ($\delta{H}_{ii}/H_{ii}$) is $0.05$.

\section{\label{app:SmallDeltaTradeoff}Trade-off in the joint estimation of $\phi$ and $\Delta$ for HB states as $\Delta\rightarrow{0}$}
\vskip-0.1in
Similarly to the approach taken when calculating the trade-off in the large phase diffusion approximation (shown in Appendix~(\ref{app:LargeDeltaTradeoff})), we use the HB density matrix expanded to the second order in $\Delta$ about $\Delta=0$ (Eqn.~\eqref{eqn:MixedHBStateTaylor2}) and a complete POVM set $\{\pi_{y}\}$ composed of projectors $|v_y\rangle=\sum_{n=0}^{K}r_{y,n}\e^{\i{X_{y,n}}}|2n,2K-2n\rangle$ to calculate the probability associated with the $y^{\text{th}}$ outcome. The probability and its corresponding derivatives with respect to $\phi$ and $\Delta$ are,
\begin{align}
\begin{split}
p_{y}=\langle{v}_{y}|\hat{\varrho}_{1,\text{S}\Delta}|v_{y}\rangle
=\sum_{n,n'=0}^{K}R_{y,n,n'}
(1-2\Delta^2(n-n')^2)\cos{(\Theta_{y,n,n'})},\\
\end{split}
\end{align}
\begin{equation}
\begin{split}
\partial_{\phi}{p_y}=-2\sum_{n,n'=0}^{K}R_{y,n,n'}(1-2\Delta^{2}(n-n')^2)
(n-n')\sin{\left(\Theta_{y,n,n'}\right)}
\end{split}
\end{equation}
and
\begin{equation}
\begin{split}
\partial_{\Delta}{p_y}=&-4\Delta\sum_{n,n'=0}^{K}R_{y,n,n'}(n-n')^2\cos{\left(\Theta_{y,n,n'}\right)},
\end{split}
\end{equation}
where $\Theta_{y,n,n'}=\left(X_{y,n'}-X_{y,n}+2\phi{(n-n')}\right)$ and $R_{y,n,n'}=r_{y,n}r_{y,n'}b_{n}b_{n'}$. The associated diagonal elements of the Fisher information matrix, $F_{11}$ and $F_{22}$, corresponding to the individual estimation of $\phi$ and $\Delta$ respectively are,
\begin{widetext}
\begin{equation}
\begin{split}
F_{11}&=\sum_{y=0}^{K}\left(\frac{4\left(\sum_{n,n'=0}^{K}R_{y,n,n'}\sin{\left(\Theta_{y,n,n'}\right)}\left(n-n'\right)\left(1-2\Delta^2\left(n-n'\right)^2\right)\right)^2}{\sum_{n,n'=0}^{K}R_{y,n,n'}\cos{\left(\Theta_{y,n,n'}\right)}\left(1-2\Delta^2\left(n-n'\right)^2\right)}\right)\\
&\approx\sum_{y=0}^{K}\left(\frac{4\left(\sum_{n,n'=0}^{K}R_{y,n,n'}\sin{\left(\Theta_{y,n,n'}\right)}\left(n-n'\right)\right)^2}{\sum_{n,n'=0}^{K}R_{y,n,n'}\cos{\left(\Theta_{y,n,n'}\right)}}\right)\\
\end{split}
\end{equation} 

and

\begin{equation}
\begin{split}
F_{22}&=\sum_{y=0}^{K}\left(\frac{16\Delta^2\left(\sum_{n,n'=0}^{K}R_{y,n,n'}\cos{\left(\Theta_{y,n,n'}\right)}\left(n-n'\right)^2\right)^2}{\sum_{n,n'=0}^{K}R_{y,n,n'}\cos{\left(\Theta_{y,n,n'}\right)}\left(1-2\Delta^2\left(n-n'\right)^2\right)}\right)\\
&\approx\sum_{y=0}^{K}\left(\frac{16\Delta^2\left(\sum_{n,n'=0}^{K}R_{y,n,n'}\cos{\left(\Theta_{y,n,n'}\right)}\left(n-n'\right)^2\right)^2}{\sum_{n,n'=0}^{K}R_{y,n,n'}\cos{\left(\Theta_{y,n,n'}\right)}}\right),\\
\end{split}
\end{equation}
where the approximations are made assuming that $\Delta\rightarrow{0}$ so that terms containing $2\Delta^2\left(n-n'\right)^2$ are negligible compared to $1$. Using the QFI expressions found in Eqn.~\eqref{eqn:H11SmallDelta} and Eqn.~\eqref{eqn:H22SmallDelta} and assuming $\Delta\rightarrow{0}$ so that terms containing $2K(K+1)\Delta^2$ are negligible compared to $1$, we can compute $\operatorname{Tr}\left[\textbf{F}\textbf{H}^{-1}\right]$
\begin{equation}
\begin{split}
&\operatorname{Tr}\left[\textbf{F}\textbf{H}^{-1}\right]=\\
&=\frac{2}{K(K+1)}\sum_{y=0}^{K}\frac{\left(\sum_{n,n'=0}^{K}R_{y,n,n'}\sin{\left(\Theta_{y,n,n'}\right)}\left(n-n'\right)\right)^2+4\Delta^2\left(\sum_{n,n'=0}^{K}R_{y,n,n'}\cos{\left(\Theta_{y,n,n'}\right)}\left(n-n'\right)^2\right)^2}{\sum_{n,n'=0}^{K}R_{y,n,n'}\cos{\left(\Theta_{y,n,n'}\right)}}.\\
\label{eqn:SmallDeltaTradeOff}
\end{split}
\end{equation}

\end{widetext}
The $\Delta$ term appearing in the trade-off in Eqn.~\eqref{eqn:SmallDeltaTradeOff} cannot be simply neglected (by invoking $\Delta\rightarrow{0}$) because we require to compare terms of the form: $\sin{\left(\Theta_{y,n,n'}\right)}\sin{\left(\Theta_{y,m,m'}\right)}$ and $4\Delta^2\cos{\left(\Theta_{y,n,n'}\right)}\cos{\left(\Theta_{y,m,m'}\right)}\left(n-n'\right)\left(m-m'\right)$. If the phase of the POVM has some $\Delta$ dependence then it is not straightforward which term dominates. In fact, to obtain significant Fisher information for the diffusion parameter for decreasing values of $\Delta$, we would expect some $\Delta$ dependence in the POVM to reduce the effect of the $\Delta^2$ factor. Without knowing further structure of the optimal POVM, standard optimisation is a difficult task to perform. However, the numerical results based on the simulated annealing algorithm suggest that as $\Delta\rightarrow{0}$ the trade-off approaches $1$ (see Fig.~(\ref{fig:BoundMaxSmallDelta}) of the main text). 

\bibliographystyle{aipnum4-1}
\bibliography{BibliographyJointEstimationOfPhaseAndPhaseDiffusion}

\begin{thebibliography}{59}%
\makeatletter
\providecommand \@ifxundefined [1]{%
 \@ifx{#1\undefined}
}%
\providecommand \@ifnum [1]{%
 \ifnum #1\expandafter \@firstoftwo
 \else \expandafter \@secondoftwo
 \fi
}%
\providecommand \@ifx [1]{%
 \ifx #1\expandafter \@firstoftwo
 \else \expandafter \@secondoftwo
 \fi
}%
\providecommand \natexlab [1]{#1}%
\providecommand \enquote  [1]{``#1''}%
\providecommand \bibnamefont  [1]{#1}%
\providecommand \bibfnamefont [1]{#1}%
\providecommand \citenamefont [1]{#1}%
\providecommand \href@noop [0]{\@secondoftwo}%
\providecommand \href [0]{\begingroup \@sanitize@url \@href}%
\providecommand \@href[1]{\@@startlink{#1}\@@href}%
\providecommand \@@href[1]{\endgroup#1\@@endlink}%
\providecommand \@sanitize@url [0]{\catcode `\\12\catcode `\$12\catcode
  `\&12\catcode `\#12\catcode `\^12\catcode `\_12\catcode `\%12\relax}%
\providecommand \@@startlink[1]{}%
\providecommand \@@endlink[0]{}%
\providecommand \url  [0]{\begingroup\@sanitize@url \@url }%
\providecommand \@url [1]{\endgroup\@href {#1}{\urlprefix }}%
\providecommand \urlprefix  [0]{URL }%
\providecommand \Eprint [0]{\href }%
\providecommand \doibase [0]{http://dx.doi.org/}%
\providecommand \selectlanguage [0]{\@gobble}%
\providecommand \bibinfo  [0]{\@secondoftwo}%
\providecommand \bibfield  [0]{\@secondoftwo}%
\providecommand \translation [1]{[#1]}%
\providecommand \BibitemOpen [0]{}%
\providecommand \bibitemStop [0]{}%
\providecommand \bibitemNoStop [0]{.\EOS\space}%
\providecommand \EOS [0]{\spacefactor3000\relax}%
\providecommand \BibitemShut  [1]{\csname bibitem#1\endcsname}%
\let\auto@bib@innerbib\@empty
\bibitem [{\citenamefont {Caves}\ \emph {et~al.}(1980)\citenamefont {Caves},
  \citenamefont {Thorne}, \citenamefont {Drever}, \citenamefont {Sandberg},\
  and\ \citenamefont {Zimmermann}}]{CavesRMP1980}%
  \BibitemOpen
  \bibfield  {author} {\bibinfo {author} {\bibfnamefont {C.~M.}\ \bibnamefont
  {Caves}}, \bibinfo {author} {\bibfnamefont {K.~S.}\ \bibnamefont {Thorne}},
  \bibinfo {author} {\bibfnamefont {R.~W.~P.}\ \bibnamefont {Drever}}, \bibinfo
  {author} {\bibfnamefont {V.~D.}\ \bibnamefont {Sandberg}}, \ and\ \bibinfo
  {author} {\bibfnamefont {M.}~\bibnamefont {Zimmermann}},\ }\href {\doibase
  10.1103/RevModPhys.52.341} {\bibfield  {journal} {\bibinfo  {journal} {Rev.
  Mod. Phys.}\ }\textbf {\bibinfo {volume} {52}},\ \bibinfo {pages} {341}
  (\bibinfo {year} {1980})}\BibitemShut {NoStop}%
\bibitem [{\citenamefont {Caves}(1981)}]{Caves1981}%
  \BibitemOpen
  \bibfield  {author} {\bibinfo {author} {\bibfnamefont {C.~M.}\ \bibnamefont
  {Caves}},\ }\href {http://link.aps.org/doi/10.1103/PhysRevD.23.1693}
  {\bibfield  {journal} {\bibinfo  {journal} {Phys. Rev. D}\ }\textbf {\bibinfo
  {volume} {23}},\ \bibinfo {pages} {1693} (\bibinfo {year}
  {1981})}\BibitemShut {NoStop}%
\bibitem [{\citenamefont {Blair}\ \emph {et~al.}(2015)\citenamefont {Blair},
  \citenamefont {Ju}, \citenamefont {Zhao}, \citenamefont {Wen},\ and\
  \citenamefont {et. al}}]{Blair2016}%
  \BibitemOpen
  \bibfield  {author} {\bibinfo {author} {\bibfnamefont {D.}~\bibnamefont
  {Blair}}, \bibinfo {author} {\bibfnamefont {L.}~\bibnamefont {Ju}}, \bibinfo
  {author} {\bibfnamefont {C.}~\bibnamefont {Zhao}}, \bibinfo {author}
  {\bibfnamefont {L.}~\bibnamefont {Wen}}, \ and\ \bibinfo {author}
  {\bibnamefont {et. al}},\ }\href {\doibase 10.1007/s11433-015-5748-6}
  {\bibfield  {journal} {\bibinfo  {journal} {Sci. China-Phys. Mech. Astron.}\
  }\textbf {\bibinfo {volume} {58}},\ \bibinfo {pages} {120402} (\bibinfo
  {year} {2015})}\BibitemShut {NoStop}%
\bibitem [{\citenamefont {Taylor}\ and\ \citenamefont
  {Bowen}(2016)}]{Taylor2014}%
  \BibitemOpen
  \bibfield  {author} {\bibinfo {author} {\bibfnamefont {M.~A.}\ \bibnamefont
  {Taylor}}\ and\ \bibinfo {author} {\bibfnamefont {W.~P.}\ \bibnamefont
  {Bowen}},\ }\href {\doibase https://doi.org/10.1016/j.physrep.2015.12.002}
  {\bibfield  {journal} {\bibinfo  {journal} {Phys. Rep.}\ }\textbf {\bibinfo
  {volume} {615}},\ \bibinfo {pages} {1 } (\bibinfo {year} {2016})}\BibitemShut
  {NoStop}%
\bibitem [{\citenamefont {Giovannetti}, \citenamefont {Lloyd},\ and\
  \citenamefont {Maccone}(2004)}]{Giovannetti2004}%
  \BibitemOpen
  \bibfield  {author} {\bibinfo {author} {\bibfnamefont {V.}~\bibnamefont
  {Giovannetti}}, \bibinfo {author} {\bibfnamefont {S.}~\bibnamefont {Lloyd}},
  \ and\ \bibinfo {author} {\bibfnamefont {L.}~\bibnamefont {Maccone}},\ }\href
  {http://www.sciencemag.org/content/306/5700/1330.full.pdf} {\bibfield
  {journal} {\bibinfo  {journal} {Science}\ }\textbf {\bibinfo {volume}
  {306}},\ \bibinfo {pages} {1330} (\bibinfo {year} {2004})}\BibitemShut
  {NoStop}%
\bibitem [{\citenamefont {Giovannetti}, \citenamefont {Lloyd},\ and\
  \citenamefont {Maccone}(2011)}]{Giovannetti2011}%
  \BibitemOpen
  \bibfield  {author} {\bibinfo {author} {\bibfnamefont {V.}~\bibnamefont
  {Giovannetti}}, \bibinfo {author} {\bibfnamefont {S.}~\bibnamefont {Lloyd}},
  \ and\ \bibinfo {author} {\bibfnamefont {L.}~\bibnamefont {Maccone}},\ }\href
  {http://www.nature.com/doifinder/10.1038/nphoton.2011.35} {\bibfield
  {journal} {\bibinfo  {journal} {Nat. Photon.}\ }\textbf {\bibinfo {volume}
  {5}},\ \bibinfo {pages} {222} (\bibinfo {year} {2011})}\BibitemShut {NoStop}%
\bibitem [{\citenamefont {Demkowicz-Dobrza{\'n}ski}, \citenamefont {Jarzyna},\
  and\ \citenamefont {Ko\l{}ody\ifmmode~\acute{n}\else
  \'{n}\fi{}ski}(2015)}]{Demkowicz-Dobrzanski2014}%
  \BibitemOpen
  \bibfield  {author} {\bibinfo {author} {\bibfnamefont {R.}~\bibnamefont
  {Demkowicz-Dobrza{\'n}ski}}, \bibinfo {author} {\bibfnamefont
  {M.}~\bibnamefont {Jarzyna}}, \ and\ \bibinfo {author} {\bibfnamefont
  {J.}~\bibnamefont {Ko\l{}ody\ifmmode~\acute{n}\else \'{n}\fi{}ski}},\ }\href
  {\doibase http://dx.doi.org/10.1016/bs.po.2015.02.003} {\bibfield  {journal}
  {\bibinfo  {journal} {Progress in Optics}\ }\textbf {\bibinfo {volume}
  {60}},\ \bibinfo {pages} {345 } (\bibinfo {year} {2015})}\BibitemShut
  {NoStop}%
\bibitem [{\citenamefont {Escher}, \citenamefont {de~Matos~Filho},\ and\
  \citenamefont {Davidovich}(2011)}]{EscherNatPhys2011}%
  \BibitemOpen
  \bibfield  {author} {\bibinfo {author} {\bibfnamefont {B.~M.}\ \bibnamefont
  {Escher}}, \bibinfo {author} {\bibfnamefont {R.~L.}\ \bibnamefont
  {de~Matos~Filho}}, \ and\ \bibinfo {author} {\bibfnamefont {L.}~\bibnamefont
  {Davidovich}},\ }\href {\doibase 10.1038/nphys1958} {\bibfield  {journal}
  {\bibinfo  {journal} {Nat. Phys.}\ }\textbf {\bibinfo {volume} {7}},\
  \bibinfo {pages} {406} (\bibinfo {year} {2011})}\BibitemShut {NoStop}%
\bibitem [{\citenamefont {Demkowicz-Dobrza{\'n}ski}, \citenamefont
  {Ko{\l}ody{\'n}ski},\ and\ \citenamefont
  {Gu{\c{t}}{\u{a}}}(2012)}]{Demkowicz-DobrzanskiNatComm2012}%
  \BibitemOpen
  \bibfield  {author} {\bibinfo {author} {\bibfnamefont {R.}~\bibnamefont
  {Demkowicz-Dobrza{\'n}ski}}, \bibinfo {author} {\bibfnamefont
  {J.}~\bibnamefont {Ko{\l}ody{\'n}ski}}, \ and\ \bibinfo {author}
  {\bibfnamefont {M.}~\bibnamefont {Gu{\c{t}}{\u{a}}}},\ }\href {\doibase
  10.1038/ncomms2067} {\bibfield  {journal} {\bibinfo  {journal} {Nat.
  Commun.}\ }\textbf {\bibinfo {volume} {3}},\ \bibinfo {pages} {1063}
  (\bibinfo {year} {2012})}\BibitemShut {NoStop}%
\bibitem [{\citenamefont {Demkowicz-Dobrza{\'n}ski}\ \emph
  {et~al.}(2009)\citenamefont {Demkowicz-Dobrza{\'n}ski}, \citenamefont
  {Dorner}, \citenamefont {Smith}, \citenamefont {Lundeen}, \citenamefont
  {Wasilewski}, \citenamefont {Banaszek},\ and\ \citenamefont
  {Walmsley}}]{Demkowicz-Dobrzanski2009}%
  \BibitemOpen
  \bibfield  {author} {\bibinfo {author} {\bibfnamefont {R.}~\bibnamefont
  {Demkowicz-Dobrza{\'n}ski}}, \bibinfo {author} {\bibfnamefont
  {U.}~\bibnamefont {Dorner}}, \bibinfo {author} {\bibfnamefont {B.~J.}\
  \bibnamefont {Smith}}, \bibinfo {author} {\bibfnamefont {J.~S.}\ \bibnamefont
  {Lundeen}}, \bibinfo {author} {\bibfnamefont {W.}~\bibnamefont {Wasilewski}},
  \bibinfo {author} {\bibfnamefont {K.}~\bibnamefont {Banaszek}}, \ and\
  \bibinfo {author} {\bibfnamefont {I.~A.}\ \bibnamefont {Walmsley}},\ }\href
  {\doibase 10.1103/PhysRevA.80.013825} {\bibfield  {journal} {\bibinfo
  {journal} {Phys. Rev. A}\ }\textbf {\bibinfo {volume} {80}},\ \bibinfo
  {pages} {013825} (\bibinfo {year} {2009})}\BibitemShut {NoStop}%
\bibitem [{\citenamefont {Dorner}\ \emph {et~al.}(2009)\citenamefont {Dorner},
  \citenamefont {Demkowicz-Dobrza{\'n}ski}, \citenamefont {Smith},
  \citenamefont {Lundeen}, \citenamefont {Wasilewski}, \citenamefont
  {Banaszek},\ and\ \citenamefont {Walmsley}}]{Dorner2009}%
  \BibitemOpen
  \bibfield  {author} {\bibinfo {author} {\bibfnamefont {U.}~\bibnamefont
  {Dorner}}, \bibinfo {author} {\bibfnamefont {R.}~\bibnamefont
  {Demkowicz-Dobrza{\'n}ski}}, \bibinfo {author} {\bibfnamefont {B.~J.}\
  \bibnamefont {Smith}}, \bibinfo {author} {\bibfnamefont {J.~S.}\ \bibnamefont
  {Lundeen}}, \bibinfo {author} {\bibfnamefont {W.}~\bibnamefont {Wasilewski}},
  \bibinfo {author} {\bibfnamefont {K.}~\bibnamefont {Banaszek}}, \ and\
  \bibinfo {author} {\bibfnamefont {I.~A.}\ \bibnamefont {Walmsley}},\ }\href
  {\doibase 10.1103/PhysRevLett.102.040403} {\bibfield  {journal} {\bibinfo
  {journal} {Phys. Rev. Lett.}\ }\textbf {\bibinfo {volume} {102}},\ \bibinfo
  {pages} {040403} (\bibinfo {year} {2009})}\BibitemShut {NoStop}%
\bibitem [{\citenamefont {Kacprowicz}\ \emph {et~al.}(2009)\citenamefont
  {Kacprowicz}, \citenamefont {Demkowicz-Dobrza{\'n}ski}, \citenamefont
  {Wasilewski}, \citenamefont {Banaszek},\ and\ \citenamefont
  {Walmsley}}]{Kacprowicz2009}%
  \BibitemOpen
  \bibfield  {author} {\bibinfo {author} {\bibfnamefont {M.}~\bibnamefont
  {Kacprowicz}}, \bibinfo {author} {\bibfnamefont {R.}~\bibnamefont
  {Demkowicz-Dobrza{\'n}ski}}, \bibinfo {author} {\bibfnamefont
  {W.}~\bibnamefont {Wasilewski}}, \bibinfo {author} {\bibfnamefont
  {K.}~\bibnamefont {Banaszek}}, \ and\ \bibinfo {author} {\bibfnamefont
  {I.~A.}\ \bibnamefont {Walmsley}},\ }\href {\doibase 10.1038/nphoton.2010.39}
  {\bibfield  {journal} {\bibinfo  {journal} {Nat. Photon.}\ }\textbf {\bibinfo
  {volume} {4}},\ \bibinfo {pages} {357} (\bibinfo {year} {2009})}\BibitemShut
  {NoStop}%
\bibitem [{\citenamefont {Datta}\ \emph {et~al.}(2011)\citenamefont {Datta},
  \citenamefont {Zhang}, \citenamefont {Thomas-Peter}, \citenamefont {Dorner},
  \citenamefont {Smith},\ and\ \citenamefont {Walmsley}}]{Datta2011}%
  \BibitemOpen
  \bibfield  {author} {\bibinfo {author} {\bibfnamefont {A.}~\bibnamefont
  {Datta}}, \bibinfo {author} {\bibfnamefont {L.}~\bibnamefont {Zhang}},
  \bibinfo {author} {\bibfnamefont {N.}~\bibnamefont {Thomas-Peter}}, \bibinfo
  {author} {\bibfnamefont {U.}~\bibnamefont {Dorner}}, \bibinfo {author}
  {\bibfnamefont {B.~J.}\ \bibnamefont {Smith}}, \ and\ \bibinfo {author}
  {\bibfnamefont {I.~A.}\ \bibnamefont {Walmsley}},\ }\href {\doibase
  10.1103/PhysRevA.83.063836} {\bibfield  {journal} {\bibinfo  {journal} {Phys.
  Rev. A - Atomic, Molecular, and Optical Physics}\ }\textbf {\bibinfo {volume}
  {83}},\ \bibinfo {pages} {1} (\bibinfo {year} {2011})}\BibitemShut {NoStop}%
\bibitem [{\citenamefont {Brivio}\ \emph {et~al.}(2010)\citenamefont {Brivio},
  \citenamefont {Cialdi}, \citenamefont {Vezzoli}, \citenamefont {Gebrehiwot},
  \citenamefont {Genoni}, \citenamefont {Olivares},\ and\ \citenamefont
  {Paris}}]{Brivio2010}%
  \BibitemOpen
  \bibfield  {author} {\bibinfo {author} {\bibfnamefont {D.}~\bibnamefont
  {Brivio}}, \bibinfo {author} {\bibfnamefont {S.}~\bibnamefont {Cialdi}},
  \bibinfo {author} {\bibfnamefont {S.}~\bibnamefont {Vezzoli}}, \bibinfo
  {author} {\bibfnamefont {B.~T.}\ \bibnamefont {Gebrehiwot}}, \bibinfo
  {author} {\bibfnamefont {M.~G.}\ \bibnamefont {Genoni}}, \bibinfo {author}
  {\bibfnamefont {S.}~\bibnamefont {Olivares}}, \ and\ \bibinfo {author}
  {\bibfnamefont {M.~G.~A.}\ \bibnamefont {Paris}},\ }\href {\doibase
  10.1103/PhysRevA.81.012305} {\bibfield  {journal} {\bibinfo  {journal} {Phys.
  Rev. A}\ }\textbf {\bibinfo {volume} {81}},\ \bibinfo {pages} {012305}
  (\bibinfo {year} {2010})}\BibitemShut {NoStop}%
\bibitem [{\citenamefont {Genoni}, \citenamefont {Olivares},\ and\
  \citenamefont {Paris}(2011)}]{Genoni2011}%
  \BibitemOpen
  \bibfield  {author} {\bibinfo {author} {\bibfnamefont {M.~G.}\ \bibnamefont
  {Genoni}}, \bibinfo {author} {\bibfnamefont {S.}~\bibnamefont {Olivares}}, \
  and\ \bibinfo {author} {\bibfnamefont {M.~G.~A.}\ \bibnamefont {Paris}},\
  }\href {\doibase 10.1103/PhysRevLett.106.153603} {\bibfield  {journal}
  {\bibinfo  {journal} {Phys. Rev. Lett.}\ }\textbf {\bibinfo {volume} {106}},\
  \bibinfo {pages} {153603} (\bibinfo {year} {2011})}\BibitemShut {NoStop}%
\bibitem [{\citenamefont {Genoni}\ \emph {et~al.}(2012)\citenamefont {Genoni},
  \citenamefont {Olivares}, \citenamefont {Brivio}, \citenamefont {Cialdi},
  \citenamefont {Cipriani}, \citenamefont {Santamato}, \citenamefont
  {Vezzoli},\ and\ \citenamefont {Paris}}]{Genoni2012}%
  \BibitemOpen
  \bibfield  {author} {\bibinfo {author} {\bibfnamefont {M.~G.}\ \bibnamefont
  {Genoni}}, \bibinfo {author} {\bibfnamefont {S.}~\bibnamefont {Olivares}},
  \bibinfo {author} {\bibfnamefont {D.}~\bibnamefont {Brivio}}, \bibinfo
  {author} {\bibfnamefont {S.}~\bibnamefont {Cialdi}}, \bibinfo {author}
  {\bibfnamefont {D.}~\bibnamefont {Cipriani}}, \bibinfo {author}
  {\bibfnamefont {A.}~\bibnamefont {Santamato}}, \bibinfo {author}
  {\bibfnamefont {S.}~\bibnamefont {Vezzoli}}, \ and\ \bibinfo {author}
  {\bibfnamefont {M.~G.~A.}\ \bibnamefont {Paris}},\ }\href {\doibase
  10.1103/PhysRevA.85.043817} {\bibfield  {journal} {\bibinfo  {journal} {Phys.
  Rev. A}\ }\textbf {\bibinfo {volume} {85}},\ \bibinfo {pages} {043817}
  (\bibinfo {year} {2012})}\BibitemShut {NoStop}%
\bibitem [{\citenamefont {Degen}, \citenamefont {Reinhard},\ and\ \citenamefont
  {Cappellaro}(2017)}]{Degen2016}%
  \BibitemOpen
  \bibfield  {author} {\bibinfo {author} {\bibfnamefont {C.~L.}\ \bibnamefont
  {Degen}}, \bibinfo {author} {\bibfnamefont {F.}~\bibnamefont {Reinhard}}, \
  and\ \bibinfo {author} {\bibfnamefont {P.}~\bibnamefont {Cappellaro}},\
  }\href {\doibase 10.1103/RevModPhys.89.035002} {\bibfield  {journal}
  {\bibinfo  {journal} {Rev. Mod. Phys.}\ }\textbf {\bibinfo {volume} {89}},\
  \bibinfo {pages} {035002} (\bibinfo {year} {2017})}\BibitemShut {NoStop}%
\bibitem [{\citenamefont {Escher}\ \emph {et~al.}(2012)\citenamefont {Escher},
  \citenamefont {Davidovich}, \citenamefont {Zagury},\ and\ \citenamefont
  {de~Matos~Filho}}]{Escher2012}%
  \BibitemOpen
  \bibfield  {author} {\bibinfo {author} {\bibfnamefont {B.~M.}\ \bibnamefont
  {Escher}}, \bibinfo {author} {\bibfnamefont {L.}~\bibnamefont {Davidovich}},
  \bibinfo {author} {\bibfnamefont {N.}~\bibnamefont {Zagury}}, \ and\ \bibinfo
  {author} {\bibfnamefont {R.~L.}\ \bibnamefont {de~Matos~Filho}},\ }\href
  {\doibase 10.1103/PhysRevLett.109.190404} {\bibfield  {journal} {\bibinfo
  {journal} {Phys. Rev. Lett.}\ }\textbf {\bibinfo {volume} {109}},\ \bibinfo
  {pages} {190404} (\bibinfo {year} {2012})}\BibitemShut {NoStop}%
\bibitem [{\citenamefont {Cole}\ and\ \citenamefont
  {Hollenberg}(2009)}]{Cole2009}%
  \BibitemOpen
  \bibfield  {author} {\bibinfo {author} {\bibfnamefont {J.~H.}\ \bibnamefont
  {Cole}}\ and\ \bibinfo {author} {\bibfnamefont {L.~C.~L.}\ \bibnamefont
  {Hollenberg}},\ }\href {http://stacks.iop.org/0957-4484/20/i=49/a=495401}
  {\bibfield  {journal} {\bibinfo  {journal} {Nanotechnology}\ }\textbf
  {\bibinfo {volume} {20}},\ \bibinfo {pages} {495401} (\bibinfo {year}
  {2009})}\BibitemShut {NoStop}%
\bibitem [{\citenamefont {Schirhagl}\ \emph {et~al.}(2014)\citenamefont
  {Schirhagl}, \citenamefont {Chang}, \citenamefont {Loretz},\ and\
  \citenamefont {Degen}}]{Schirhagl2014}%
  \BibitemOpen
  \bibfield  {author} {\bibinfo {author} {\bibfnamefont {R.}~\bibnamefont
  {Schirhagl}}, \bibinfo {author} {\bibfnamefont {K.}~\bibnamefont {Chang}},
  \bibinfo {author} {\bibfnamefont {M.}~\bibnamefont {Loretz}}, \ and\ \bibinfo
  {author} {\bibfnamefont {C.~L.}\ \bibnamefont {Degen}},\ }\href {\doibase
  10.1146/annurev-physchem-040513-103659} {\bibfield  {journal} {\bibinfo
  {journal} {Annu. Rev. Phys. Chem.}\ }\textbf {\bibinfo {volume} {65}},\
  \bibinfo {pages} {83} (\bibinfo {year} {2014})}\BibitemShut {NoStop}%
\bibitem [{\citenamefont {Stace}(2010)}]{Stace2010}%
  \BibitemOpen
  \bibfield  {author} {\bibinfo {author} {\bibfnamefont {T.~M.}\ \bibnamefont
  {Stace}},\ }\href {\doibase 10.1103/PhysRevA.82.011611} {\bibfield  {journal}
  {\bibinfo  {journal} {Phys. Rev. A}\ }\textbf {\bibinfo {volume} {82}},\
  \bibinfo {pages} {011611} (\bibinfo {year} {2010})}\BibitemShut {NoStop}%
\bibitem [{\citenamefont {Kucsko}\ \emph {et~al.}(2013)\citenamefont {Kucsko},
  \citenamefont {Maurer}, \citenamefont {Yao}, \citenamefont {Kubo},
  \citenamefont {Noh}, \citenamefont {Lo}, \citenamefont {Park},\ and\
  \citenamefont {Lukin}}]{Kucsko2013}%
  \BibitemOpen
  \bibfield  {author} {\bibinfo {author} {\bibfnamefont {G.}~\bibnamefont
  {Kucsko}}, \bibinfo {author} {\bibfnamefont {P.}~\bibnamefont {Maurer}},
  \bibinfo {author} {\bibfnamefont {N.~Y.}\ \bibnamefont {Yao}}, \bibinfo
  {author} {\bibfnamefont {M.}~\bibnamefont {Kubo}}, \bibinfo {author}
  {\bibfnamefont {H.}~\bibnamefont {Noh}}, \bibinfo {author} {\bibfnamefont
  {P.}~\bibnamefont {Lo}}, \bibinfo {author} {\bibfnamefont {H.}~\bibnamefont
  {Park}}, \ and\ \bibinfo {author} {\bibfnamefont {M.~D.}\ \bibnamefont
  {Lukin}},\ }\href {\doibase 10.1038/nature12373} {\bibfield  {journal}
  {\bibinfo  {journal} {Nature}\ }\textbf {\bibinfo {volume} {500}},\ \bibinfo
  {pages} {54} (\bibinfo {year} {2013})}\BibitemShut {NoStop}%
\bibitem [{\citenamefont {Toyli}\ \emph {et~al.}(2013)\citenamefont {Toyli},
  \citenamefont {de~las Casas}, \citenamefont {Christle}, \citenamefont
  {Dobrovitski},\ and\ \citenamefont {Awschalom}}]{Toyli2013}%
  \BibitemOpen
  \bibfield  {author} {\bibinfo {author} {\bibfnamefont {D.~M.}\ \bibnamefont
  {Toyli}}, \bibinfo {author} {\bibfnamefont {C.~F.}\ \bibnamefont {de~las
  Casas}}, \bibinfo {author} {\bibfnamefont {D.~J.}\ \bibnamefont {Christle}},
  \bibinfo {author} {\bibfnamefont {V.~V.}\ \bibnamefont {Dobrovitski}}, \ and\
  \bibinfo {author} {\bibfnamefont {D.~D.}\ \bibnamefont {Awschalom}},\ }\href
  {\doibase 10.1073/pnas.1306825110} {\bibfield  {journal} {\bibinfo  {journal}
  {PNAS}\ }\textbf {\bibinfo {volume} {110}},\ \bibinfo {pages} {8417}
  (\bibinfo {year} {2013})}\BibitemShut {NoStop}%
\bibitem [{\citenamefont {Xie}, \citenamefont {Xu},\ and\ \citenamefont
  {Wang}(2017)}]{Xie2016}%
  \BibitemOpen
  \bibfield  {author} {\bibinfo {author} {\bibfnamefont {D.}~\bibnamefont
  {Xie}}, \bibinfo {author} {\bibfnamefont {C.}~\bibnamefont {Xu}}, \ and\
  \bibinfo {author} {\bibfnamefont {A.~M.}\ \bibnamefont {Wang}},\ }\href
  {\doibase 10.1007/s11128-017-1605-z} {\bibfield  {journal} {\bibinfo
  {journal} {Quantum Information Processing}\ }\textbf {\bibinfo {volume}
  {16}},\ \bibinfo {pages} {155} (\bibinfo {year} {2017})}\BibitemShut
  {NoStop}%
\bibitem [{\citenamefont {Aspelmeyer}, \citenamefont {Kippenberg},\ and\
  \citenamefont {Marquardt}(2014)}]{Aspelmeyer2014}%
  \BibitemOpen
  \bibfield  {author} {\bibinfo {author} {\bibfnamefont {M.}~\bibnamefont
  {Aspelmeyer}}, \bibinfo {author} {\bibfnamefont {T.~J.}\ \bibnamefont
  {Kippenberg}}, \ and\ \bibinfo {author} {\bibfnamefont {F.}~\bibnamefont
  {Marquardt}},\ }\href {\doibase 10.1103/RevModPhys.86.1391} {\bibfield
  {journal} {\bibinfo  {journal} {Rev. Mod. Phys.}\ }\textbf {\bibinfo {volume}
  {86}},\ \bibinfo {pages} {1391} (\bibinfo {year} {2014})}\BibitemShut
  {NoStop}%
\bibitem [{\citenamefont {Brawley}\ \emph {et~al.}(2016)\citenamefont
  {Brawley}, \citenamefont {Vanner}, \citenamefont {Larsen}, \citenamefont
  {Schmid}, \citenamefont {Boisen},\ and\ \citenamefont {Bowen}}]{Brawley2016}%
  \BibitemOpen
  \bibfield  {author} {\bibinfo {author} {\bibfnamefont {G.~A.}\ \bibnamefont
  {Brawley}}, \bibinfo {author} {\bibfnamefont {M.~R.}\ \bibnamefont {Vanner}},
  \bibinfo {author} {\bibfnamefont {P.~E.}\ \bibnamefont {Larsen}}, \bibinfo
  {author} {\bibfnamefont {S.}~\bibnamefont {Schmid}}, \bibinfo {author}
  {\bibfnamefont {A.}~\bibnamefont {Boisen}}, \ and\ \bibinfo {author}
  {\bibfnamefont {W.~P.}\ \bibnamefont {Bowen}},\ }\href {\doibase
  10.1038/ncomms10988} {\bibfield  {journal} {\bibinfo  {journal} {Nat.
  Commun.}\ }\textbf {\bibinfo {volume} {7}},\ \bibinfo {pages} {10988}
  (\bibinfo {year} {2016})}\BibitemShut {NoStop}%
\bibitem [{\citenamefont {Cronin}, \citenamefont {Schmiedmayer},\ and\
  \citenamefont {Pritchard}(2009)}]{Cronin2009}%
  \BibitemOpen
  \bibfield  {author} {\bibinfo {author} {\bibfnamefont {A.~D.}\ \bibnamefont
  {Cronin}}, \bibinfo {author} {\bibfnamefont {J.}~\bibnamefont
  {Schmiedmayer}}, \ and\ \bibinfo {author} {\bibfnamefont {D.~E.}\
  \bibnamefont {Pritchard}},\ }\href {\doibase 10.1103/RevModPhys.81.1051}
  {\bibfield  {journal} {\bibinfo  {journal} {Rev. Mod. Phys.}\ }\textbf
  {\bibinfo {volume} {81}},\ \bibinfo {pages} {1051} (\bibinfo {year}
  {2009})}\BibitemShut {NoStop}%
\bibitem [{\citenamefont {Musha}, \citenamefont {ichi Kamimura},\ and\
  \citenamefont {Nakazawa}(1982)}]{Musha1982}%
  \BibitemOpen
  \bibfield  {author} {\bibinfo {author} {\bibfnamefont {T.}~\bibnamefont
  {Musha}}, \bibinfo {author} {\bibfnamefont {J.}~\bibnamefont {ichi
  Kamimura}}, \ and\ \bibinfo {author} {\bibfnamefont {M.}~\bibnamefont
  {Nakazawa}},\ }\href {\doibase 10.1364/AO.21.000694} {\bibfield  {journal}
  {\bibinfo  {journal} {Appl. Opt.}\ }\textbf {\bibinfo {volume} {21}},\
  \bibinfo {pages} {694} (\bibinfo {year} {1982})}\BibitemShut {NoStop}%
\bibitem [{\citenamefont {Du}\ and\ \citenamefont {He}(2013)}]{Du2013}%
  \BibitemOpen
  \bibfield  {author} {\bibinfo {author} {\bibfnamefont {J.}~\bibnamefont
  {Du}}\ and\ \bibinfo {author} {\bibfnamefont {Z.}~\bibnamefont {He}},\ }\href
  {\doibase 10.1364/OE.21.027111} {\bibfield  {journal} {\bibinfo  {journal}
  {Opt. Express}\ }\textbf {\bibinfo {volume} {21}},\ \bibinfo {pages} {27111}
  (\bibinfo {year} {2013})}\BibitemShut {NoStop}%
\bibitem [{\citenamefont {Berrada}\ \emph {et~al.}(2013)\citenamefont
  {Berrada}, \citenamefont {van Frank}, \citenamefont {Schumm}, \citenamefont
  {Schaff},\ and\ \citenamefont {Schmiedmayer}}]{Berrada2013}%
  \BibitemOpen
  \bibfield  {author} {\bibinfo {author} {\bibfnamefont {T.}~\bibnamefont
  {Berrada}}, \bibinfo {author} {\bibfnamefont {S.}~\bibnamefont {van Frank}},
  \bibinfo {author} {\bibfnamefont {T.}~\bibnamefont {Schumm}}, \bibinfo
  {author} {\bibfnamefont {J.-F.}\ \bibnamefont {Schaff}}, \ and\ \bibinfo
  {author} {\bibfnamefont {J.}~\bibnamefont {Schmiedmayer}},\ }\href {\doibase
  10.1038/ncomms3077} {\bibfield  {journal} {\bibinfo  {journal} {Nat.
  Commun.}\ }\textbf {\bibinfo {volume} {4}},\ \bibinfo {pages} {2077}
  (\bibinfo {year} {2013})}\BibitemShut {NoStop}%
\bibitem [{\citenamefont {Fang}\ and\ \citenamefont {Qin}(2012)}]{Fang2012}%
  \BibitemOpen
  \bibfield  {author} {\bibinfo {author} {\bibfnamefont {J.~C.}\ \bibnamefont
  {Fang}}\ and\ \bibinfo {author} {\bibfnamefont {J.}~\bibnamefont {Qin}},\
  }\href {\doibase 10.3390/s120506331} {\bibfield  {journal} {\bibinfo
  {journal} {Sensors}\ }\textbf {\bibinfo {volume} {12}},\ \bibinfo {pages}
  {6331} (\bibinfo {year} {2012})}\BibitemShut {NoStop}%
\bibitem [{\citenamefont {Rosi}\ \emph {et~al.}(2014)\citenamefont {Rosi},
  \citenamefont {Sorrentino}, \citenamefont {Cacciapuoti}, \citenamefont
  {Prevedelli},\ and\ \citenamefont {Tino}}]{Rosi2014}%
  \BibitemOpen
  \bibfield  {author} {\bibinfo {author} {\bibfnamefont {G.}~\bibnamefont
  {Rosi}}, \bibinfo {author} {\bibfnamefont {F.}~\bibnamefont {Sorrentino}},
  \bibinfo {author} {\bibfnamefont {L.}~\bibnamefont {Cacciapuoti}}, \bibinfo
  {author} {\bibfnamefont {M.}~\bibnamefont {Prevedelli}}, \ and\ \bibinfo
  {author} {\bibfnamefont {G.~M.}\ \bibnamefont {Tino}},\ }\href {\doibase
  10.1038/nature13433} {\bibfield  {journal} {\bibinfo  {journal} {Nature}\
  }\textbf {\bibinfo {volume} {510}},\ \bibinfo {pages} {518} (\bibinfo {year}
  {2014})}\BibitemShut {NoStop}%
\bibitem [{\citenamefont {Pikovski}, \citenamefont {Zych},\ and\ \citenamefont
  {Costa}(2015)}]{Pikovski2015}%
  \BibitemOpen
  \bibfield  {author} {\bibinfo {author} {\bibfnamefont {I.}~\bibnamefont
  {Pikovski}}, \bibinfo {author} {\bibfnamefont {M.}~\bibnamefont {Zych}}, \
  and\ \bibinfo {author} {\bibfnamefont {F.}~\bibnamefont {Costa}},\ }\href
  {\doibase 10.1038/NPHYS3366} {\bibfield  {journal} {\bibinfo  {journal}
  {Nature Phys.}\ }\textbf {\bibinfo {volume} {11}},\ \bibinfo {pages} {668}
  (\bibinfo {year} {2015})}\BibitemShut {NoStop}%
\bibitem [{\citenamefont {Oniga}\ and\ \citenamefont {Wang}(2016)}]{Oniga2016}%
  \BibitemOpen
  \bibfield  {author} {\bibinfo {author} {\bibfnamefont {T.}~\bibnamefont
  {Oniga}}\ and\ \bibinfo {author} {\bibfnamefont {C.~H.-T.}\ \bibnamefont
  {Wang}},\ }\href {\doibase 10.1103/PhysRevD.93.044027} {\bibfield  {journal}
  {\bibinfo  {journal} {Phys. Rev. D}\ }\textbf {\bibinfo {volume} {93}},\
  \bibinfo {pages} {044027} (\bibinfo {year} {2016})}\BibitemShut {NoStop}%
\bibitem [{\citenamefont {Genoni}, \citenamefont {Duarte},\ and\ \citenamefont
  {Serafini}(2016)}]{Genoni2016}%
  \BibitemOpen
  \bibfield  {author} {\bibinfo {author} {\bibfnamefont {M.~G.}\ \bibnamefont
  {Genoni}}, \bibinfo {author} {\bibfnamefont {O.~S.}\ \bibnamefont {Duarte}},
  \ and\ \bibinfo {author} {\bibfnamefont {A.}~\bibnamefont {Serafini}},\
  }\href {http://stacks.iop.org/1367-2630/18/i=10/a=103040} {\bibfield
  {journal} {\bibinfo  {journal} {New J. Phys.}\ }\textbf {\bibinfo {volume}
  {18}},\ \bibinfo {pages} {103040} (\bibinfo {year} {2016})}\BibitemShut
  {NoStop}%
\bibitem [{\citenamefont {Riedel}(2013)}]{Riedel2013}%
  \BibitemOpen
  \bibfield  {author} {\bibinfo {author} {\bibfnamefont {C.~J.}\ \bibnamefont
  {Riedel}},\ }\href {\doibase 10.1103/PhysRevD.88.116005} {\bibfield
  {journal} {\bibinfo  {journal} {Phys. Rev. D}\ }\textbf {\bibinfo {volume}
  {88}},\ \bibinfo {pages} {116005} (\bibinfo {year} {2013})}\BibitemShut
  {NoStop}%
\bibitem [{\citenamefont {Knysh}\ and\ \citenamefont
  {Durkin}(2013)}]{Knysh2013}%
  \BibitemOpen
  \bibfield  {author} {\bibinfo {author} {\bibfnamefont {S.~I.}\ \bibnamefont
  {Knysh}}\ and\ \bibinfo {author} {\bibfnamefont {G.~A.}\ \bibnamefont
  {Durkin}},\ }\href {http://arxiv.org/abs/1307.0470} {\bibfield  {journal}
  {\bibinfo  {journal} {arXiv:1307.0470}\ } (\bibinfo {year}
  {2013})}\BibitemShut {NoStop}%
\bibitem [{\citenamefont {Vidrighin}\ \emph {et~al.}(2014)\citenamefont
  {Vidrighin}, \citenamefont {Donati}, \citenamefont {Genoni}, \citenamefont
  {Jin}, \citenamefont {Kolthammer}, \citenamefont {Kim}, \citenamefont
  {Datta}, \citenamefont {Barbieri},\ and\ \citenamefont
  {Walmsley}}]{Vidrighin2014}%
  \BibitemOpen
  \bibfield  {author} {\bibinfo {author} {\bibfnamefont {M.~D.}\ \bibnamefont
  {Vidrighin}}, \bibinfo {author} {\bibfnamefont {G.}~\bibnamefont {Donati}},
  \bibinfo {author} {\bibfnamefont {M.~G.}\ \bibnamefont {Genoni}}, \bibinfo
  {author} {\bibfnamefont {X.-M.}\ \bibnamefont {Jin}}, \bibinfo {author}
  {\bibfnamefont {W.~S.}\ \bibnamefont {Kolthammer}}, \bibinfo {author}
  {\bibfnamefont {M.~S.}\ \bibnamefont {Kim}}, \bibinfo {author} {\bibfnamefont
  {A.}~\bibnamefont {Datta}}, \bibinfo {author} {\bibfnamefont
  {M.}~\bibnamefont {Barbieri}}, \ and\ \bibinfo {author} {\bibfnamefont
  {I.~A.}\ \bibnamefont {Walmsley}},\ }\href
  {http://www.ncbi.nlm.nih.gov/pubmed/24727938} {\bibfield  {journal} {\bibinfo
   {journal} {Nat. Commun.}\ }\textbf {\bibinfo {volume} {5}},\ \bibinfo
  {pages} {3532} (\bibinfo {year} {2014})}\BibitemShut {NoStop}%
\bibitem [{\citenamefont {Altorio}\ \emph {et~al.}(2015)\citenamefont
  {Altorio}, \citenamefont {Genoni}, \citenamefont {Vidrighin}, \citenamefont
  {Somma},\ and\ \citenamefont {Barbieri}}]{Altorio2015}%
  \BibitemOpen
  \bibfield  {author} {\bibinfo {author} {\bibfnamefont {M.}~\bibnamefont
  {Altorio}}, \bibinfo {author} {\bibfnamefont {M.~G.}\ \bibnamefont {Genoni}},
  \bibinfo {author} {\bibfnamefont {M.~D.}\ \bibnamefont {Vidrighin}}, \bibinfo
  {author} {\bibfnamefont {F.}~\bibnamefont {Somma}}, \ and\ \bibinfo {author}
  {\bibfnamefont {M.}~\bibnamefont {Barbieri}},\ }\href {\doibase
  10.1103/PhysRevA.92.032114} {\bibfield  {journal} {\bibinfo  {journal} {Phys.
  Rev. A}\ }\textbf {\bibinfo {volume} {92}},\ \bibinfo {pages} {032114}
  (\bibinfo {year} {2015})}\BibitemShut {NoStop}%
\bibitem [{\citenamefont {Ragy}, \citenamefont {Jarzyna},\ and\ \citenamefont
  {Demkowicz-Dobrza\ifmmode~\acute{n}\else \'{n}\fi{}ski}(2016)}]{Ragy2016}%
  \BibitemOpen
  \bibfield  {author} {\bibinfo {author} {\bibfnamefont {S.}~\bibnamefont
  {Ragy}}, \bibinfo {author} {\bibfnamefont {M.}~\bibnamefont {Jarzyna}}, \
  and\ \bibinfo {author} {\bibfnamefont {R.}~\bibnamefont
  {Demkowicz-Dobrza\ifmmode~\acute{n}\else \'{n}\fi{}ski}},\ }\href {\doibase
  10.1103/PhysRevA.94.052108} {\bibfield  {journal} {\bibinfo  {journal} {Phys.
  Rev. A}\ }\textbf {\bibinfo {volume} {94}},\ \bibinfo {pages} {052108}
  (\bibinfo {year} {2016})}\BibitemShut {NoStop}%
\bibitem [{\citenamefont {Szczykulska}, \citenamefont {Baumgratz},\ and\
  \citenamefont {Datta}(2016)}]{Szczykulska2016}%
  \BibitemOpen
  \bibfield  {author} {\bibinfo {author} {\bibfnamefont {M.}~\bibnamefont
  {Szczykulska}}, \bibinfo {author} {\bibfnamefont {T.}~\bibnamefont
  {Baumgratz}}, \ and\ \bibinfo {author} {\bibfnamefont {A.}~\bibnamefont
  {Datta}},\ }\href {\doibase 10.1080/23746149.2016.1230476} {\bibfield
  {journal} {\bibinfo  {journal} {Advances in Physics: X}\ }\textbf {\bibinfo
  {volume} {1}},\ \bibinfo {pages} {621} (\bibinfo {year} {2016})}\BibitemShut
  {NoStop}%
\bibitem [{\citenamefont {Helstrom}(1976)}]{Helstrom1976}%
  \BibitemOpen
  \bibfield  {author} {\bibinfo {author} {\bibfnamefont {C.}~\bibnamefont
  {Helstrom}},\ }\href@noop {} {\emph {\bibinfo {title} {Quantum Detection and
  Estimation Theory}}},\ Mathematics in Science and Engineering : a series of
  monographs and textbooks\ (\bibinfo  {publisher} {London: Academic Press
  Inc.},\ \bibinfo {year} {1976})\BibitemShut {NoStop}%
\bibitem [{\citenamefont {Braunstein}\ and\ \citenamefont
  {Caves}(1994)}]{Braunstein1994}%
  \BibitemOpen
  \bibfield  {author} {\bibinfo {author} {\bibfnamefont {S.~L.}\ \bibnamefont
  {Braunstein}}\ and\ \bibinfo {author} {\bibfnamefont {C.~M.}\ \bibnamefont
  {Caves}},\ }\href {\doibase 10.1103/PhysRevLett.72.3439} {\bibfield
  {journal} {\bibinfo  {journal} {Phys. Rev. Lett.}\ }\textbf {\bibinfo
  {volume} {72}},\ \bibinfo {pages} {3439} (\bibinfo {year}
  {1994})}\BibitemShut {NoStop}%
\bibitem [{\citenamefont {Gill}\ and\ \citenamefont {Massar}(2000)}]{Gill2000}%
  \BibitemOpen
  \bibfield  {author} {\bibinfo {author} {\bibfnamefont {R.~D.}\ \bibnamefont
  {Gill}}\ and\ \bibinfo {author} {\bibfnamefont {S.}~\bibnamefont {Massar}},\
  }\href {\doibase 10.1103/PhysRevA.61.042312} {\bibfield  {journal} {\bibinfo
  {journal} {Phys. Rev. A}\ }\textbf {\bibinfo {volume} {61}},\ \bibinfo
  {pages} {042312} (\bibinfo {year} {2000})}\BibitemShut {NoStop}%
\bibitem [{\citenamefont {Li}\ \emph {et~al.}(2016)\citenamefont {Li},
  \citenamefont {Ferrie}, \citenamefont {Gross}, \citenamefont {Kalev},\ and\
  \citenamefont {Caves}}]{Li2016}%
  \BibitemOpen
  \bibfield  {author} {\bibinfo {author} {\bibfnamefont {N.}~\bibnamefont
  {Li}}, \bibinfo {author} {\bibfnamefont {C.}~\bibnamefont {Ferrie}}, \bibinfo
  {author} {\bibfnamefont {J.~A.}\ \bibnamefont {Gross}}, \bibinfo {author}
  {\bibfnamefont {A.}~\bibnamefont {Kalev}}, \ and\ \bibinfo {author}
  {\bibfnamefont {C.~M.}\ \bibnamefont {Caves}},\ }\href {\doibase
  10.1103/PhysRevLett.116.180402} {\bibfield  {journal} {\bibinfo  {journal}
  {Phys. Rev. Lett.}\ }\textbf {\bibinfo {volume} {116}},\ \bibinfo {pages}
  {180402} (\bibinfo {year} {2016})}\BibitemShut {NoStop}%
\bibitem [{\citenamefont {Holland}\ and\ \citenamefont
  {Burnett}(1993)}]{Holland1993}%
  \BibitemOpen
  \bibfield  {author} {\bibinfo {author} {\bibfnamefont {M.~J.}\ \bibnamefont
  {Holland}}\ and\ \bibinfo {author} {\bibfnamefont {K.}~\bibnamefont
  {Burnett}},\ }\href {\doibase 10.1103/PhysRevLett.71.1355} {\bibfield
  {journal} {\bibinfo  {journal} {Phys. Rev. Lett.}\ }\textbf {\bibinfo
  {volume} {71}},\ \bibinfo {pages} {1355} (\bibinfo {year}
  {1993})}\BibitemShut {NoStop}%
\bibitem [{\citenamefont {Jarzyna}\ and\ \citenamefont
  {Demkowicz-Dobrza\ifmmode~\acute{n}\else \'{n}\fi{}ski}(2015)}]{Jarzyna2015}%
  \BibitemOpen
  \bibfield  {author} {\bibinfo {author} {\bibfnamefont {M.}~\bibnamefont
  {Jarzyna}}\ and\ \bibinfo {author} {\bibfnamefont {R.}~\bibnamefont
  {Demkowicz-Dobrza\ifmmode~\acute{n}\else \'{n}\fi{}ski}},\ }\href
  {http://stacks.iop.org/1367-2630/17/i=1/a=013010} {\bibfield  {journal}
  {\bibinfo  {journal} {New Journal of Physics}\ }\textbf {\bibinfo {volume}
  {17}},\ \bibinfo {pages} {013010} (\bibinfo {year} {2015})}\BibitemShut
  {NoStop}%
\bibitem [{\citenamefont {Bohachevsky}, \citenamefont {Johnson},\ and\
  \citenamefont {Stein}(1986)}]{Bohachevsky1986}%
  \BibitemOpen
  \bibfield  {author} {\bibinfo {author} {\bibfnamefont {I.}~\bibnamefont
  {Bohachevsky}}, \bibinfo {author} {\bibfnamefont {M.}~\bibnamefont
  {Johnson}}, \ and\ \bibinfo {author} {\bibfnamefont {M.}~\bibnamefont
  {Stein}},\ }\href {\doibase 10.2307/1269076} {\bibfield  {journal} {\bibinfo
  {journal} {Technometrics}\ }\textbf {\bibinfo {volume} {28}},\ \bibinfo
  {pages} {209} (\bibinfo {year} {1986})}\BibitemShut {NoStop}%
\bibitem [{\citenamefont {Cover}\ and\ \citenamefont
  {Thomas}(2006)}]{Cover2006}%
  \BibitemOpen
  \bibfield  {author} {\bibinfo {author} {\bibfnamefont {T.~M.}\ \bibnamefont
  {Cover}}\ and\ \bibinfo {author} {\bibfnamefont {J.~A.}\ \bibnamefont
  {Thomas}},\ }\href@noop {} {\emph {\bibinfo {title} {Elements of Information
  Theory}}},\ Wiley Series in Telecommunications and Signal Processing\
  (\bibinfo  {publisher} {Hoboken, NJ: Wiley},\ \bibinfo {year}
  {2006})\BibitemShut {NoStop}%
\bibitem [{\citenamefont {Paris}(2009)}]{Paris2009}%
  \BibitemOpen
  \bibfield  {author} {\bibinfo {author} {\bibfnamefont {M.~G.~A.}\
  \bibnamefont {Paris}},\ }\href {\doibase
  http://dx.doi.org/10.1142/S0219749909004839} {\bibfield  {journal} {\bibinfo
  {journal} {Int. J. Quantum Inf.}\ }\textbf {\bibinfo {volume} {07}},\
  \bibinfo {pages} {125} (\bibinfo {year} {2009})}\BibitemShut {NoStop}%
\bibitem [{\citenamefont {Roccia}\ \emph {et~al.}(2017)\citenamefont {Roccia},
  \citenamefont {Gianani}, \citenamefont {Mancino}, \citenamefont {Sbroscia},
  \citenamefont {Somma}, \citenamefont {Genoni},\ and\ \citenamefont
  {Barbieri}}]{Roccia2017}%
  \BibitemOpen
  \bibfield  {author} {\bibinfo {author} {\bibfnamefont {E.}~\bibnamefont
  {Roccia}}, \bibinfo {author} {\bibfnamefont {I.}~\bibnamefont {Gianani}},
  \bibinfo {author} {\bibfnamefont {L.}~\bibnamefont {Mancino}}, \bibinfo
  {author} {\bibfnamefont {M.}~\bibnamefont {Sbroscia}}, \bibinfo {author}
  {\bibfnamefont {F.}~\bibnamefont {Somma}}, \bibinfo {author} {\bibfnamefont
  {M.~G.}\ \bibnamefont {Genoni}}, \ and\ \bibinfo {author} {\bibfnamefont
  {M.}~\bibnamefont {Barbieri}},\ }\href {https://arxiv.org/abs/1704.03327}
  {\bibfield  {journal} {\bibinfo  {journal} {arXiv:1704.03327}\ } (\bibinfo
  {year} {2017})}\BibitemShut {NoStop}%
\bibitem [{\citenamefont {Dempster}, \citenamefont {Laird},\ and\ \citenamefont
  {Rubin}(1977)}]{Dempster1977}%
  \BibitemOpen
  \bibfield  {author} {\bibinfo {author} {\bibfnamefont {A.~P.}\ \bibnamefont
  {Dempster}}, \bibinfo {author} {\bibfnamefont {N.~M.}\ \bibnamefont {Laird}},
  \ and\ \bibinfo {author} {\bibfnamefont {D.~B.}\ \bibnamefont {Rubin}},\
  }\href {http://www.jstor.org/stable/2984875} {\bibfield  {journal} {\bibinfo
  {journal} {J. Roy. Statist. Soc. (B)}\ }\textbf {\bibinfo {volume} {39}},\
  \bibinfo {pages} {1} (\bibinfo {year} {1977})}\BibitemShut {NoStop}%
\bibitem [{\citenamefont {Wiseman}\ and\ \citenamefont
  {Killip}(1998)}]{Wiseman1998}%
  \BibitemOpen
  \bibfield  {author} {\bibinfo {author} {\bibfnamefont {H.~M.}\ \bibnamefont
  {Wiseman}}\ and\ \bibinfo {author} {\bibfnamefont {R.~B.}\ \bibnamefont
  {Killip}},\ }\href {\doibase 10.1103/PhysRevA.57.2169} {\bibfield  {journal}
  {\bibinfo  {journal} {Phys. Rev. A}\ }\textbf {\bibinfo {volume} {57}},\
  \bibinfo {pages} {2169} (\bibinfo {year} {1998})}\BibitemShut {NoStop}%
\bibitem [{\citenamefont {Pellonp\"a\"a}\ and\ \citenamefont
  {Schultz}(2013)}]{Schultz2013}%
  \BibitemOpen
  \bibfield  {author} {\bibinfo {author} {\bibfnamefont {J.-P.}\ \bibnamefont
  {Pellonp\"a\"a}}\ and\ \bibinfo {author} {\bibfnamefont {J.}~\bibnamefont
  {Schultz}},\ }\href {\doibase 10.1103/PhysRevA.88.012121} {\bibfield
  {journal} {\bibinfo  {journal} {Phys. Rev. A}\ }\textbf {\bibinfo {volume}
  {88}},\ \bibinfo {pages} {012121} (\bibinfo {year} {2013})}\BibitemShut
  {NoStop}%
\bibitem [{\citenamefont {Crowley}\ \emph {et~al.}(2014)\citenamefont
  {Crowley}, \citenamefont {Datta}, \citenamefont {Barbieri},\ and\
  \citenamefont {Walmsley}}]{Crowley2014}%
  \BibitemOpen
  \bibfield  {author} {\bibinfo {author} {\bibfnamefont {P.~J.~D.}\
  \bibnamefont {Crowley}}, \bibinfo {author} {\bibfnamefont {A.}~\bibnamefont
  {Datta}}, \bibinfo {author} {\bibfnamefont {M.}~\bibnamefont {Barbieri}}, \
  and\ \bibinfo {author} {\bibfnamefont {I.~A.}\ \bibnamefont {Walmsley}},\
  }\href {http://link.aps.org/doi/10.1103/PhysRevA.89.023845} {\bibfield
  {journal} {\bibinfo  {journal} {Physical Review A - Atomic, Molecular, and
  Optical Physics}\ }\textbf {\bibinfo {volume} {89}},\ \bibinfo {pages} {1}
  (\bibinfo {year} {2014})}\BibitemShut {NoStop}%
\bibitem [{\citenamefont {Ballester}(2004)}]{Ballester2004}%
  \BibitemOpen
  \bibfield  {author} {\bibinfo {author} {\bibfnamefont {M.~A.}\ \bibnamefont
  {Ballester}},\ }\href {\doibase 10.1103/PhysRevA.69.022303} {\bibfield
  {journal} {\bibinfo  {journal} {Phys. Rev. A}\ }\textbf {\bibinfo {volume}
  {69}},\ \bibinfo {pages} {022303} (\bibinfo {year} {2004})}\BibitemShut
  {NoStop}%
\bibitem [{\citenamefont {Humphreys}\ \emph {et~al.}(2013)\citenamefont
  {Humphreys}, \citenamefont {Barbieri}, \citenamefont {Datta},\ and\
  \citenamefont {Walmsley}}]{Humphreys2013}%
  \BibitemOpen
  \bibfield  {author} {\bibinfo {author} {\bibfnamefont {P.~C.}\ \bibnamefont
  {Humphreys}}, \bibinfo {author} {\bibfnamefont {M.}~\bibnamefont {Barbieri}},
  \bibinfo {author} {\bibfnamefont {A.}~\bibnamefont {Datta}}, \ and\ \bibinfo
  {author} {\bibfnamefont {I.~A.}\ \bibnamefont {Walmsley}},\ }\href
  {http://link.aps.org/doi/10.1103/PhysRevLett.111.070403} {\bibfield
  {journal} {\bibinfo  {journal} {Phys. Rev. Lett.}\ }\textbf {\bibinfo
  {volume} {111}},\ \bibinfo {pages} {1} (\bibinfo {year} {2013})}\BibitemShut
  {NoStop}%
\bibitem [{\citenamefont {Baumgratz}\ and\ \citenamefont
  {Datta}(2016)}]{Baumgratz2016}%
  \BibitemOpen
  \bibfield  {author} {\bibinfo {author} {\bibfnamefont {T.}~\bibnamefont
  {Baumgratz}}\ and\ \bibinfo {author} {\bibfnamefont {A.}~\bibnamefont
  {Datta}},\ }\href {\doibase 10.1103/PhysRevLett.116.030801} {\bibfield
  {journal} {\bibinfo  {journal} {Phys. Rev. Lett.}\ }\textbf {\bibinfo
  {volume} {116}},\ \bibinfo {pages} {030801} (\bibinfo {year}
  {2016})}\BibitemShut {NoStop}%
\bibitem [{\citenamefont {Gagatsos}, \citenamefont {Branford},\ and\
  \citenamefont {Datta}(2016)}]{Gagatsos2016}%
  \BibitemOpen
  \bibfield  {author} {\bibinfo {author} {\bibfnamefont {C.}~\bibnamefont
  {Gagatsos}}, \bibinfo {author} {\bibfnamefont {D.}~\bibnamefont {Branford}},
  \ and\ \bibinfo {author} {\bibfnamefont {A.}~\bibnamefont {Datta}},\ }\href
  {http://link.aps.org/doi/10.1103/PhysRevA.94.042342} {\bibfield  {journal}
  {\bibinfo  {journal} {Phys. Rev. A}\ }\textbf {\bibinfo {volume} {94}},\
  \bibinfo {pages} {042342} (\bibinfo {year} {2016})}\BibitemShut {NoStop}%
\end{thebibliography}%
\maketitle
\end{document}